\documentclass{article}
\usepackage{arxiv}

\usepackage{textcomp}
\usepackage{lscape}
\usepackage{pdflscape}
\usepackage{graphicx}

\usepackage{pdfpages}

\usepackage{multibib}

\usepackage[utf8]{inputenc} 
\usepackage[T1]{fontenc}    

\usepackage {hyperref}       
\hypersetup{colorlinks,citecolor=black,linkcolor=black, urlcolor=black}

\usepackage{url}            
\usepackage{booktabs}       
\usepackage{amsfonts}       
\usepackage{nicefrac}       
\usepackage{microtype}      
\usepackage{lipsum}        

\usepackage{fancyhdr}
\title{Computer-Aided Data Mining: Automating a Novel Knowledge Discovery and Data mining Process Model for Metabolomics}

\author{
  Ahmed ~BaniMustafa\thanks{Corresponding author \texttt{banimustafa@gmail.com}. The authors would like to thank the department of computer science at Aberystwyth university for supporting this research} \\
  Department of Computer Science\\
  American University of Madaba\\
  Kings Highway, Madaba, Jordan \\
   \And
  Nigel Hardy \\
  Department of Computer Science\\
  Aberystwyth University, Penglais\\
  Aberystwyth, Ceredigion, UK\\
}

\begin{document}
\maketitle

\begin{abstract}
This work presents MeKDDaM-SAGA, computer-aided automation software for implementing a novel knowledge discovery and data mining process model that was designed for performing justifiable, traceable and reproducible metabolomics data analysis. The process model focuses on achieving metabolomics analytical objectives and on considering the nature of its involved data. MeKDDaM-SAGA was successfully used for guiding the process model execution in a number of metabolomics applications. It satisfies the requirements of the proposed process model design and execution. The software realises the process model layout, structure and flow and it enables its execution externally using various data mining and machine learning tools or internally using a number of embedded facilities that were built for performing a number of automated activities such as data preprocessing, data exploration, data acclimatization, modelling, evaluation and visualization. MeKDDaM-SAGA was developed using object-oriented software engineering methodology and was constructed in Java. It consists of 241 design classes that were designed to implement 27 use-cases. The software uses an XML database to guarantee portability and uses a GUI interface to ensure its user-friendliness. It implements an internal embedded version control system that is used to realise and manage the process flow, feedback and iterations and to enable undoing and redoing the execution of the process phases, activities, and the internal tasks within its phases.
\end{abstract}


\flushbottom
\maketitle
\thispagestyle{empty}
\noindent

\section{Introduction}
\emph{\textbf{Metabolomics}} ~is defined as \textgravedbl \emph{the study of all low molecule weight chemicals (metabolites) which are involved in metabolism, either as an end product or as necessary chemicals for metabolism}\textacutedbl \cite{Mal04,Det04,Dun05}. Metabolomics data consists of both metabolomics dataset that is generated by the data acquisition instruments and their associated meta-data. The dataset format varies depending on the data acquisition instruments \cite{Xia09}. It takes the format of metabolites signal intensities, spectra, spectra bins, spectra peaks, and interferogram. Metabolomics meta-data refers to data that are collected in metabolomics assay and that might influence the dataset such as bio-source, sample preparation, instrument and assay parameters in addition to other administrative data \cite{Goo07a,Spa06,Sum07b,Jen05}.

\emph{\textbf{Knowledge discovery}} ~is defined as \textgravedbl \emph{Non-trivial process of identifying valid, novel, potentially useful, and ultimately understandable patterns in data}\textacutedbl, while data mining is considered as a step in the knowledge discovery process \cite{Fra92,Fay96a}. Data mining has two joint aspects: techniques and processes. Data mining techniques concern applying a variety of algorithms adopted from statistics, machine learning and pattern recognition, while data mining process concern the set of activities that are required for applying data mining techniques. A data mining process model provides a systematic approach for conducting data mining activities and procedures, which can take the form of a formalised framework that provides effective guidelines. Data mining is used in a variety of metabolomics applications. Examples for these applications can be found in \cite{Ban19d,Wis08a,Rou10,Tay02,Wis08b} and in many other publications.

This paper presents software that provides computer-aided automation, realization and guidance for a novel knowledge discovery and data mining process model that was proposed for performing metabolomics data analysis (MeKDDaM). The software was named MeKDDaM-SAGA. The realised process model was developed to enable carrying out a justifiable, traceable and reproducible data analysis to achieve the analytical objectives of metabolomics studies and suiting the nature of the data. MeKDDaM offers improvements to existing data mining process models regarding their layout structure and scientific orientation based largely on the principles of Scientific Methodology, Process Engineering, Software Engineering. The idea of providing a software realisation of data mining process model is inspired by the success of the Rational Unified Process (RUP) and the integrated tools available for software development \cite{Kru04}. Yet, the importance of automation for data mining process has attracted the attention of practitioners for some time \cite{Bra94} as the need for an integrated environment was emphasised in order to enable the users to apply complex data mining process in several application domains. However, until now, no serious efforts have been made in addressing the automation of a data mining process.

The paper provides an overview of MeKDDaM-SAGA software environment features and provides documentation for its development process that covers its requirements, analysis, design, construction, and testing. It also presents snapshots for its execution that was demonstrated using four metabolomics applications that were reported in \cite{Ban12b}. MeKDDaM-SAGA software environment was developed to realise and manage the process flow and iteration and to guide the execution of its phases and activities. It manages the process inputs, outcomes and deliveries and provides support to its practical consideration including management, standardisation, quality assurance and process-human interaction in addition to a number of other desired features such as \textbf{data exploration}, \textbf{knowledge representation}, \textbf{visualization} and \textbf{automation support}.

The software realization of the proposed process enables both external and internal execution of the proposed model. MeKDDaM-SAGA enables the internal executions of data preprocessing, \textbf{data exploration} and \textbf{data acclimatisation} in addition a number of internal facilities that enable the building, evaluating and visualizing the generated data mining models using a number of embedded machine learning algorithms. However, the external execution of the process phases and activities is also supported by the software implementation through recording the results of their execution and importing their outcomes into the software. MeKDDaM-SAGA implements an embedded version control system that is used to realise, organise and manage the process flow, feedback and iterations using a number of rollbacks and resuming mechanisms that enable undoing and redoing the process-level and phase-level activities execution.

MeKDDaM-SAGA was developed using object-oriented software engineering methodology and was implemented in Java. The software realises the functionality of 27 use-cases using 241 design classes which were organized over 26 packages. It uses an XML database that stores and manages the process execution data in order to guarantee the process execution portability and reduce its reliance on other software installations and technologies. The user interface was designed as GUI in order to ensure its ease of use and user-friendliness. MeKDDaM-SAGA software is freely and publicly available on Github software repositories \cite{Ban19c}. The paper uses a number of UML diagrams to describe the software data model and refers to screenshots that provide examples for the process model execution in the four demonstrated real-life applications that are reported in \cite{Ban19b}.

The second section provides a brief overview of the implemented process model, while the third section discusses the software development methodology which covers the requirements of the software development and its design, construction and testing. The fourth section provides an evaluation of the software features and other aspects, in addition to its satisfaction of the process model design requirements. The last provides a discussion for the automation results and also a conclusion. Snapshots for the process execution using MeKDDaM-SAGA software is also provided in the appendices as well as samples for its generated deliveries in XML format.

\section{Overview of MeKDDaM Process Model}
\begin{figure*}[htb!h] 
\centerline{\includegraphics[width=1\textwidth]{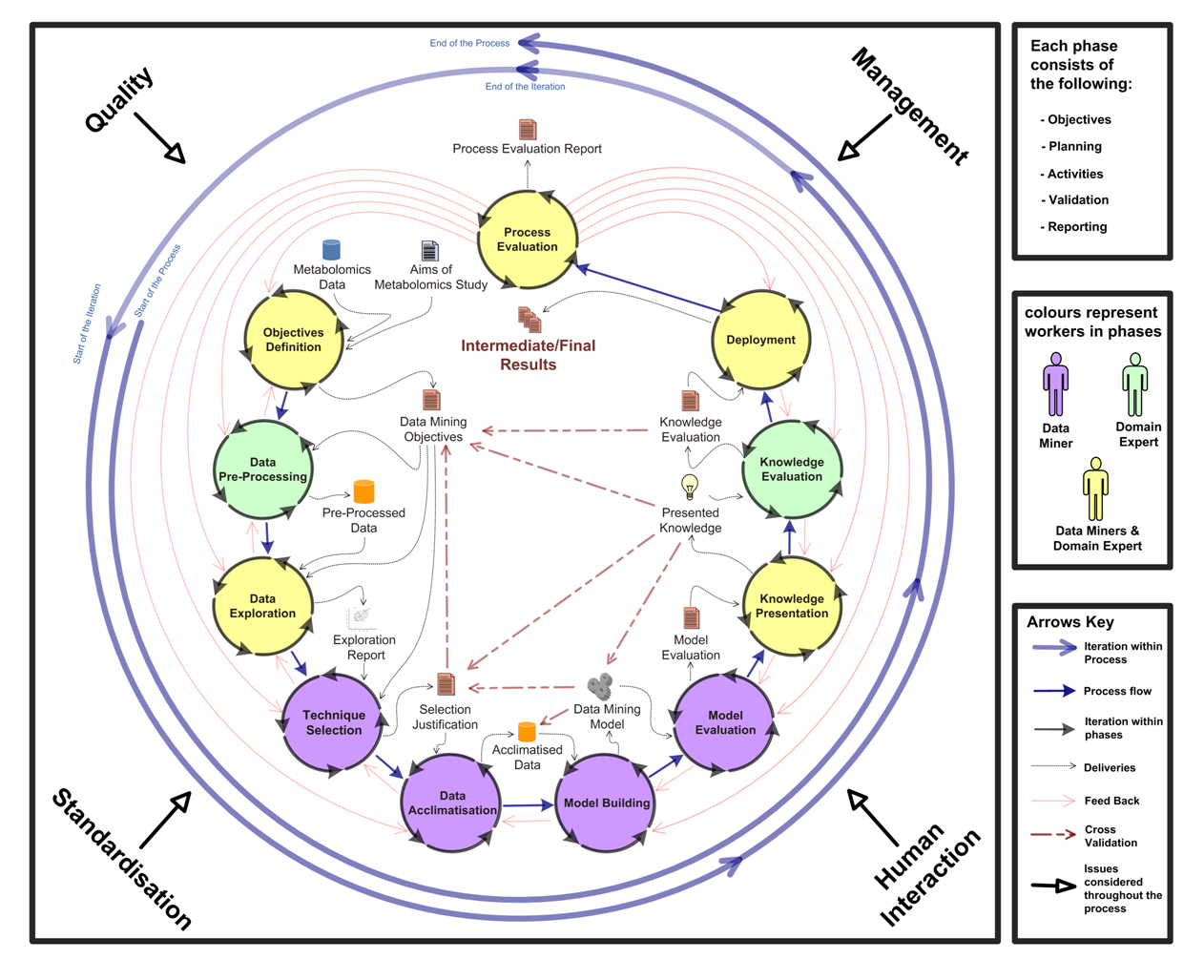}}
\centering
\caption{\textbf{MeKDDaM Process Model}-
         the graphical representation of the process model illustrates the process model structure and the flow of its phases. It also defines the inputs, deliveries and the participants in each phase.}
\label{fig:process2}
\end{figure*}

The proposed process model (MeKDDaM) consists of eleven phases that are illustrated in Figure~\ref{fig:process2}. All the process phases have a template for the execution of their internal tasks which covers: prerequisites, objectives, activities planning, performing, validation and reporting. The process model defines the structure and layout of its phases organization. It defines the normal flow of its phases' execution and all the possible iteration and feedback between its phases. The process model defines the inputs of the process execution which takes the form of both the metabolomics dataset and its associated meta-data as well as the aims of metabolomics study. It also defines the inputs and outputs of each of the process phases and the final output of the process model instantiated execution.

The phases are briefly described as follows:
\begin{enumerate}
\item \textbf{Objectives Definition}
provides a mechanism for defining the process objectives by matching the data mining approaches, goals, and tasks to aims of the metabolomics study and the goals of its original investigation.

\item \textbf{Data Pre-Processing}
is an optional phase which aims to clean the raw data acquired by the assay instruments. The scale of this phase, and its extent and the procedures applied, depends on the data acquisition instruments and techniques and on the procedures that are already performed during data acquisition.

\item \textbf{Data Exploration}
involves performing a set of activities which aim to get insight into the data through: investigation, understanding and prospecting in order to assess the potential of the data. The output of this phase takes the form of a report, which contains details regarding the activities that are performed in the phase and their outcomes.

\item \textbf{Technique Selection}
provides a strategy for selecting and justifying the selection of a data mining technique that should
achieve the process objectives and suit the nature of the targeted data. The strategy considers the requirements and feasibility of the selected technique and defines its performance measurability and success criteria. The results of this phase are the selected technique and its justification, as well as a record of factors, considerations and assessments involved. A detailed description of the strategy proposed for technique selection is available in \cite{Ban12a}.

\item \textbf{Data Acclimatisation}
involves processing and preparing the dataset(s) for \textbf{model building} and evaluation using the selected data mining technique and the applied tools. This phase generates one or more datasets, which can be used for \textbf{model building}, training and testing.

\item \textbf{Model Building}
involves building and training a data mining model that fulfills the defined process objectives by applying the selected data mining modelling technique on the dataset acclimatised in the previous phase.

\item \textbf{Model Evaluation}
involves testing, validating and evaluating the model based on the defined objectives and using measurement criteria for the technique applied. Model evaluation is usually performed using a separated data split which must
be allocated for model validation during the \textbf{data acclimatisation} phase.

\item \textbf{Knowledge Presentation}
involves presenting the model built and validated in the previous phases in a form which presents the acquired metabolomics knowledge. Knowledge presentation might require complex visualisation techniques in order to facilitate interactive presentation of knowledge.

\item \textbf{Knowledge Evaluation}
evaluates the knowledge acquired and presented earlier from a metabolomics perspective. This
is performed in terms of its fulfillment of the objectives defined in the first phase, as well as in terms of its validity as a metabolomics knowledge, based on the background knowledge.

\item \textbf{Deployment}
aims to deploy the acquired knowledge through a mechanism that enables effective knowledge utilisation. It involves selecting appropriate deployment mechanisms in the light of the defined process objectives and within the available resources, as well as the selection of the particular deliveries which must be deployed with the knowledge.

\item \textbf{Process Evaluation} \label{subsect:Process Evaluation}
concerns evaluating the execution of MeKDDaM process model in terms of the flow of its phases and the validity of the tasks applied within the performed phases. It also ensure the quality of the process deliveries through the defined mechanisms of cross-deliveries validation. Cross-delivery validation is a mechanism for evaluating process deliveries. This mechanism aims at providing a high level quality assurance on the level of process deliveries. Its results can cause a feedback to an earlier phase either to resolve inconsistencies or may cause a process iteration. Possible cross-delivery validations between the process phase deliveries are shown in Figure~\ref{fig:process2}.

\end{enumerate}

The process model phases allow the iterative execution of a number of internal tasks which include:
\begin{itemize}
\item \textbf{Phase Prerequisites} involves confirming the validity of process inputs, phase deliveries, and other relevant information which concerns phase customisation, implementation and running. Phase prerequisites must be sufficient, specific, relevant and valid in order to perform the phase activities with no additional resources or information. This helps the justifiability and traceability of the phase results, as well as the reproducibility of its deliveries.

\item \textbf{Phase Objectives} defines the objectives which the phase is expected to achieve, the deliveries it must generate and the desired attributes and characteristics of those objectives in addition to the measurement criteria of these objectives.

\item \textbf{Phase Planning} maps the defined phase objectives to a set of practical actions designed to fulfil them. The planned activities must be consistent with phase prerequisites and it must comply with data mining and metabolomics procedural standards in addition to project management principles and human interaction best practices.

\item \textbf{Phase Performing} involves carrying out the planned phase activities and recording and justifying its execution. Possible problems, gaps and limitations during the phase execution must be recorded and reported as it may be used later for the purpose of phase validation, process iteration or cross-delivery validation.

\item \textbf{Phase Validation:} aims at ensuring the validity of the phase execution and the quality of the data involved and its compliance with the adopted standards. This validation is performed based on the planned phase activities and its identified objectives.

\item \textbf{Phase Reporting} concerns the generation of the phase outcomes including the tasks performed, deliveries, processed data and report of running which must conform to the relevant standards in terms of their contents, format and structure.
\end{itemize}

\begin{figure}[htb!h]
\centerline{\includegraphics[width=0.9\textwidth]{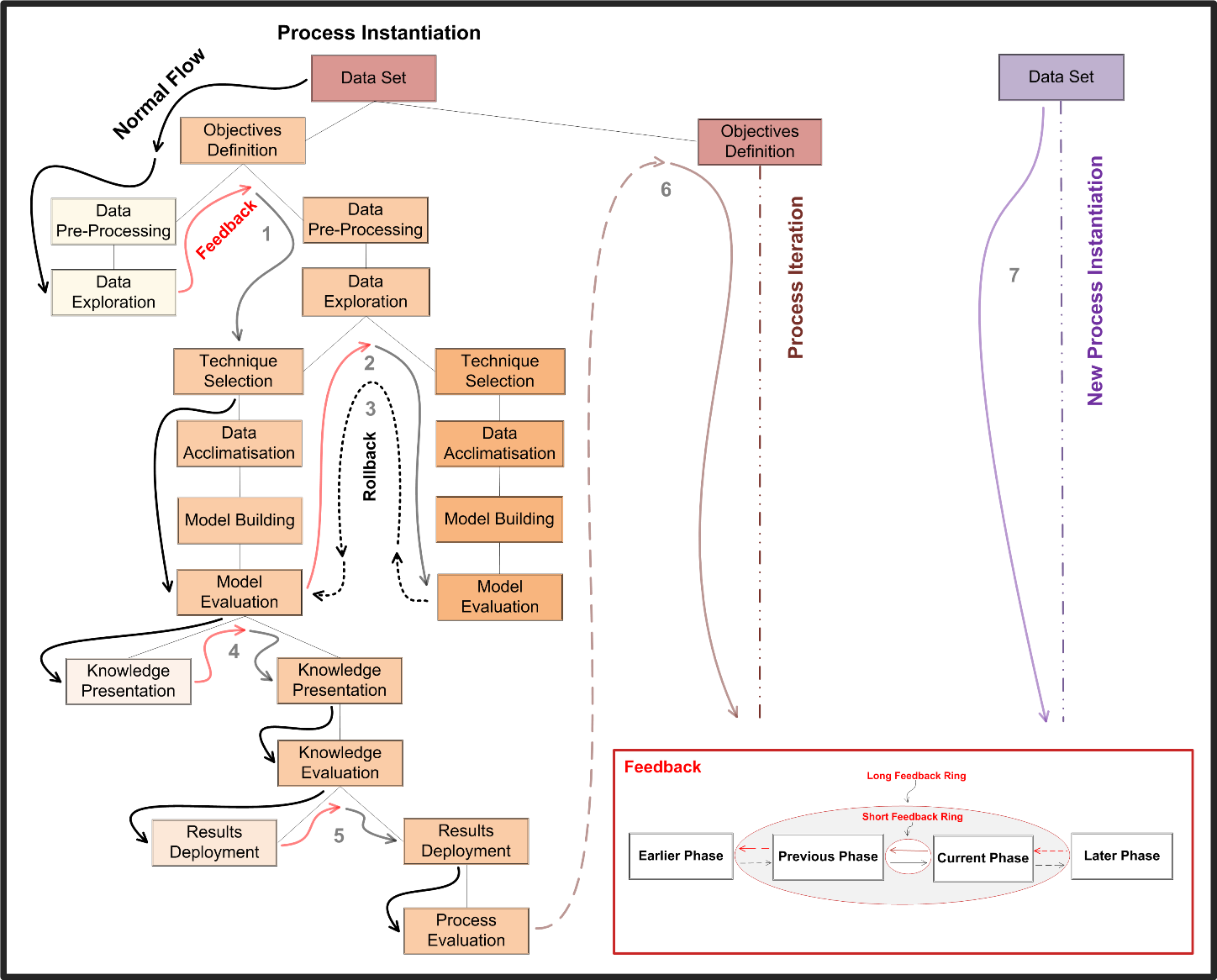}}
\centering
\caption{\textbf{Process Execution Scenarios (Repeated for Reference only:} The tree-like graph shows seven process execution scenarios, which illustrates examples of MeKDDaM process model execution that demonstrates the process feedback, rollback, phase iteration, and process iteration mechanisms.
\label{fig:process_execution_scenarios}}
\end{figure}

The process execution is carried out by running the internal tasks of the process phases and performing its planned activities in order to generate its defined outcomes and deliveries either as part of the process normal flow, feedback or a process iteration. Normal flow involves executing the process phases as defined by the process model, while iteration involves the repetitive execution of all phases maintaining the flow of their execution as defined by the process model and into consideration the inputs and deliveries of each iteration. An iteration might be triggered due to a significant change in process objectives, or in order to formulate and test a new hypothesis, to achieve different analytical objectives, to answer a fresh or propagated question, to improve the process execution results, or to resolve major problems in the process execution. A feedback ring is a micro scale iteration that involves two or more of the process phases. It involves the re-running of all phases inside the feedback ring. Feedback is usually triggered as a response to poor phase outcomes, problems with the running of the internal tasks within the phase, or due to the inadequacy of the phase prerequisites or inputs. The process model also supports rollback, which enables undoing or discarding phase or process iteration when it fails to achieve its purpose or intended objectives e.g. building a better model. Figure~\ref{fig:process_execution_scenarios} provides examples for the process execution scenarios that illustrate the process normal flow as well as concepts of iteration, feedback, and rollback.

\section{Methodology of Software Development}
The development of MeKDDaM-SAGA has been conducted based on object-oriented software engineering methodology using Rational Unified Process (RUP) as described in \cite{Kru04}.

\subsection{MeKDDaM-SAGA Requirements Analysis}
The requirements of the software features were identified and analysed based on proposed process model requirements, foundations, enhancements and design features as described in \cite{Ban12b}. Section~\ref{sec:description} provides a detailed description of the software designed features that satisfy each of these requirements.

\begin{figure*}[htb!h] 
\centerline{\includegraphics[width=1\textwidth]{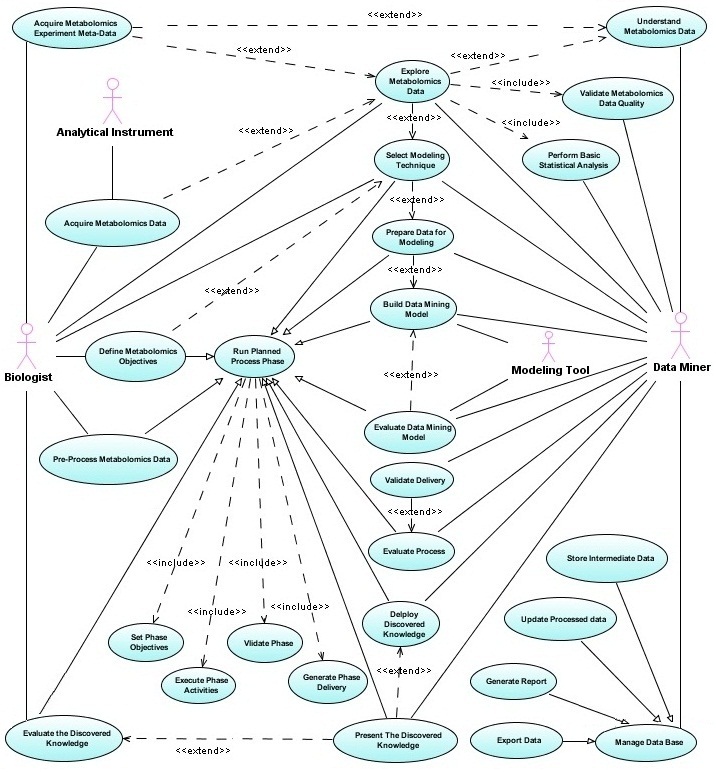}}
\centering
\caption{\textbf{Process Use-Case Model}: A UML Use-Case Model showing interaction between human users and the process}
\label{fig:UseCase}
\end{figure*}

Figure~\ref{fig:UseCase} illustrates the software use-case model, which was built based on the proposed process model requirements and design features. Each of the use-cases in the model was then described using a use-case description scenario which illustrated its preconditions, normal and alternative flows and also its actors and post-conditions.

The use-case description of each use case was then analysed in order to extract the requirements analysis classes: Entity, control, utility, and interface classes. The analysis classes were then refined, structured, and detailed in order to create the current software design classes.

\subsection{MeKDDaM-SAGA Design and Construction}
MeKDDaM-SAGA implementation was constructed using 241 Java classes which were organised into two major packages. The first package was designed to implement and realise the process data model as described in Figure~\ref{fig:Packages}. It contains 163 control and entity classes that were organised across a number of sub-packages. The second package was designed to create a graphical user interface (GUI) for the software environment and to handle a number of embedded software services and to communicate with external tools and resources. This package contain 78 boundary and utility classes in addition to a number of utility classes which provided a number of generic services e.g. objects persistence, files management, XML database handling, etc.

MeKDDaM-SAGA enables the internal execution of several activities in the \textbf{data exploration} and acclimatisation phases. It also allows building, evaluating, and visualising data mining models which are built internally using a number of embedded data mining techniques. However, the software also supports the execution of all the activities in the process externally, and then allows the recording of their results and importing their outcomes. This paper focuses on the design of the process data model described in the first package, which forms the backbone of the software implementation, while the second package is demonstrated using a number of screenshots, which are provided in

\subsection{MeKDDaM-SAGA Testing}
MeKDDaM-SAGA was tested based on unit and integration testing. More importantly, the process functionality was also tested using test cases which were based on the process requirements and execution scenarios discussed. The software has also been executed using four different real-world applications which have been discussed and demonstrated in \cite{Ban12b,Ban19b}.

\section{MeKDDaM-SAGA Implemented and Realized Features} \label{sec:description}
This section demonstrates how the implementation software satisfies MeKDDaM process model features. It also refers to a number of diagrams which illustrate the software design and to a number of screenshots for the  XML files that were used to persist and store the data that are related to the process execution in a structured format including its involved inputs, outcomes and deliveries.

\subsection{Process Inputs}
This section presents an overview of the facilities provided by the software environment to support the process inputs. Figure~\ref{fig:Inputs} illustrates the design of the process \emph{\textbf{Input}} class and its relationship with relevant classes.

\subsubsection*{Metabolomics Data}
MeKDDaM-SAGA provides the option of importing the data into the system or interacting with it externally as an independent file in the file system. The data importing facilities support the targeted dataset and also its associated meta-data. The software is also capable of capturing information regarding the name and data types of the data attributes.

MeKDDaM-SAGA provides facilities for importing, exporting and converting the data using a number of common formats. In addition, it is also capable of exporting the data to an Oracle database, where the data can be manipulated using the Oracle PL/SQL language, and then be imported back by the software system. Figure~\ref{fig:DataSet_HiMetIP9} shows a screenshot of an example input data that was captured from the software GUI. Appendix~\ref{Persistence_Process_Inputs} provides examples XML files that for the process inputs.

Since most data pre-processing procedures are usually performed externally on the level of the data acquisition instruments or using dedicated external software, the software environment provides a number of facilities that can be used for handling and processing the data throughout all the process phases. It provides facilities for importing the results of \textbf{data pre-processing} and also for recording their applied procedures. Yet, MeKDDaM-SAGA is capable of performing data exploration, data acclimatization and data visualisation procedures using a number of internal facilities. Examples of these procedures include: replacing missing values by zero, median, and mean values in addition to data standardisation, normalisation, dimensionality reduction, re-sampling, randomisation, data splitting, and attribute deletion, insertion, and merging.

\subsubsection*{Study Design}
MeKDDaM-SAGA provides facilities for recording information regarding the design of the study and its original metabolomics investigation; these include the aims of the study, sample preparation, and data acquisition procedures, which are required as prerequisites for performing some of the process phases such as \textbf{objectives definition} and \textbf{data pre-processing}. Figure~\ref{fig:Study_HiMetIP9} shows a screenshot captured while recording information regarding the study of the cow diet application.

\begin{landscape}
\pagestyle {plain}
\scriptsize

\begin{figure*}[htb!h] 
\centerline{\includegraphics[width=1.55\textwidth]{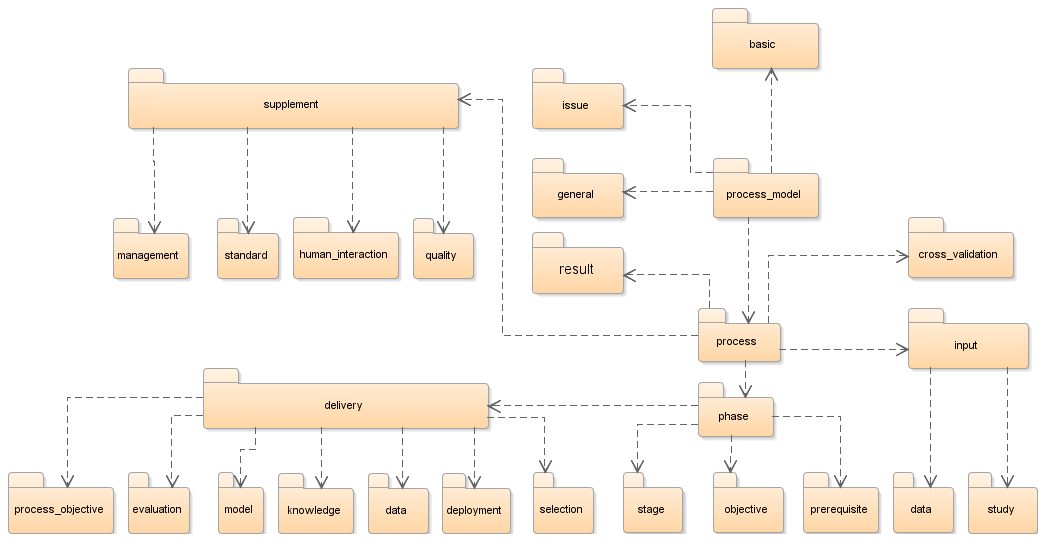}}
\centering
\caption{\textbf{Software Packages}: A UML diagram illustrating the packages in the software environment used for implementing the process}
\label{fig:Packages}
\end{figure*}

\end{landscape}

\begin{landscape}
\pagestyle {plain}
\scriptsize

\begin{figure*}[htb!h] 
\centerline{\includegraphics[width=1.25\textwidth]{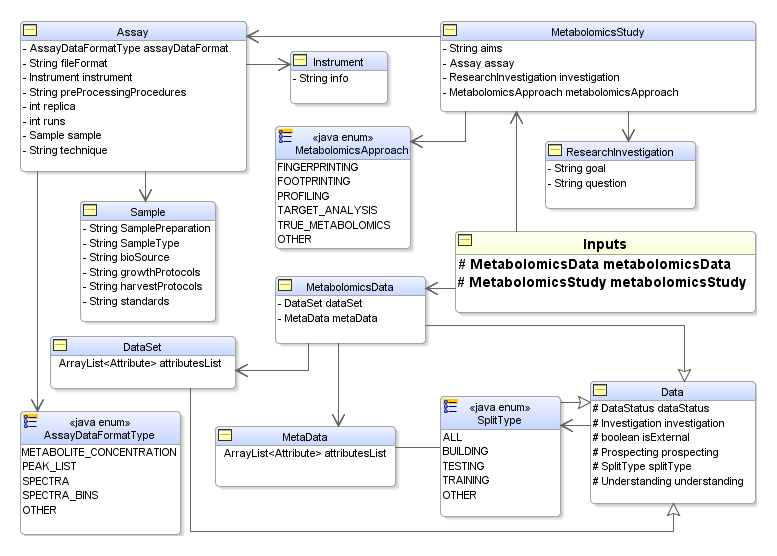}}
\centering
\caption{\textbf{Inputs Class Diagram:} {A UML class diagram for the process Inputs and their associated classes}}
\label{fig:Inputs}
\end{figure*}

\end{landscape}

\subsection{The Realisation of MeKDDaM Process Model Structure}
\begin{figure*}[htb!h] 
\centerline{\includegraphics[width=1\textwidth]{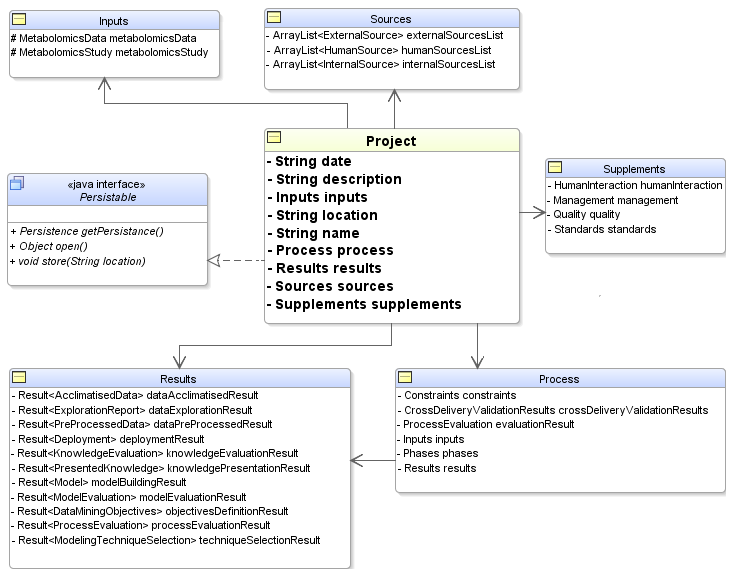}}
\centering
\caption{\textbf{Project Class Diagram:}A UML diagram representing the project and its relevant classes}
\label{fig:Project}
\end{figure*}

\begin{figure*}[htb!h] 
\centerline{\includegraphics[width=1\textwidth]{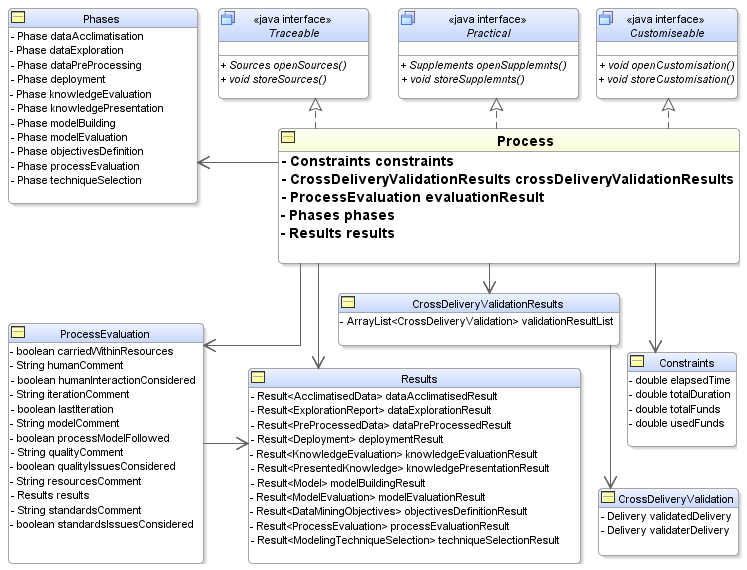}}
\centering
\caption{\textbf{Process Class Diagram:}A UML diagram representation of the process and its relevant classes}
\label{fig:Process1}
\end{figure*}

The structure of MeKDDaM is realised by a number of Java packages and classes which reflect the process hierarchy. The \emph{\textbf{Project}} class holds information regarding the process execution, results, supplements, and traceability and provides a number of functions regarding the project creation, saving, persistence, and loading. Figure~\ref{fig:Project} illustrates the \emph{\textbf{Project}} class and its relationship with the relevant classes using a UML class diagram. The \emph{\textbf{Process}} class is used to accommodate information regarding process execution and iteration covering its input data, executed phases, generated results and the outcomes of its evaluation and cross-delivery validation. This class also implements a number of functions, which allow the process iteration and rollback in addition to saving and loading of the process execution. Figure~\ref{fig:Process1} illustrates the \emph{\textbf{Process}} class and its relationship with the relevant classes using a UML class diagram, while

\subsubsection*{Process Phases}
The \emph{\textbf{Phase}} class is used for realising MeKDDaM phases and recording information regarding their execution, iteration, and feedbacks. An object of this class is created for each of the eleven process phases in addition to other copies which are instantiated, copied or cloned to resemble the phases' feedbacks and iterations.

In addition, the \emph{\textbf{Phase}} class provides functions for performing the phases internal tasks: planning, performing, validating, and reporting as well as for defining and populating their perquisites, objectives and planned activities with their preset descriptions or customisation. MeKDDaM-SAGA provides facilities for exporting and importing the phases preset description and customisation which realises the process mapping mechanism.

The \emph{\textbf{Phase}} class also implements a number of functions concerning the execution, iteration, and rollback of the process phases, in addition to their persistence, saving, and loading. Figure~\ref{fig:Phase} illustrates the \emph{\textbf{Phase}} class and its relationship with its relevant classes using a UML class diagram. Figure~\ref{fig:Phase_HiMetIP9} illustrates the phase execution of objectives definition phase, while Appendix~\ref{Persistence_Process_Phases} provides examples for the persistence of the process phases' outcomes and deliveries.

\subsubsection*{Phase Tasks}
The phase prerequisites and their preset and customised description are illustrated by Figure~\ref{fig:Phase_HiMetIP9_Prerequisites}, which also illustrates their traceability, while snapshots of the phase objectives and their preset and customised description are demonstrated in Figure~\ref{fig:Phase_HiMetIP9_Objectives}. Phase planning is demonstrated in the snapshots in Figure~\ref{fig:Phase_HiMetIP9_Planning}. The figure provides an example of the phase planning using both preset and customised description, which was linked to the phase objectives. The phase preset description is loaded into the process as an initial description for the phase generic planned activities which can be then customised and mapped to the particular application that is being conducted. It also illustrates the planned activity allocated resources and potential performer. Phase performance is demonstrated in the snapshots in Figure~\ref{fig:Phase_HiMetIP9_Performing_Justification} and Figure~\ref{fig:Phase_HiMetIP9_Performing_Problems}. The figures provide an example of performing plan activities, where it shows its performance justification, as well as its problems recording, traceability, and resolution, are shown. The figures show also the activity resources and performer. Phase validation is demonstrated in the snapshots in Figure~\ref{fig:Phase_HiMetIP9_Validating}. The figure shows an example of a phase validated activity. The phase validation then provides a mechanism for validating every aspect of the performed phase activities.

MeKDDaM-SAGA provides facilities for generating phase deliveries and reports which concern two aspects. The first concerns the execution of the phase internal tasks as illustrated in appendix~\ref{Persistence_Process_Phases} which provide examples for the phase execution reporting, while the second aspect concerns the delivery of the phase execution outcome, which is reported by instantiating the specialised subclasses \emph{\textbf{Delivery}}. An example of the GUI used for selecting the data mining technique is shown in Figure~\ref{fig:Phase_HiMetIP9_Outcome_TechniqueSelection}, while examples of the XML files that are used for the persisting of the phase outcome deliveries is illustrated in appendix~\ref{Persistence_Process_Deliveries}. The \emph{\textbf{Delivery}} class is implemented to provide a number of generic functionality and to accommodate information inherited by the process phases specific deliveries as illustrated in the class diagram, which is shown in Figure~\ref{fig:Delivery}.

\subsubsection*{Process Flow, Feedback, Rollback, and Iteration}
The software realises MeKDDaM process model flow, feedback, rollback, and iteration through a number of mechanisms. It also provides a rigorous version control mechanism, which maintains the validity of the process execution and the consistency of the performed phases and their relevant outputs and deliveries. The process flow is recorded in each of the process executions by instantiating its participant phases in sequential order and adding generated deliveries to the process outcomes. The process iteration is performed by invoking a function in the \emph{\textbf{Process}} class, where the current phase execution is backed up and a new instance of the process is instantiated. MeKDDaM-SAGA provides a facility for rolling back the execution of the process to a previous status or iteration by recovering the backed up process execution, which is also implemented as a function in the \emph{\textbf{Process}} class. A snapshot of the process iteration that was captured in the cow diet application reported is illustrated in both Figure~\ref{fig:Process_Cow_Diet1} and Figure~\ref{fig:Process_Cow_Diet2}. The software realises the process feedback mechanism based on the process model description. The phase feedback is realised by invoking a set of functions implemented in the \emph{\textbf{Phase}} class, which back up the process execution and then iterate through the internal tasks of the targeted phase and re-instantiate and execute its successor phases. The \emph{\textbf{Phase}} class also provides a rollback mechanism, which allows the recovery of the backed up phase execution and its associated process execution status during feedback. A snapshot of the process iteration that was captured in the \emph{Arabidopsis} fingerprint application is illustrated in both Figure~\ref{fig:Process_HiMet_FTIR1} and Figure~\ref{fig:Process_HiMet_FTIR2}.

\subsection{Scientific Orientation Support}
The scientific orientation of MeKDDaM process model is realised through a number of mechanisms and features; these include process activity justification and traceability as well as problem identification, traceability and resolution. Figure~\ref{fig:Supplemnts_HiMetIP9_Sources} illustrates a screenshot of the GUI used for defining the process traceability sources in the \emph{Arabidopsis} isoprenoids application.

MeKDDaM-SAGA provides traceability facilities on several levels. The prerequisites of every phase in the process are linked back to their sources and to every decision, justification, or solution. The process sources can be internal, such as other phases and deliveries or external such as the literature. Human sources are also considered and linked with human interaction facilities.

The class diagram in Figure~\ref{fig:Activity} illustrates the process model classes which are related to the process of scientific orientation. In addition, a number of snapshots of an example of the performance of a phase activity iteration that was captured during the execution of the \textbf{objectives definition} phase in the \emph{Arabidopsis} isoprenoids application.
Figure~\ref{fig:Phase_HiMetIP9_Performing_Justification} provides a screen shot of a performed activity justification, while Figure~\ref{fig:Phase_HiMetIP9_Performing_Problems} provides an example of a reported problem, which is linked to both of its reason and proposed solution.

\subsection{Process Supplements and Practical Features}
The process of supplements and practical features cover MeKDDaM process model management, human interaction, quality assurance and the process standards. This section illustrates the software realisation of these features based on the process model satisfaction the identified requirements and their realisation in the applications demonstrated in \cite{Ban12b,Ban19b}. The class diagram in Figure~\ref{fig:Supplements} illustrates the data model classes which have been used for implementing the process supplementary and practical features.

\subsubsection*{Management and planning}
The implementation realises MeKDDaM process model management and planning features, which were incorporated in response to the identified practical process requirements. MeKDDaM-SAGA provides facilities for defining the processing time and cost constraints as well as management resources which cover human expertise, and both software and hardware resources. In addition, the software also allows for the allocation of these resources during the phase execution. The process planning and management features are also implemented, where the MeKDDaM-SAGA software provides facilities for defining the objectives measurability, achievability, feasibility and other success criteria which forms the basis of the \textbf{knowledge evaluation} and also for recording the results of their assessment. Figure~\ref{fig:Phase_HiMetIP9_Outcome_ObjectivesDefinition} provides an example of \textbf{objectives definition}.

\begin{landscape}
\pagestyle {plain}
\scriptsize

\begin{figure*}[htb!h] 
\centerline{\includegraphics[width=1.4\textwidth]{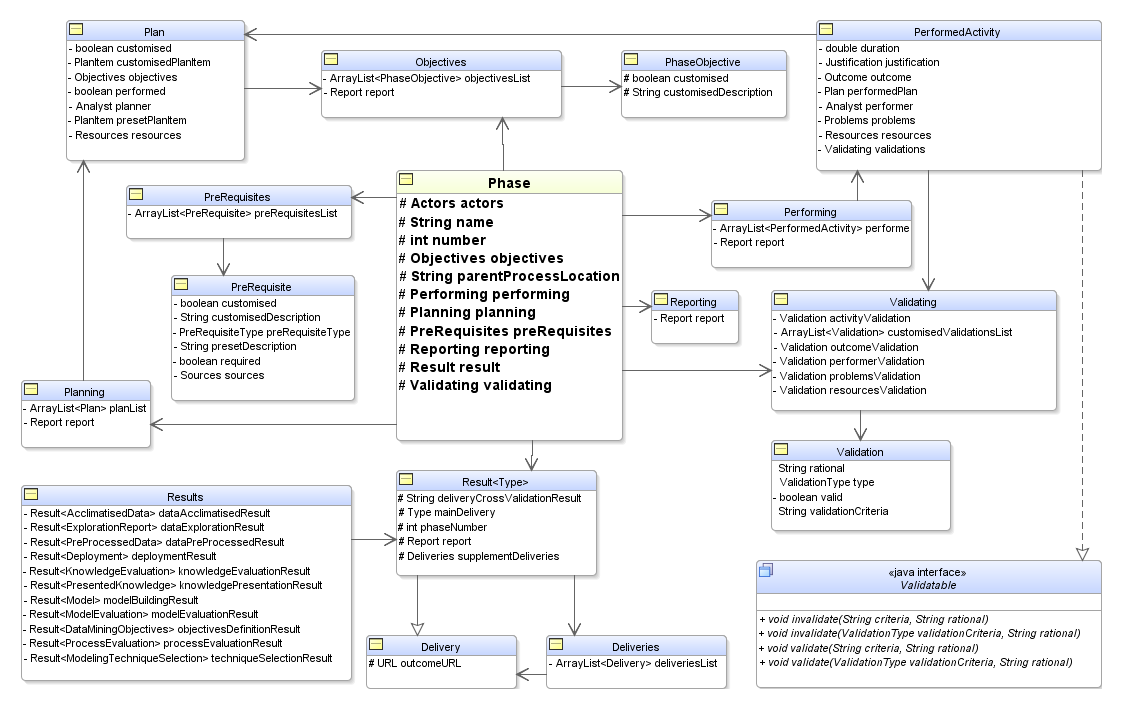}}
\centering
\caption{\textbf{Phase Class Diagram:} A UML representation of process phases and their associated classes}
\label{fig:Phase}
\end{figure*}

\end{landscape}

\begin{landscape}
\pagestyle {plain}
\scriptsize
\begin{figure*}[htb!h] 
\centerline{\includegraphics[width=1.2\textwidth]{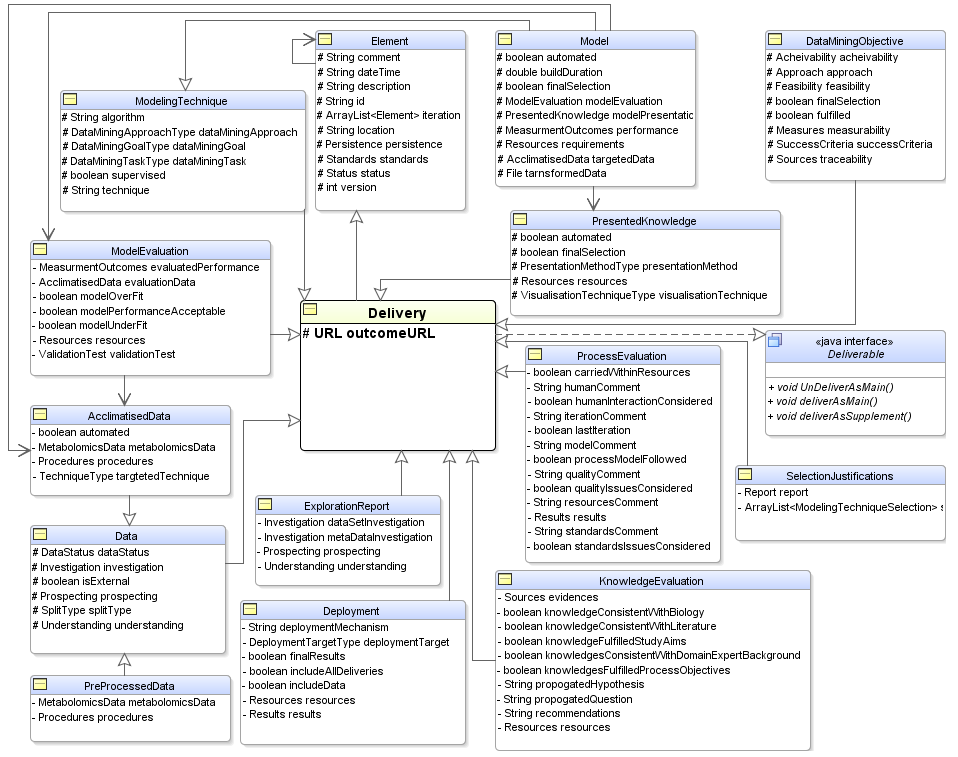}}
\centering
\caption{\textbf{Delivery Class Diagram:} A UML class diagram representing the process deliveries and their associated classes}
\label{fig:Delivery}
\end{figure*}

\end{landscape}

\begin{landscape}
\pagestyle {plain}
\scriptsize
\begin{figure*}[htb!h] 
\centerline{\includegraphics[width=1.3\textwidth]{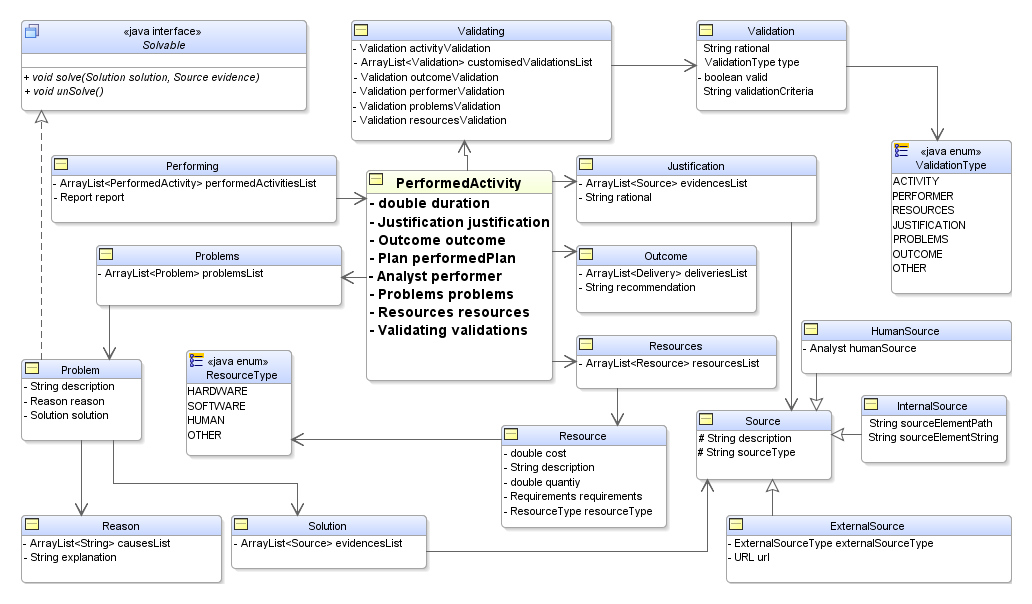}}
\centering
\caption{\textbf{Activity Class Diagram:} A UML classes diagram representing the phases activities and their considered issues and relevant classes}
\label{fig:Activity}
\end{figure*}
\end{landscape}

\begin{landscape}
\pagestyle {plain}
\scriptsize

\begin{figure*}[htb!h] 
\centerline{\includegraphics[width=1.4\textwidth]{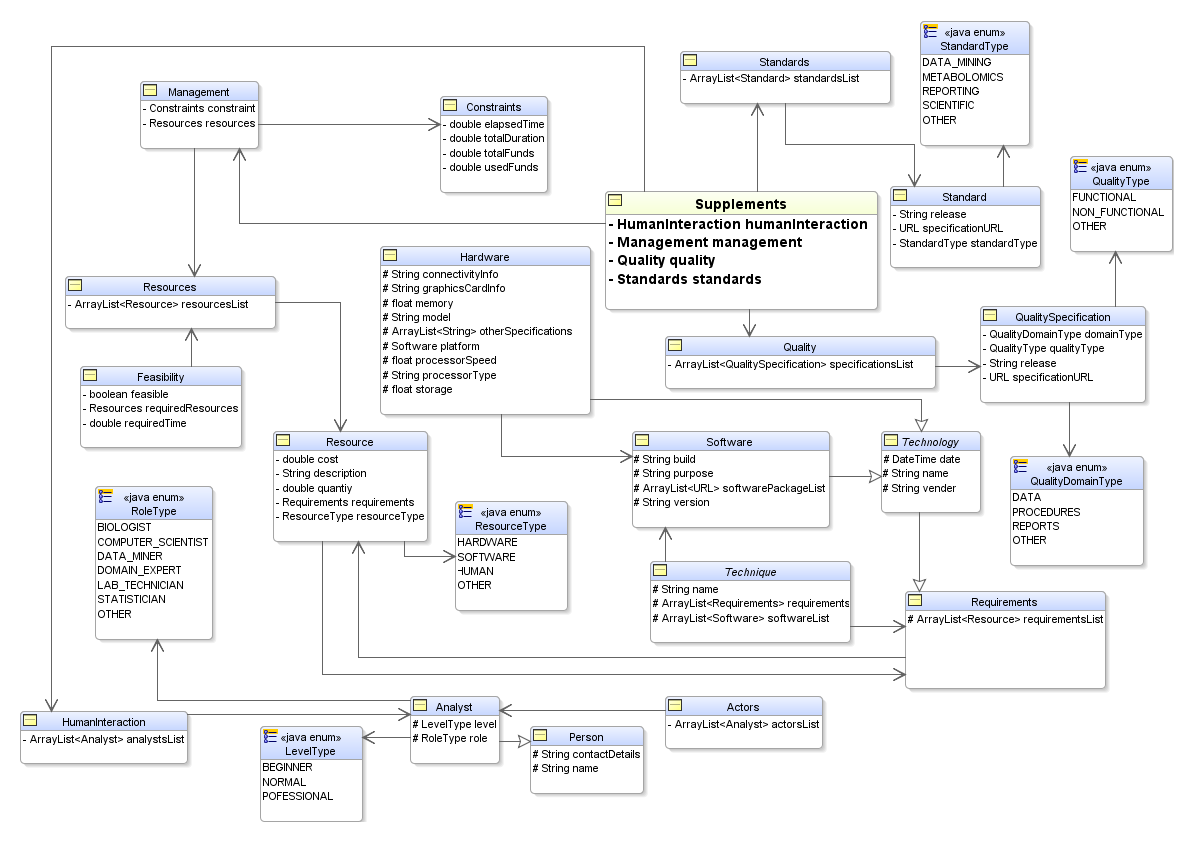}}
\centering
\caption{\textbf{Supplements Class Diagram:} A UML classes diagram showing the process supplements and their associated classes}
\label{fig:Supplements}
\end{figure*}

\end{landscape}

The same thing is also applied to the selected data mining technique, where the feasibility, measurability, and success criteria of the technique are defined. This is demonstrated in the screenshot in Figure~\ref{fig:Phase_HiMetIP9_Outcome_TechniqueSelection} that is used for evaluating the data mining model.

In addition, MeKDDaM-SAGA supports the process manageability by providing facilities for defining and customising the process phases' prerequisites, objectives, and planning and for managing its performing, validating, and reporting. Screenshots for the software management facilities are also provided. Figure~\ref{fig:Supplemnts_HiMetIP9_Management_Resources} provides an example of the process resources and their requirements, while Figure~\ref{fig:Supplemnts_HiMetIP9_Management_Constraints} provides an example of the process time and management constraints.

\subsubsection*{Human Interaction}
MeKDDaM-SAGA provides mechanisms for realising and organising human interaction throughout the process. Human interaction provides facilities for defining the participants on the level of the process as demonstrated in the appendix and assigning the performer for each individual activity during the performance of each phase in the process. The screenshots in Figure~\ref{fig:Supplemnts_HiMetIP9_HI} illustrate the GUI used for defining the process actors, while Figure~\ref{fig:Phase_HiMetIP9_Performing_Performer} illustrates the assignment of a phase activity performer.


\subsubsection*{Quality}
MeKDDaM-SAGA provides facilities for implementing the quality assurance mechanisms of the process. The software provides facilities for investigating the quality of the data within the \textbf{data exploration} phase and for enhancing its quality later during \textbf{data acclimatisation} phase internally using a number of built-in facilities e.g. missing values replacement by zero, median, and mean values in addition to data standardisation, and normalisation or externally using dedicated software tools.

MeKDDaM-SAGA is also capable of evaluating the data mining models built internally using one of the data mining techniques embedded in the software such as ANN (MLP), decision trees, random forest, MLR, and Bays Nets. In addition, the software is also capable of recording and importing the evaluation of data mining models built externally. Appendix~\ref{Persistence_Process_Deliveries} provides examples for persisting \textbf{model evaluation} and  \textbf{knowledge evaluation} and their involved deliveries. MeKDDaM-SAGA software also records the validation of the activities performed within the process phases. Figure~\ref{fig:Phase_HiMetIP9_Validating} provides an example of a process phase validation. The software also implements the \textbf{process evaluation} and provides facilities for performing cross-delivery validation. Moreover, MeKDDaM-SAGA also provides facilities for defining application-specific quality assurance policy, which can be defined on the level of the process execution. Figure~\ref{fig:Supplemnts_HiMetIP9_QA_Standards} provides a screenshot of the GUI used for defining the process quality assurance policy.

\subsubsection*{Standards}
MeKDDaM-SAGA provides support for defining application-specific standards on the level of the process execution. A GUI was used for defining process standards. The defined process standards can then be used for generating the data mining model and storing it persistently as well as for reporting the various process deliveries.

\subsection{Data Exploration Features} \label{sec:DataExplorationFeatures}
The process software implementation provides facilities for realising the metabolomics \textbf{data exploration} requirements. MeKDDaM-SAGA is capable of prospecting the data by generating basic statistics regarding its values e.g. mean, standard deviation, minimum, maximum, etc. as well as visualising the distribution of data and the correlation between its features. The software also provides facilities for recording and importing the external prospecting of the data and for data understanding and investigation. Examples of \textbf{data exploration} support provided by MeKDDaM-SAGA will be demonstrated later in the applications reported in \cite{Ban12b,Ban19b}.

\subsection{Visualisation Support Features}
The software realises MeKDDaM process model support of visualisation and provides facilities for visualising some of the models which can be built internally during the process execution e.g. decision trees and ANN. The software also provides facilities for importing the models visualised externally. The software visualisation facilities are also useful for prospecting the data, which can be performed internally by visualising the distribution of the data attributes across classes or by visualising the correlation between pairs of the data attributes using scatter plots. Examples of data and model visualisation will be demonstrated later.

\subsection{Knowledge Evaluation Features}
MeKDDaM-SAGA implements \textbf{knowledge evaluation} support which was identified as a requirement for the process development. \textbf{Knowledge evaluation} was developed as a phase in the process. The software defines a mechanism for evaluating the knowledge based on the process description.

\subsection{Automated Features} \label{sec:AutomatedFeatures}
The software environment provides an automated realization of most aspects of MeKDDaM proposed process model, It also provides internal facilities for data investigation, prospecting, and acclimatisation as well as \textbf{model building}, evaluation, and visualisation.

\subsection{Additional Features of MeKDDaM-SAGA}
Here we discuss some of the important process features and facilities, which provided support for the process configuration management and concept of mapping through implementing a number of customisation facilities which cover the process phases, practical supplements, and traceability.

\subsubsection*{Configuration Management}
The implementation considers configuration management issues and provides facilities that support version control on the level of the process as well as on the level of its subsequent phase which covers iterations and feedbacks. The realisation of the process iteration is performed by creating a clone copy from the previous process execution, making changes on this copy as required, storing, persisting and managing the previous iteration for future reference. On the other hand, the realisation of phase iteration is based on creating a copy of the current phase execution, which is used to perform the changes required by the iteration. The old copy of the phase object is then stored and managed using the process configuration management mechanism. Iteration can be performed as many times as required. The process iteration depends on how far the previous iteration has gone through creating and modifying a copy of its object.

\subsubsection*{Process Mapping and Customisation}
MeKDDaM-SAGA realises the process concept of mapping by allowing the population of the process phases with their preset generic tasks covering their objectives, prerequisites, and planning. The present tasks can then be customised according to the specific needs of the metabolomics application.

\section{Discussion}

MeKDDaM-SAGA was successful in automating MeKDDaM process model and in realising its various design aspects. The software provided a number of automated facilities that implemented the process layout, structure and flow and other practical aspects and desired features. The software realised the process model mechanism for defining measurable objectives and in providing a software implementation for the process selection strategy. MeKDDaM-SAGA also enabled the automatic generation of basic statistics regarding the process data as well as the visual prospecting of its distribution and correlation. The software allowed the internal performance of several \textbf{data acclimatisation} procedures such as missing values handling, data normalisation, and re-sampling, in addition to other data conversions, reduction, and data splitting operations in addition to attributes deletion, introduction, merging, data splitting, and class assignment. Moreover, it also allowed the automatic building and evaluation of several data mining models internally including ANN (MLP), decision trees, MLR, Bays Nets, random forest, hierarchical clustering, linear regression, and association rules as well as the visualisation of some of the models built using these techniques e.g. decision trees, ANN.

However, and despite the successful automation of the process model design aspects and its other supported features, the full automation of the data mining process seems to be far away based under the current technology. This is in fact, due to the nature of some data mining activities and procedures, particularly those involve significant human involvement or judgment such as objective setting, \textbf{technique selection}, and \textbf{knowledge evaluation}. Yet, partial automation was possible using the implemented software environment. It covered for many aspects of the data mining process such as \textbf{data exploration}, \textbf{data acclimatisation}, \textbf{model building} and evaluation. However, some aspects of the data mining process were automated by several data mining tools such as \textbf{data exploration}, \textbf{acclimatisation}, \textbf{model building} and \textbf{evaluation}.

\section{Conclusion}
This paper presented MeKDDaM-SAGA, which is a software that was developed for realising MeKDDaM process model. The paper provided an overview of the software development process that was conducted based on software engineering principles and provided a discussion for the software supported features. The software was successful in realising and automating the proposed process model design aspects and in guiding its execution.

The paper also introduced the concept of computer-aided data mining, which aims not only to automate the various aspects of the data mining process (as others have suggested), but also to realise the structure and flow of the data mining process itself and to provide support for its various practical aspects. This research calls for more attention to be paid to the software realisation of knowledge discovery and data mining process, which in turn could lead to a new branch in the field of data mining, which could be termed as Computer-Aided Data Mining analogous to Computer-Aided Software Engineering (CASE).

\bibliographystyle{unsrt}

\bibliographystyle{unsrt}

\newpage

\section{Appendix} \label{Appendix}

\subsection{Appendix: Snapshots of the Process Application} \label{ProcessApplicationSnapshots}
This appendix provides some snapshots for MeKDDaM-SAGA software environment that was used for process realization and implementation. These snapshots have been captured during the process demonstrated applications as discussed earlier in this paper.

\begin{landscape}
\pagestyle {plain}
\scriptsize
\newpage
\subsection*{A Feedback Scenario}
\label{FeedbackSnapshots}

\begin{figure*}[htb!h] 
\centerline{\includegraphics[width=1.3\textwidth]{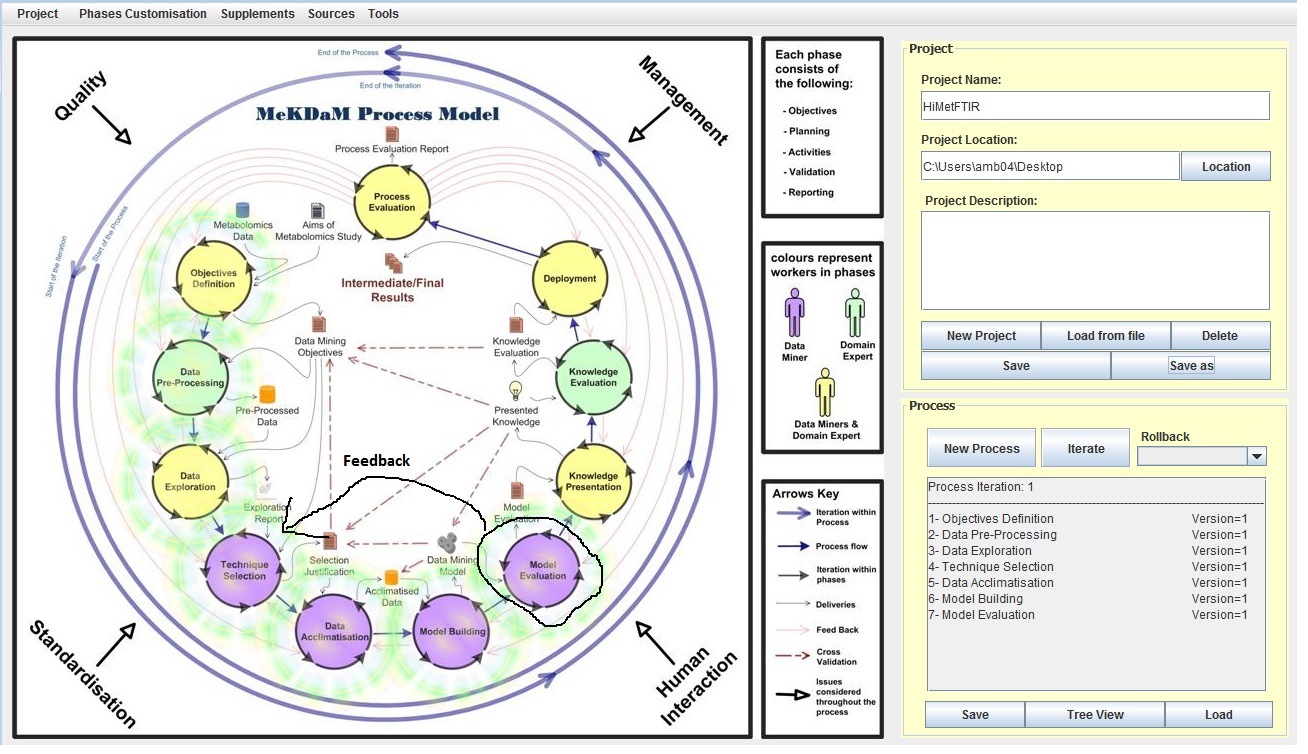}}
\centering
\caption{\textbf{Process Execution Just Before a Feedback:} A snapshot of the process execution just before performing a feedback in order to select a different modelling technique as captured in the \emph{Arabidopsis} fingerprinting application. The performed phases are shown surrounded by a green light halo}
\label{fig:Process_HiMet_FTIR1}
\end{figure*}

\end{landscape}

\begin{landscape}
\pagestyle {plain}
\scriptsize
\newpage
\label{fig:JustAfterFeedbackSnapshots}
\begin{figure*}[htb!h] 
\centerline{\includegraphics[width=1.3\textwidth]{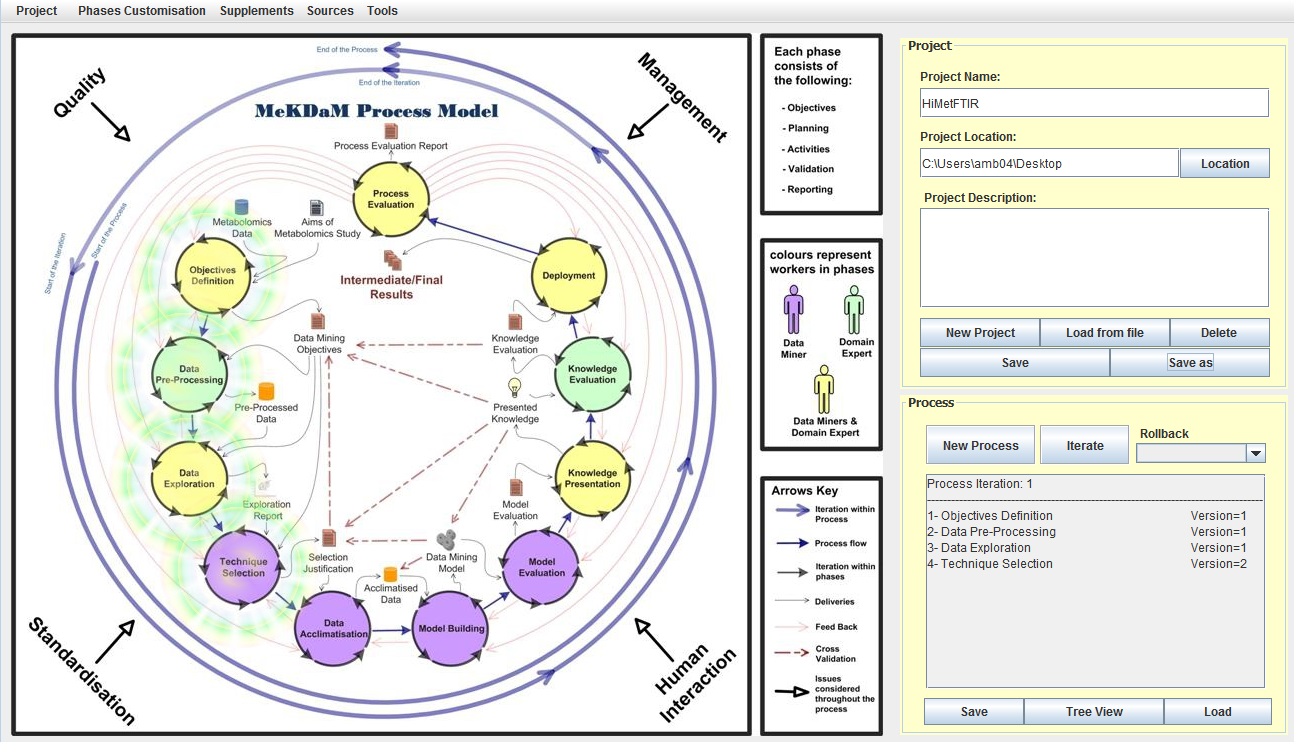}}
\centering
\caption{\textbf{Process Execution Just After a Feedback:} A snapshot of the process execution just after performing a feedback in order to select a different modelling technique. Captured in the \emph{Arabidopsis} fingerprinting application}
\label{fig:Process_HiMet_FTIR2}
\end{figure*}

\end{landscape}

\begin{landscape}
\pagestyle {plain}
\scriptsize
\newpage
\subsection*{A Process Iteration Scenario}
\label{fig:ProcessIterationSnapshots}

\begin{figure*}[htb!h] 
\centerline{\includegraphics[width=1.3\textwidth]{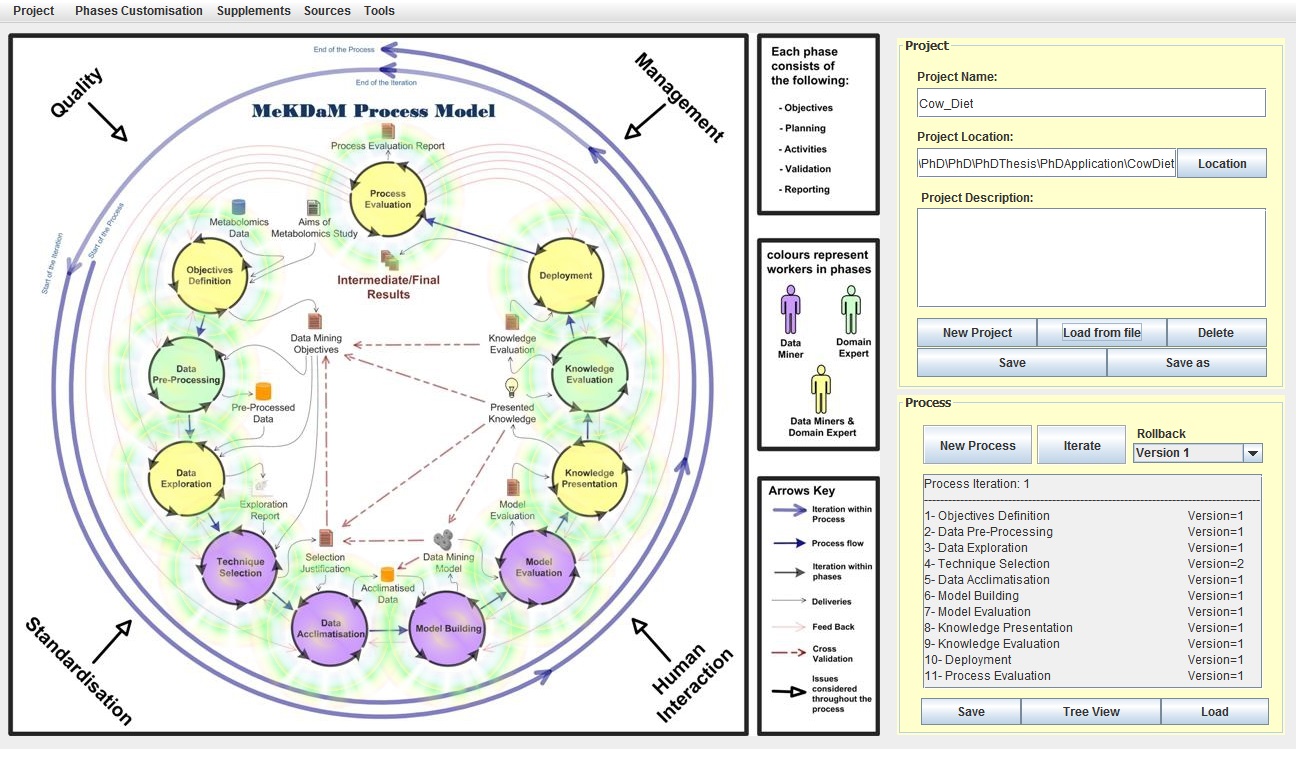}}
\centering
\caption{\textbf{Process Execution Before Process Iteration:} A snapshot of the process execution just before performing a process iteration as captured in the cow diet application}
\label{fig:Process_Cow_Diet1}
\end{figure*}

\end{landscape}

\begin{landscape}
\pagestyle {plain}
\scriptsize

\begin{figure*}[htb!h] 
\centerline{\includegraphics[width=1.3\textwidth]{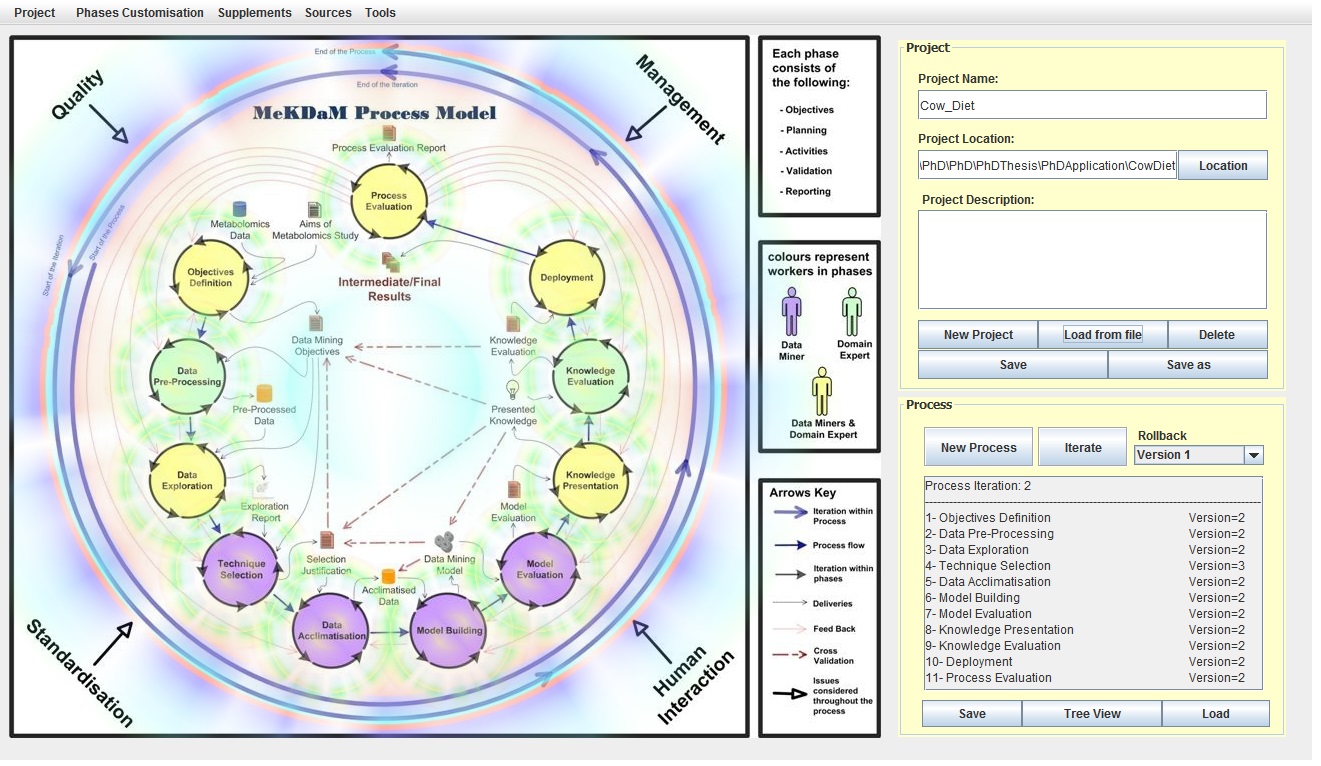}}
\centering
\caption{\textbf{Process Execution after a Second Iteration:} A snapshot of the process execution as captured in the cow diet application after the process was iterated in order to achieve new objectives. The process iteration is indicated by a light halo which surrounds the process}
\label{fig:Process_Cow_Diet2}
\end{figure*}

\end{landscape}

\begin{landscape}
\pagestyle {plain}
\scriptsize
\newpage
\subsection*{Process Inputs}
\label{fig:ProcessInputsSnapshots}

\begin{figure*}[htb!h] 
\centerline{\includegraphics[width=1.3\textwidth]{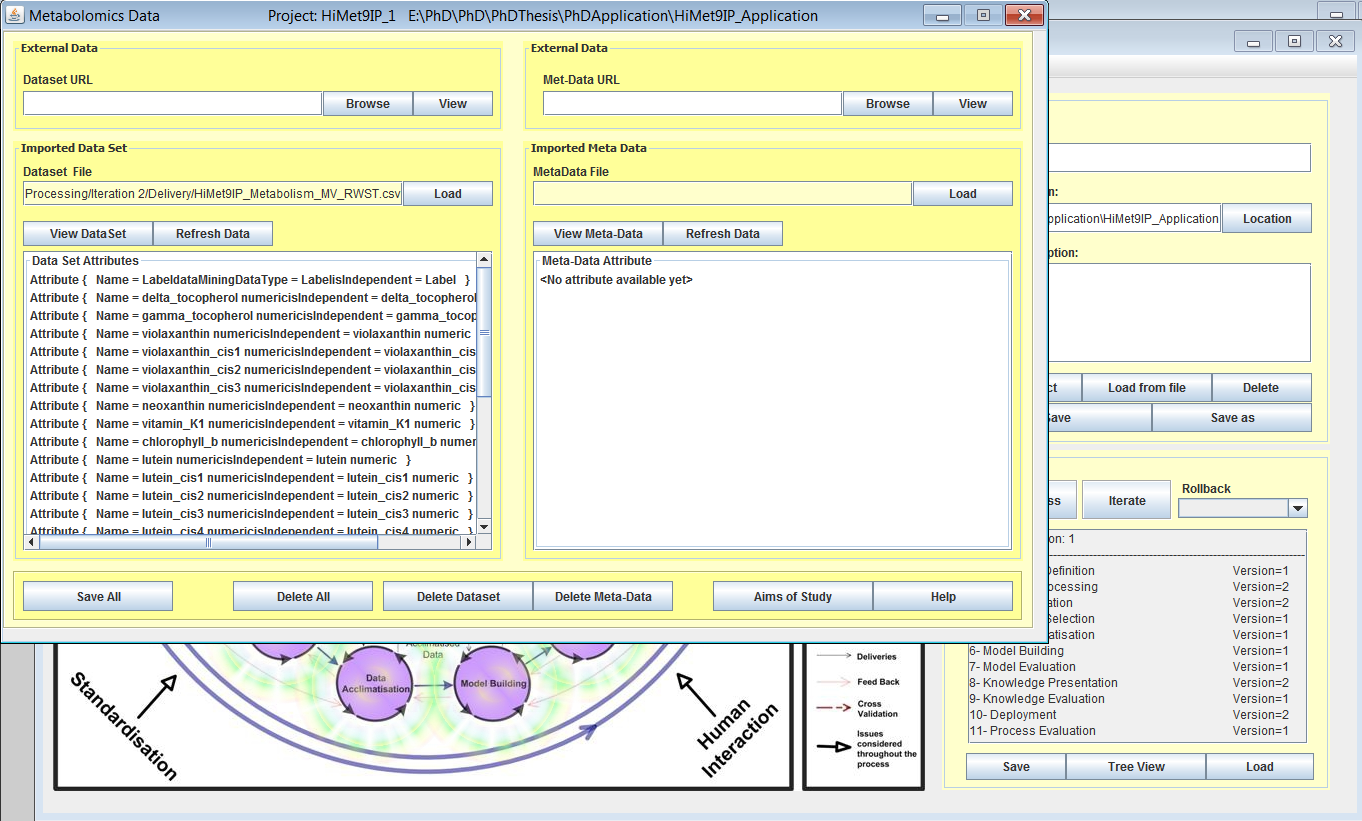}}
\centering
\caption{\textbf{Metabolomics Data:} A snapshot of metabolomics data imported to the process as captured in the \emph{Arabidopsis} isoprenoids profiling application}
\label{fig:DataSet_HiMetIP9}
\end{figure*}

\end{landscape}

\begin{landscape}
\pagestyle {plain}
\scriptsize

\begin{figure*}[htb!h] 
\centerline{\includegraphics[width=1.3\textwidth]{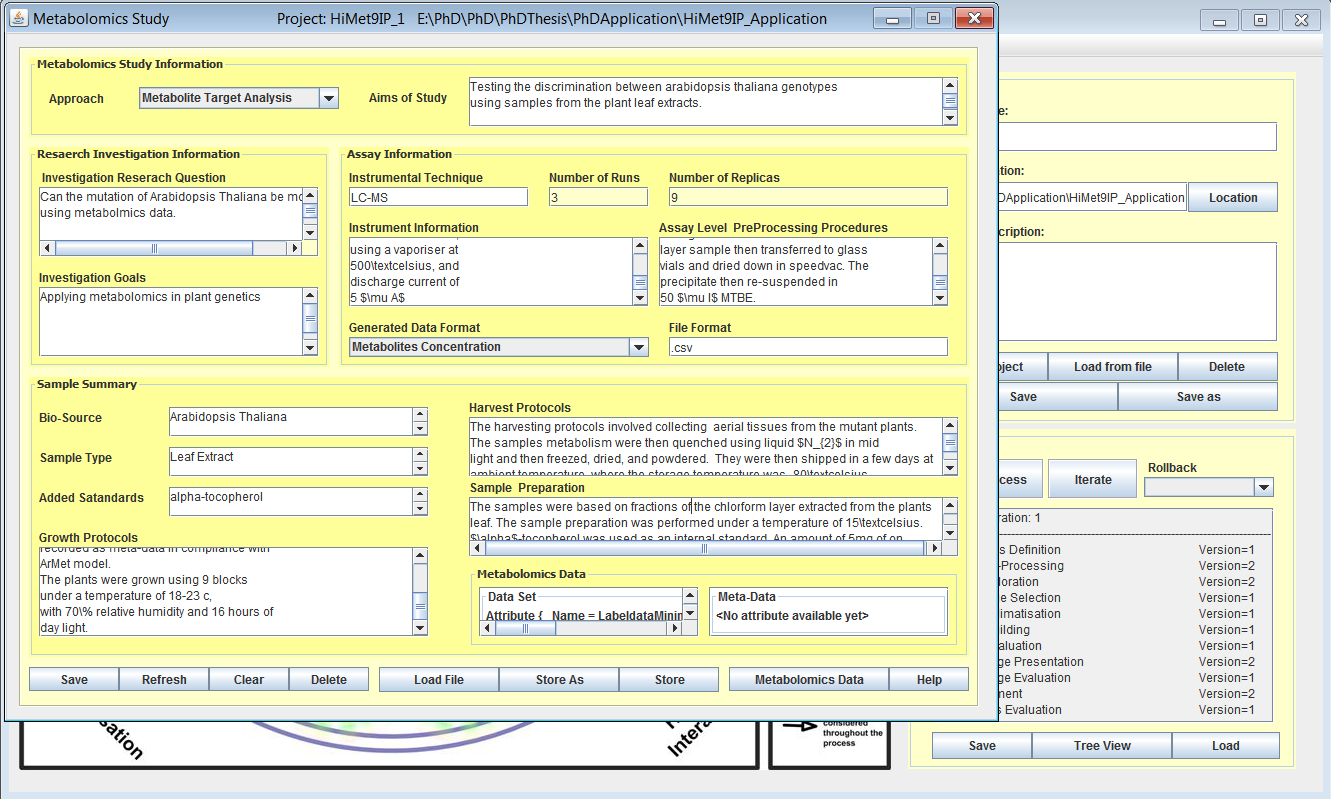}}
\centering
\caption{\textbf{Aims of a Metabolomics Study:} A snapshot of the aims of a metabolomics study as captured in the \emph{Arabidopsis} isoprenoids profiling application}
\label{fig:Study_HiMetIP9}
\end{figure*}

\end{landscape}

\begin{landscape}
\pagestyle {plain}
\scriptsize
\newpage
\subsection*{Internal Tasks within a Process Phase}
\label{fig:PhaseTasksSnapshots}

\begin{figure*}[htb!h] 
\centerline{\includegraphics[width=1.3\textwidth]{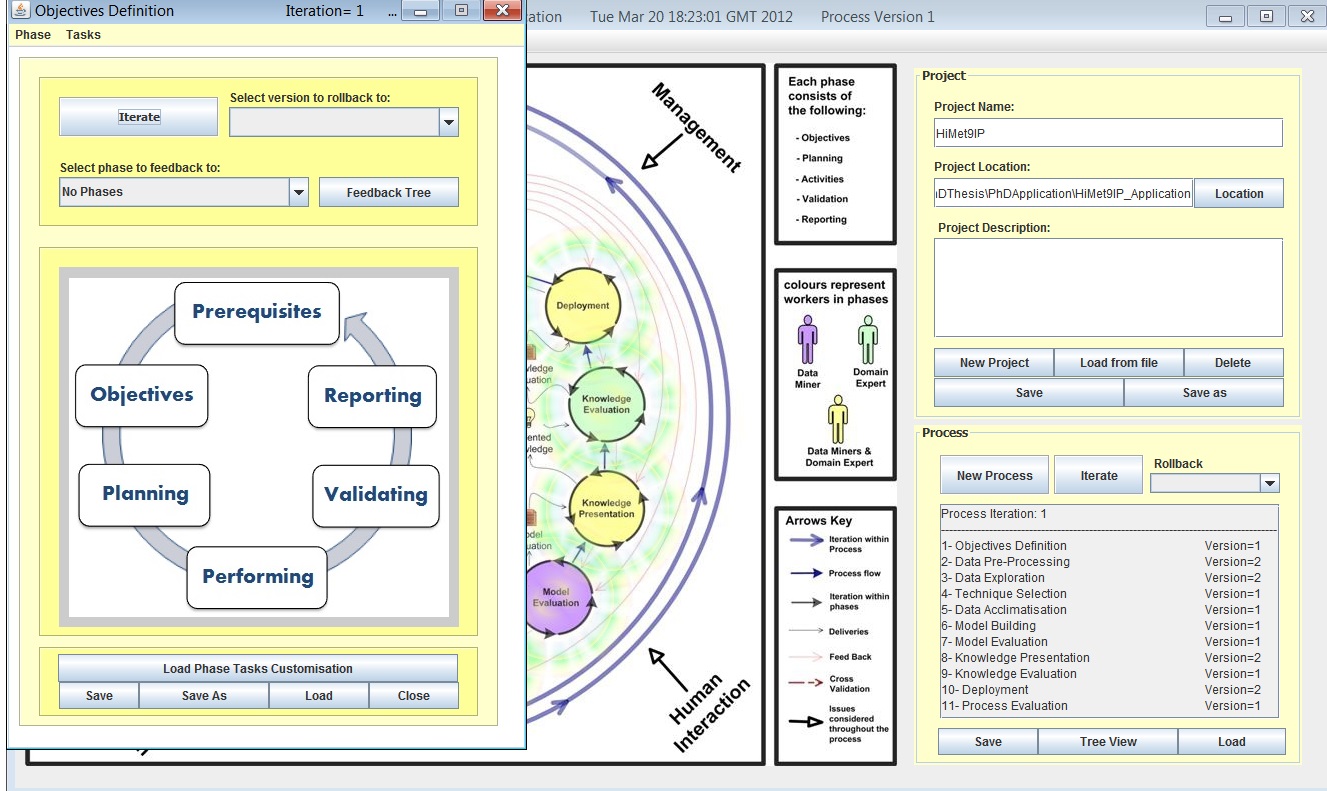}}
\centering
\caption{\textbf{Phase Running:} A snapshot of a phase running as captured in the \emph{Arabidopsis} isoprenoids application}
\label{fig:Phase_HiMetIP9}
\end{figure*}

\end{landscape}

\begin{landscape}
\pagestyle {plain}
\scriptsize

\begin{figure*}[htb!h] 
\centerline{\includegraphics[width=1.3\textwidth]{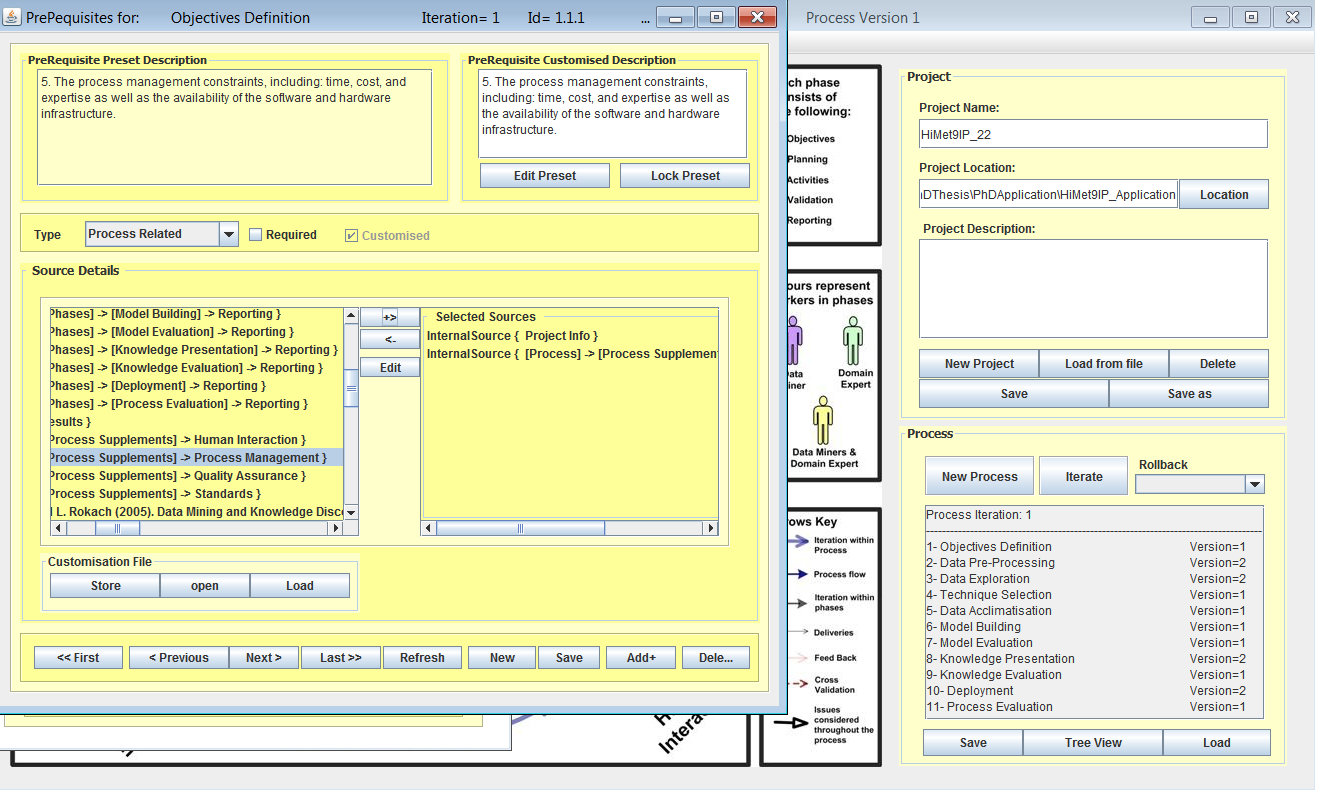}}
\centering
\caption{\textbf{Phase Prerequisites:} A snapshot of a phase prerequisites as captured in the \emph{Arabidopsis} isoprenoids profiling application}
\label{fig:Phase_HiMetIP9_Prerequisites}
\end{figure*}

\end{landscape}

\begin{landscape}
\pagestyle {plain}
\scriptsize

\begin{figure*}[htb!h] 
\centerline{\includegraphics[width=1.3\textwidth]{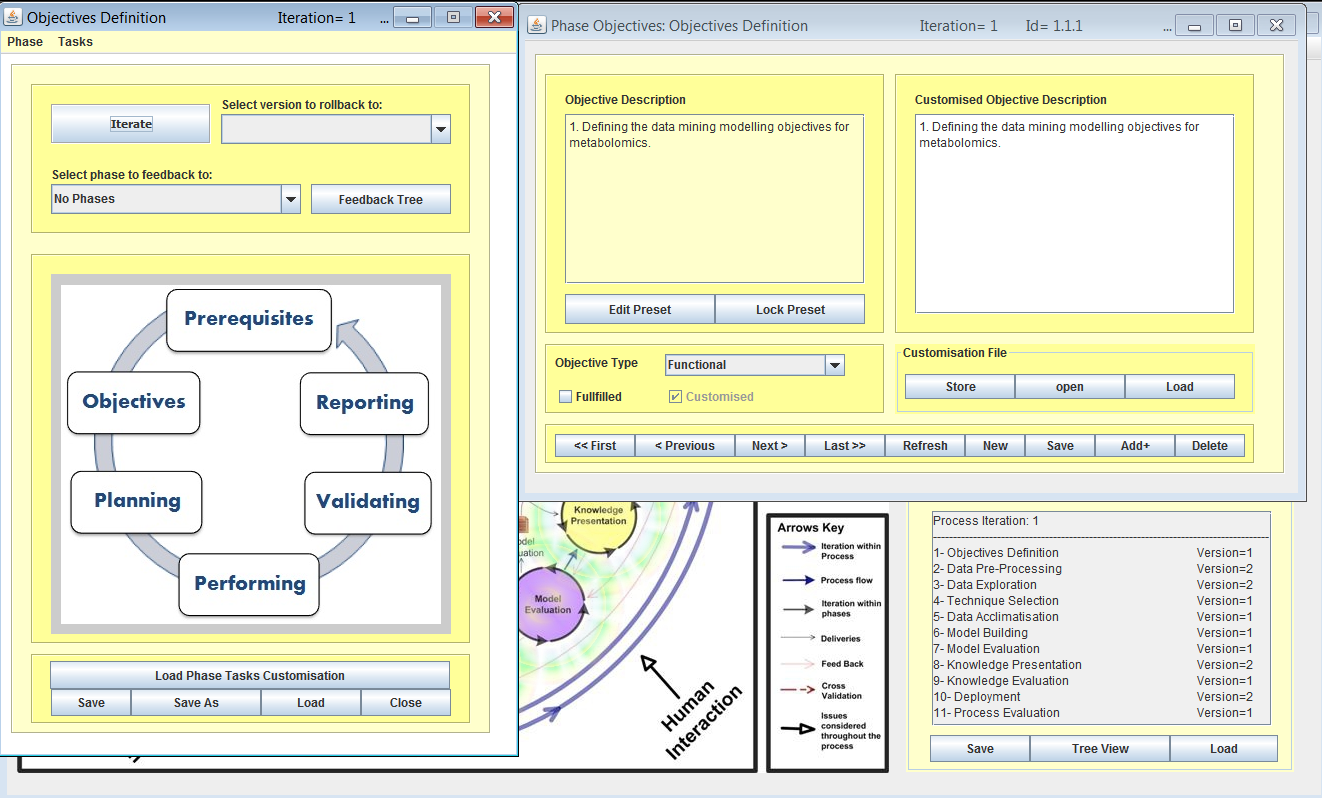}}
\centering
\caption{\textbf{Phase Objectives:} A snapshot of a phase objectives as captured in the \emph{Arabidopsis} isoprenoids profiling applications}
\label{fig:Phase_HiMetIP9_Objectives}
\end{figure*}

\end{landscape}

\begin{landscape}
\pagestyle {plain}
\scriptsize

\begin{figure*}[htb!h] 
\centerline{\includegraphics[width=1.3\textwidth]{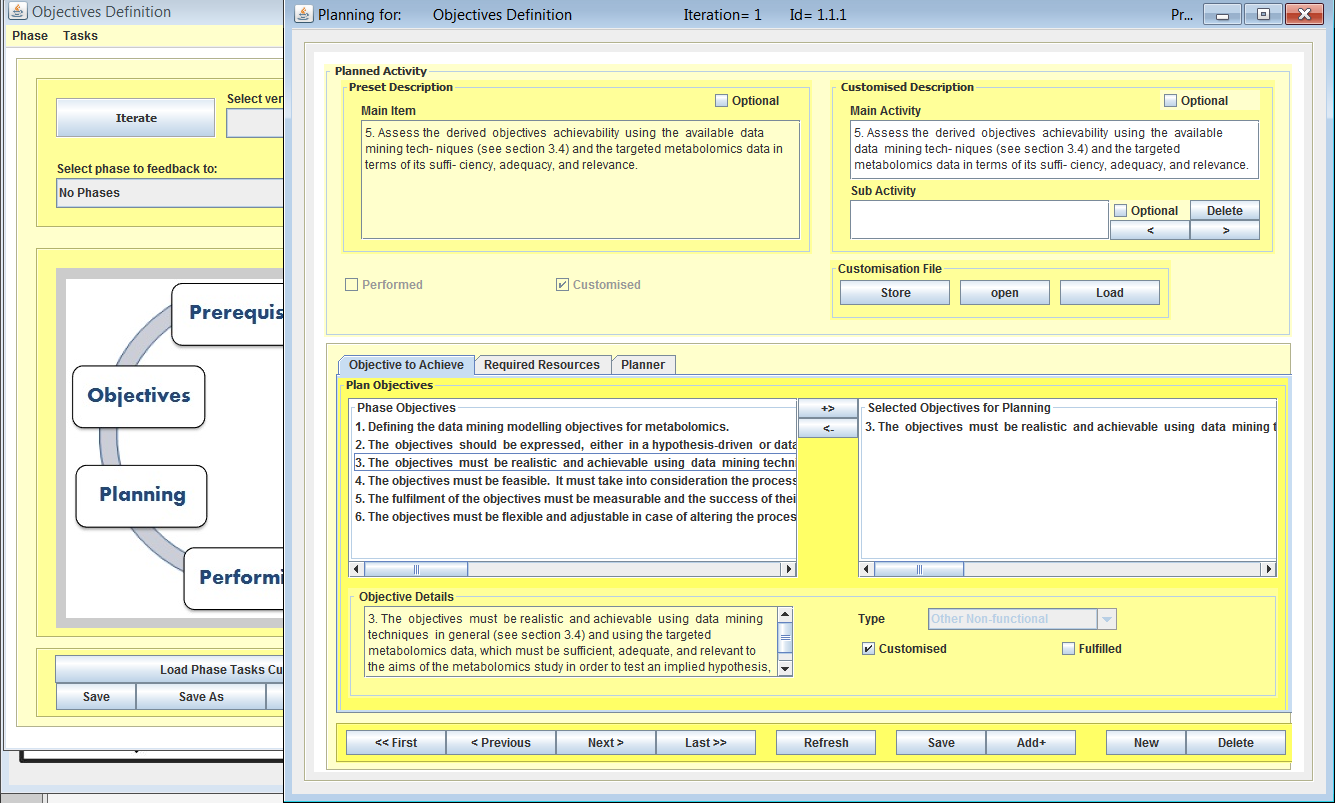}}
\centering
\caption{\textbf{Phase Planning:} A snapshot of a phase planning as captured in the \emph{Arabidopsis} isoprenoids profiling application}
\label{fig:Phase_HiMetIP9_Planning}
\end{figure*}

\end{landscape}

\begin{landscape}
\pagestyle {plain}
\scriptsize

\begin{figure*}[htb!h] 
\centerline{\includegraphics[width=1.3\textwidth]{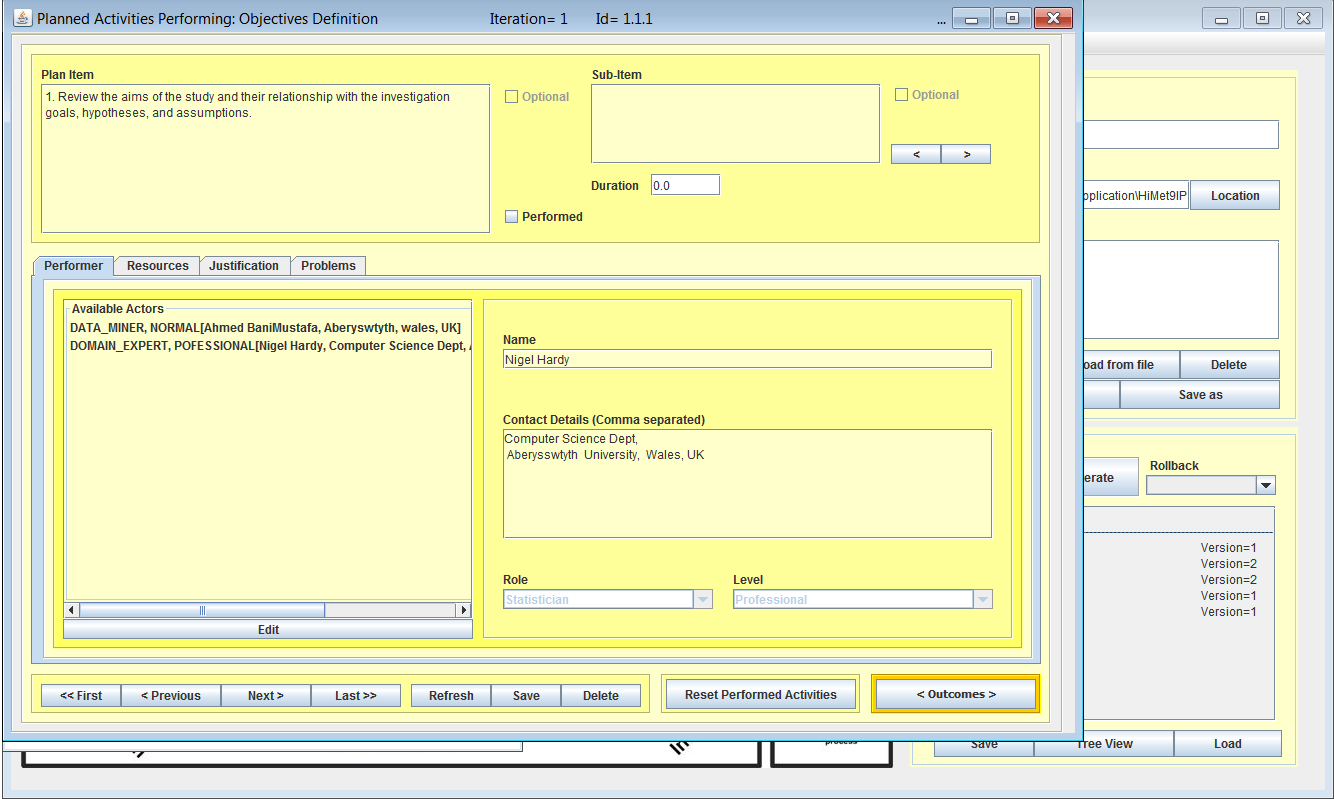}}
\centering
\caption{\textbf{Phase Activity Performer:} A snapshot of a performer, which was assigned to a phase activity in the \emph{Arabidopsis} isoprenoids application}
\label{fig:Phase_HiMetIP9_Performing_Performer}
\end{figure*}

\end{landscape}

\begin{landscape}
\pagestyle {plain}
\scriptsize

\begin{figure*}[htb!h] 
\centerline{\includegraphics[width=1.3\textwidth]{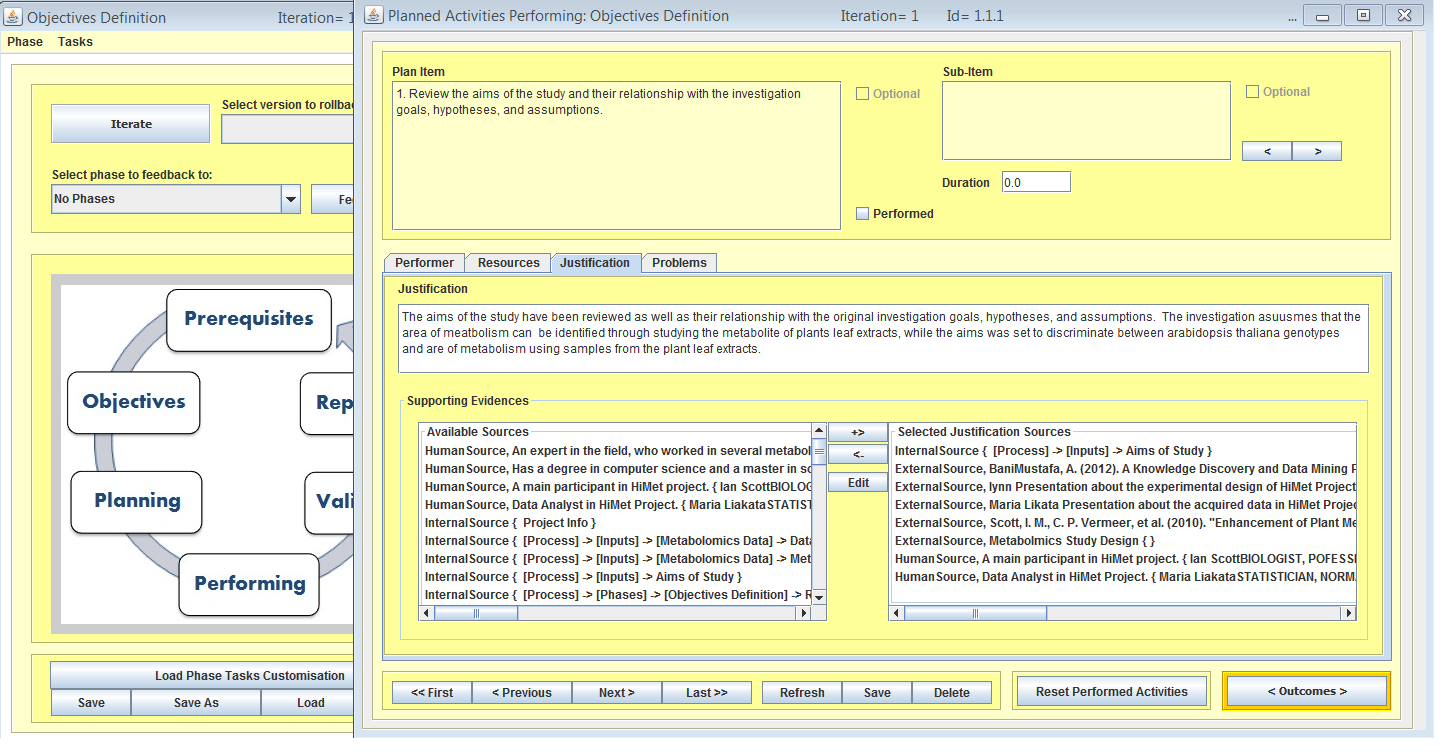}}
\centering
\caption{\textbf{Phase Performing Justification:} A snapshot of a phase performing justification as captured in the \emph{Arabidopsis} isoprenoids profiling application}
\label{fig:Phase_HiMetIP9_Performing_Justification}
\end{figure*}

\end{landscape}

\begin{landscape}
\pagestyle {plain}
\scriptsize

\begin{figure*}[htb!h] 
\centerline{\includegraphics[width=1.3\textwidth]{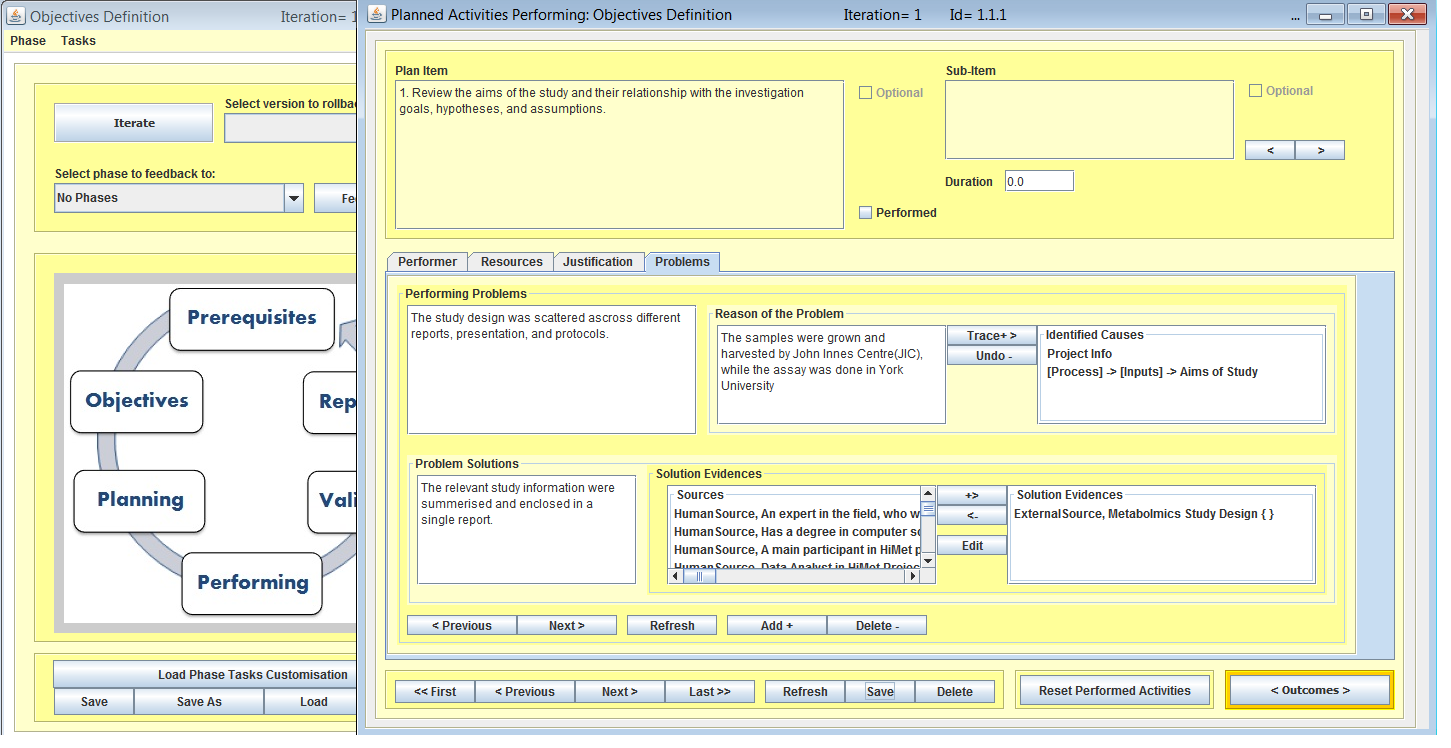}}
\centering
\caption{\textbf{Phase Performing Problem:} A snapshot of a phase performing problem as captured in the \emph{Arabidopsis} isoprenoids profiling application}
\label{fig:Phase_HiMetIP9_Performing_Problems}
\end{figure*}

\end{landscape}

\begin{landscape}
\pagestyle {plain}
\scriptsize

\begin{figure*}[htb!h] 
\centerline{\includegraphics[width=1.3\textwidth]{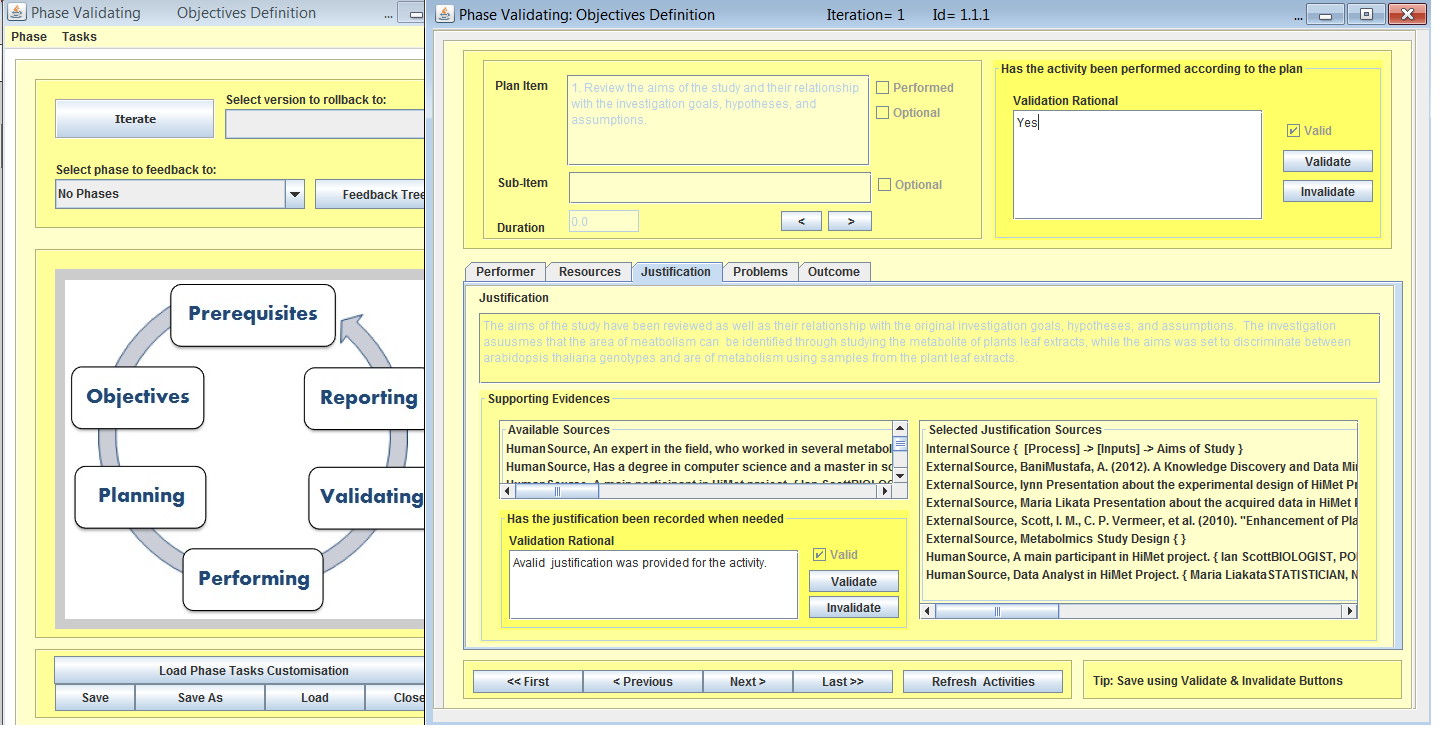}}
\centering
\caption{\textbf{Phase Validating:} A snapshot of a phase activity validation as captured in the \emph{Arabidopsis} isoprenoids profiling application}
\label{fig:Phase_HiMetIP9_Validating}
\end{figure*}

\end{landscape}

\begin{landscape}
\pagestyle {plain}
\scriptsize
\subsection*{Example Outcomes of the Process Phases}
\label{fig:PhaseOutcomeSnapshots}

\begin{figure*}[htb!h] 
\centerline{\includegraphics[width=1.3\textwidth]{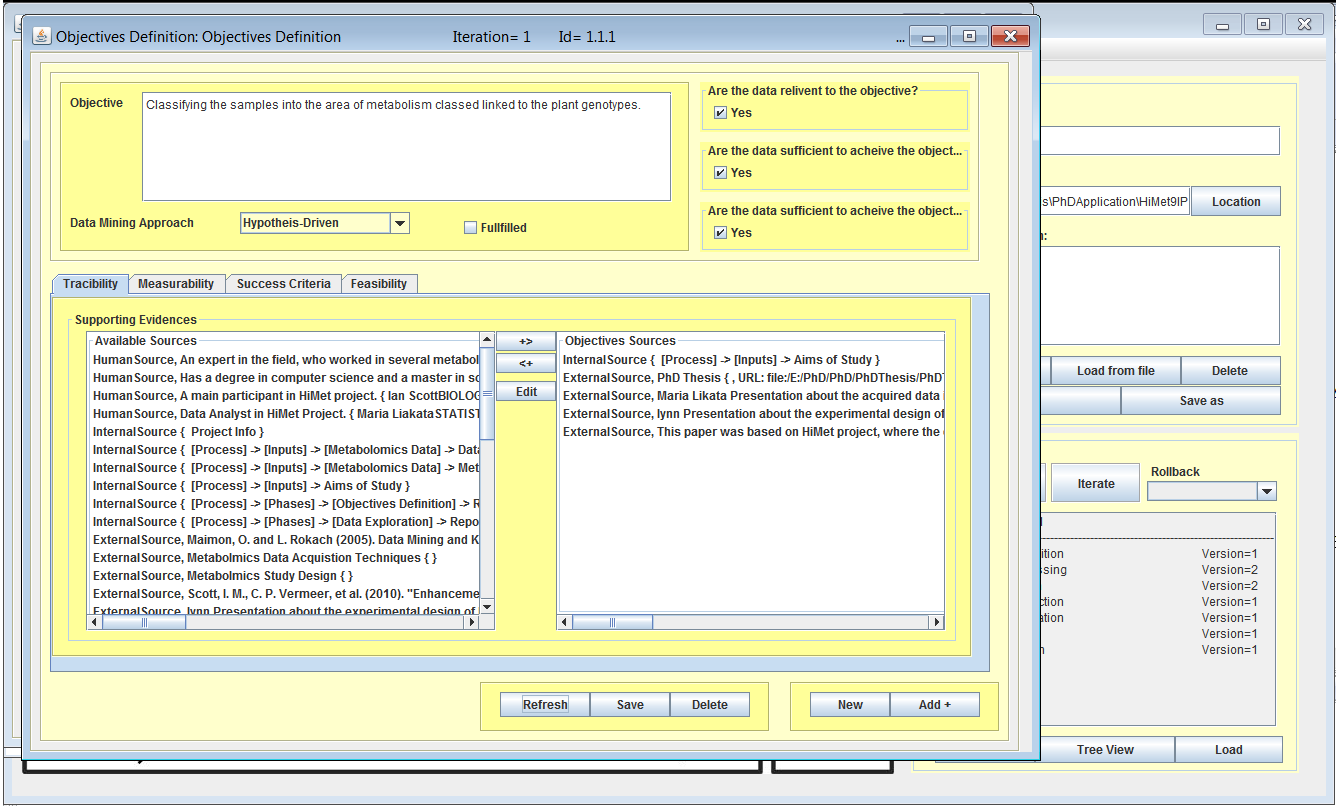}}
\centering
\caption{\textbf{Objectives Definition Phase Outcome:} A snapshot of the objectives definition phase Outcome as captured in the \emph{Arabidopsis} isoprenoids profiling application}
\label{fig:Phase_HiMetIP9_Outcome_ObjectivesDefinition}
\end{figure*}
\end{landscape}

\begin{landscape}
\pagestyle {plain}
\scriptsize

\begin{figure*}[htb!h] 
\centerline{\includegraphics[width=1.3\textwidth]{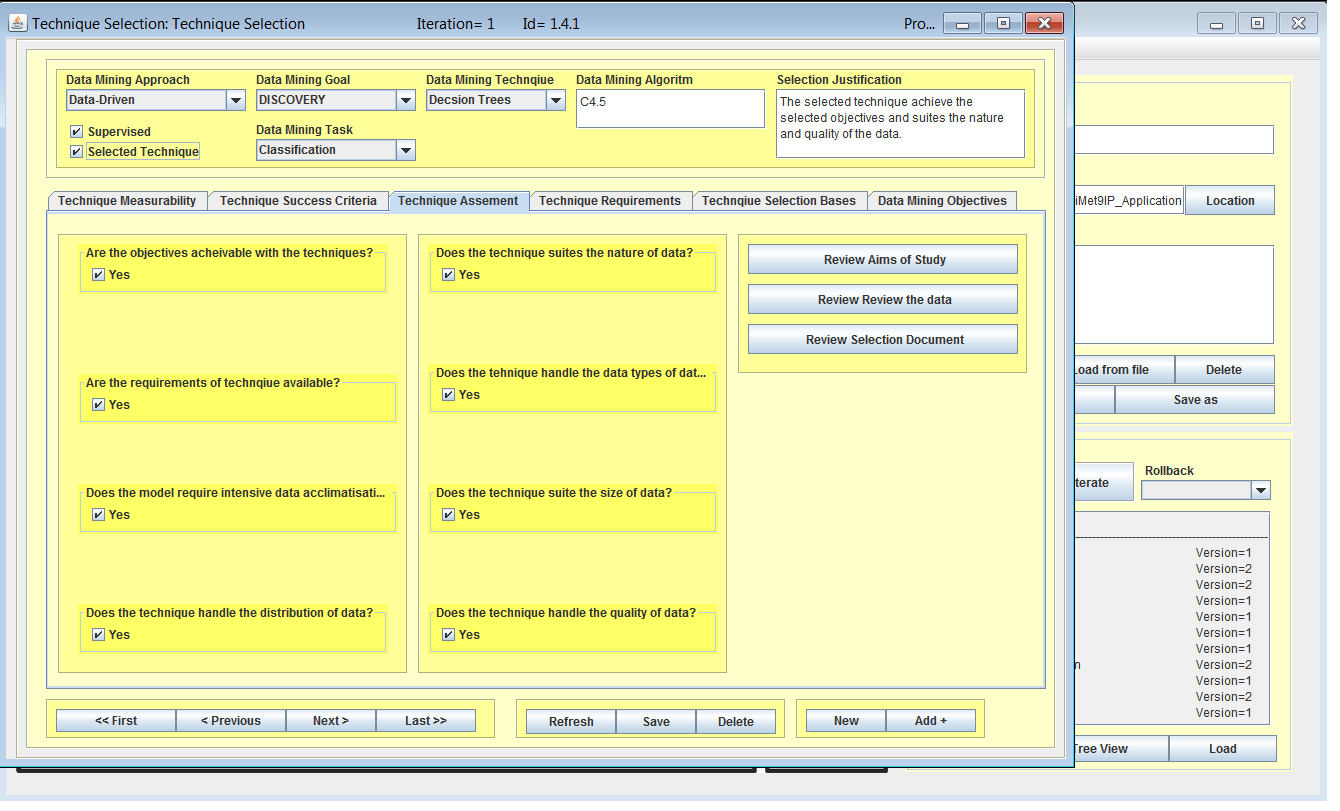}}
\centering
\caption{\textbf{Technique Selection Phase Outcome:} A snapshot of the technique selection phase outcome as captured in the \emph{Arabidopsis} isoprenoids profiling application}
\label{fig:Phase_HiMetIP9_Outcome_TechniqueSelection}
\end{figure*}
\end{landscape}

\begin{landscape}
\pagestyle {plain}
\scriptsize
\newpage
\subsection*{Practical Supplements \& Traceability}
\label{fig:ProcessSupplementsSnapshots}

\begin{figure*}[htb!h] 
\centerline{\includegraphics[width=1.3\textwidth]{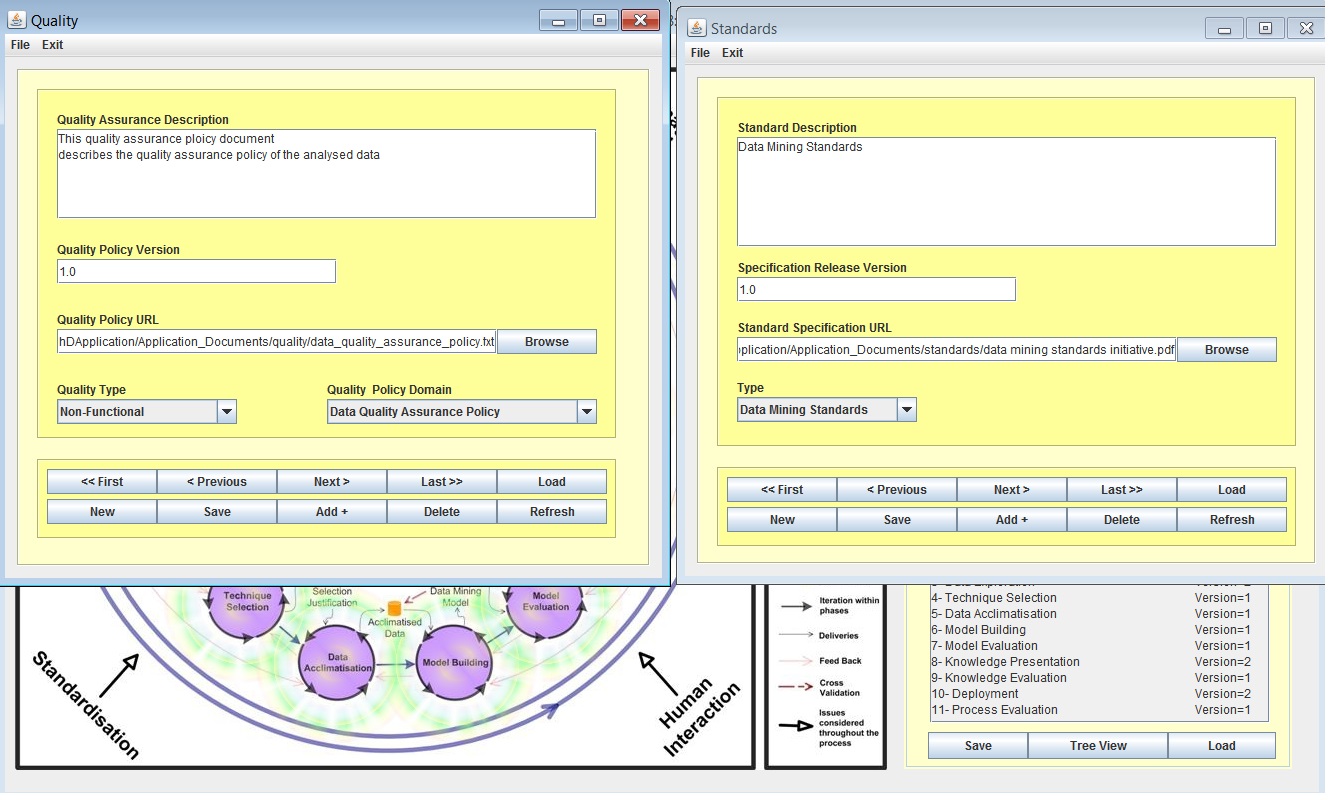}}
\centering
\caption{\textbf{Process Quality Assurance and Standards:} A snapshot of the process quality assurance and standards as captured \emph{Arabidopsis} isoprenoids profiling application}
\label{fig:Supplemnts_HiMetIP9_QA_Standards}
\end{figure*}

\end{landscape} 
\begin{landscape}
\pagestyle {plain}
\scriptsize
\newpage
\label{fig:ProcessStandardsSnapshots}

\begin{figure*}[htb!h] 
\centerline{\includegraphics[width=1.3\textwidth]{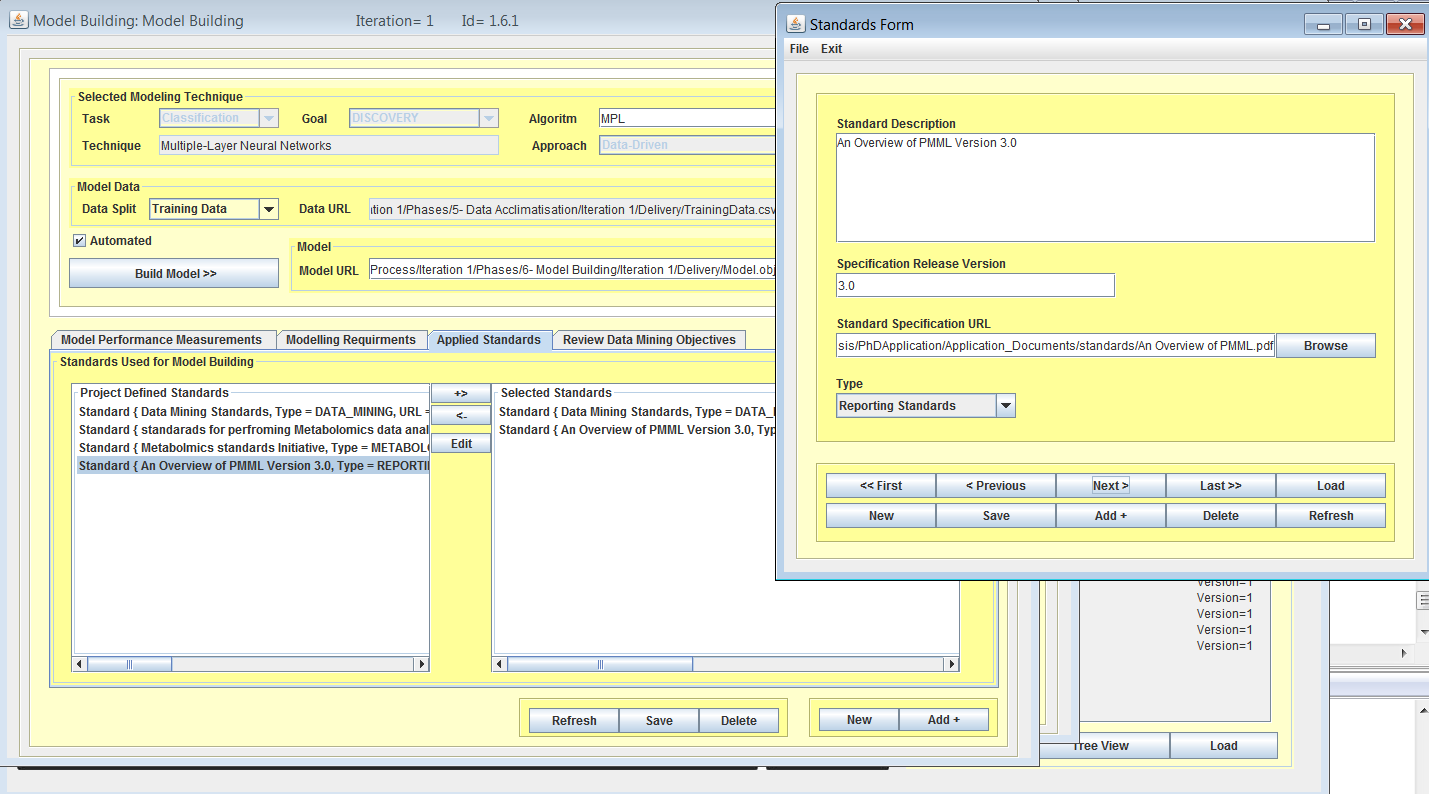}}
\centering
\caption{\textbf{The Assignment of Process Standards:} A snapshot of the assignment of standards when delivering the model in the \emph{Arabidopsis} isoprenoids application}
\label{fig:Supplemnts_HiMetIP9_Standards}
\end{figure*}

\end{landscape}

\begin{landscape}
\pagestyle {plain}
\scriptsize

\begin{figure*}[htb!h] 
\centerline{\includegraphics[width=1.3\textwidth]{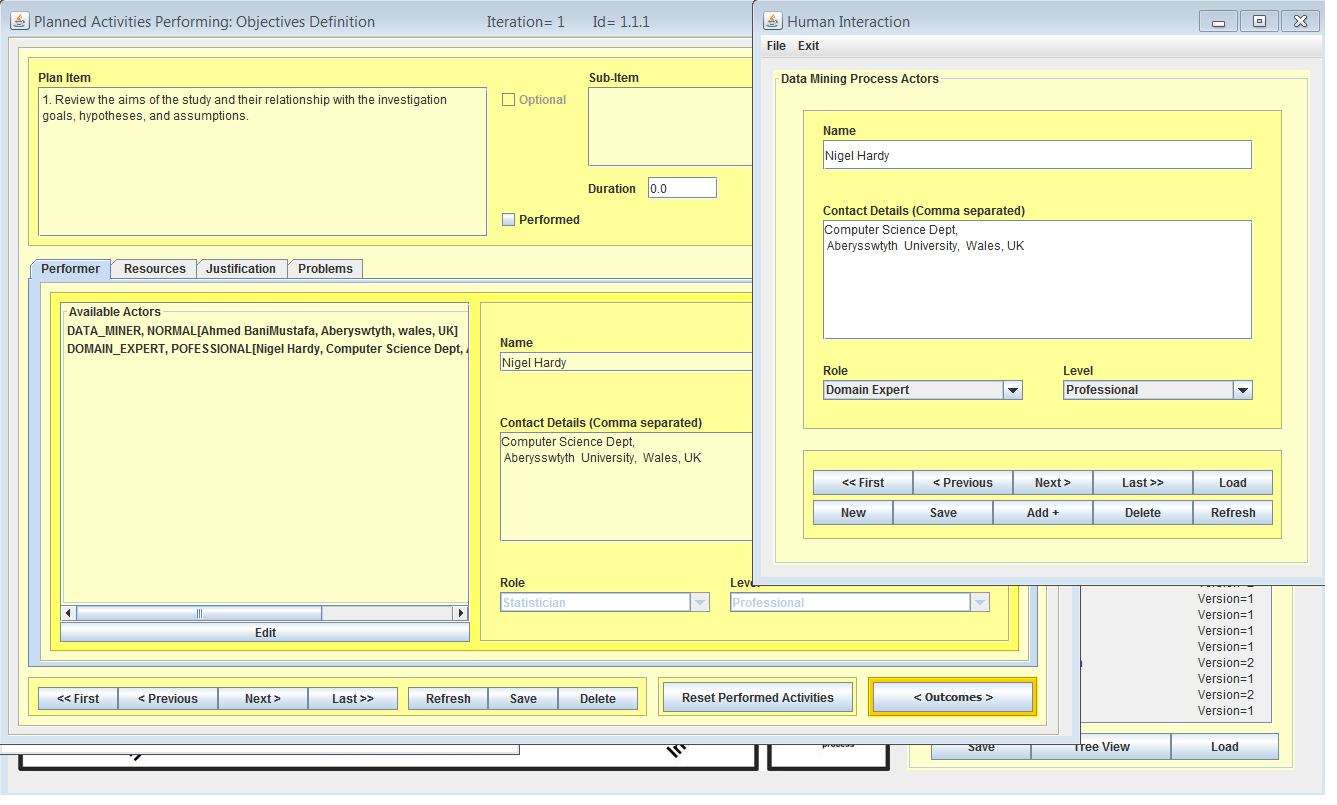}}
\centering
\caption{\textbf{Process Human Interaction:} A snapshot of the process human interaction as captured in the \emph{Arabidopsis} isoprenoids profiling application}
\label{fig:Supplemnts_HiMetIP9_HI}
\end{figure*}

\end{landscape}

\begin{landscape}
\pagestyle {plain}
\scriptsize

\begin{figure*}[htb!h] 
\centerline{\includegraphics[width=1.3\textwidth]{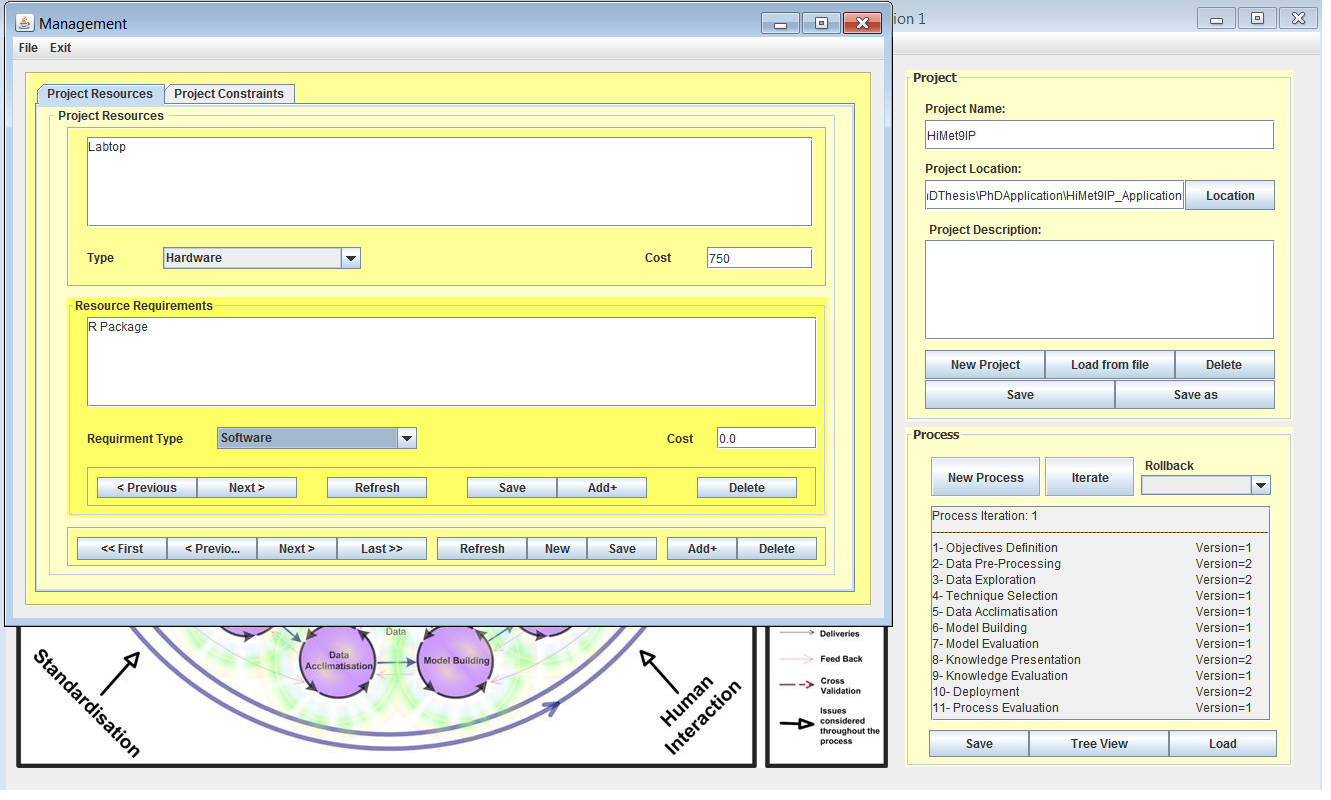}}
\centering
\caption{\textbf{Process Resources Management:} A snapshot of the process management resources as captured in the \emph{Arabidopsis} isoprenoids profiling application}
\label{fig:Supplemnts_HiMetIP9_Management_Resources}
\end{figure*}

\end{landscape}

\begin{landscape}
\pagestyle {plain}
\scriptsize
\begin{figure*}[htb!h] 
\centerline{\includegraphics[width=1.3\textwidth]{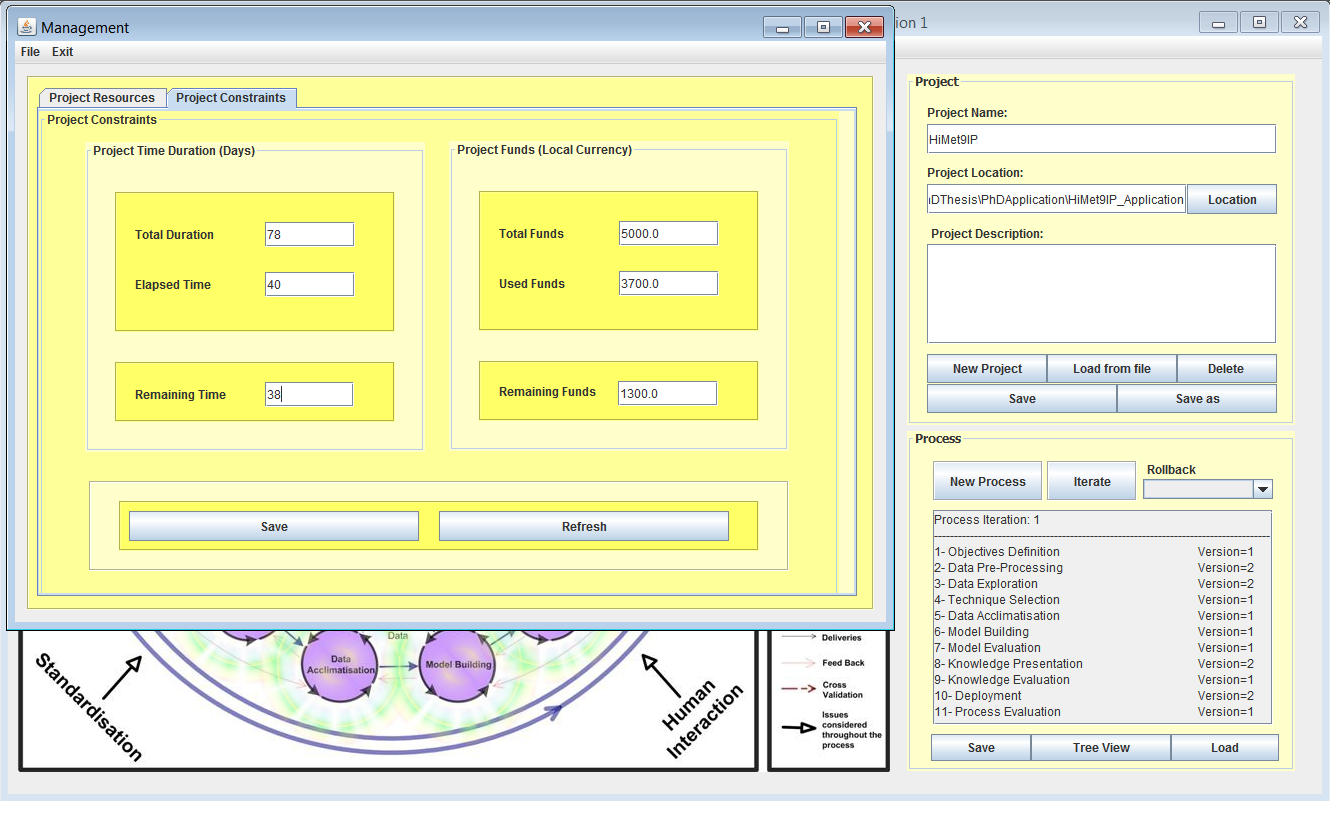}}
\centering
\caption{\textbf{Process Management Constraints:} A snapshot of the process management constraints as captured in the \emph{Arabidopsis} isoprenoids profiling application}
\label{fig:Supplemnts_HiMetIP9_Management_Constraints}
\end{figure*}

\end{landscape}

\begin{landscape}
\pagestyle {plain}
\scriptsize

\begin{figure*}[htb!h] 
\centerline{\includegraphics[width=1.3\textwidth]{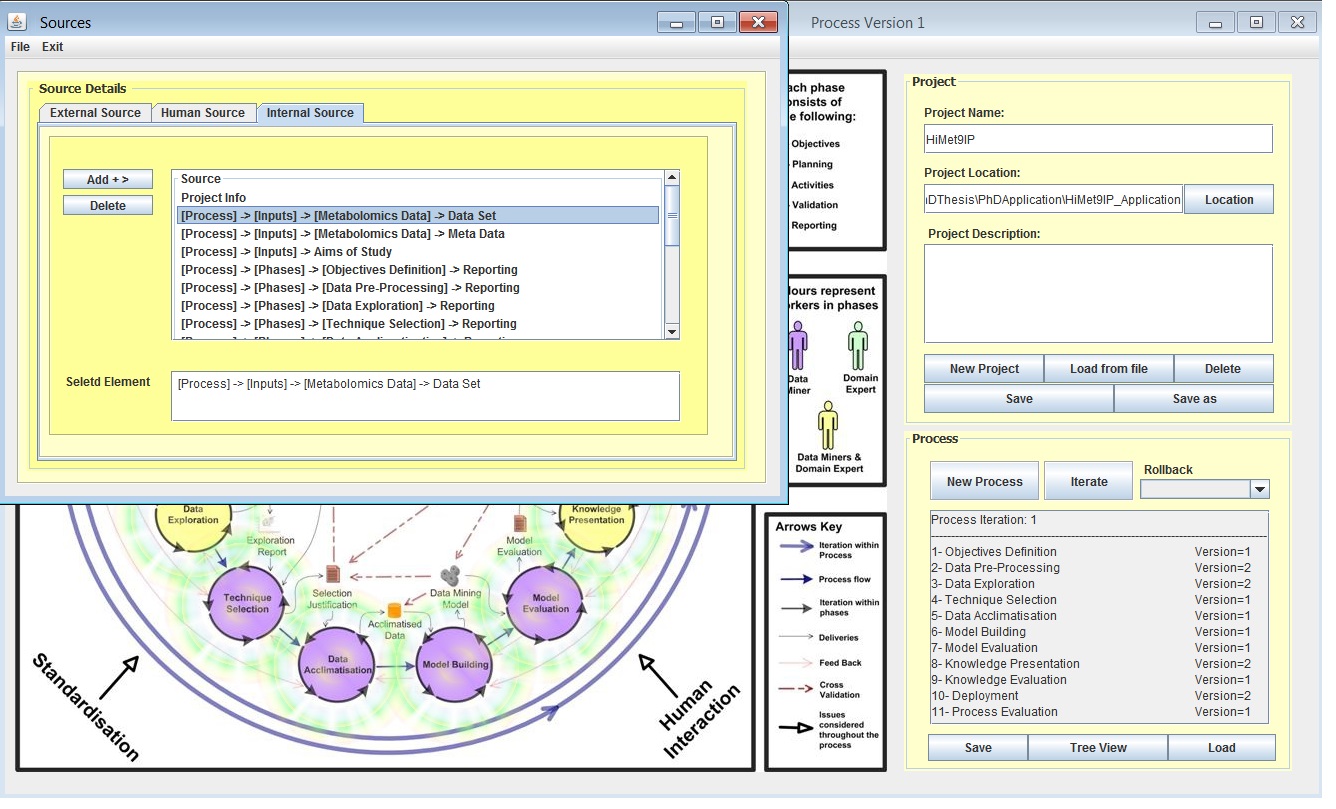}}
\centering
\caption{\textbf{Process traceability:}A snapshot of the process traceability information as captured in the \emph{Arabidopsis} isoprenoids profiling application}
\label{fig:Supplemnts_HiMetIP9_Sources}
\end{figure*}

\end{landscape}

\begin{landscape}
\pagestyle {plain}
\scriptsize
\subsection*{Process Customisation}
\label{fig:ProcessCustomisationSnapshots}
\begin{figure*}[htb!h] 
\centerline{\includegraphics[width=1.1\textwidth]{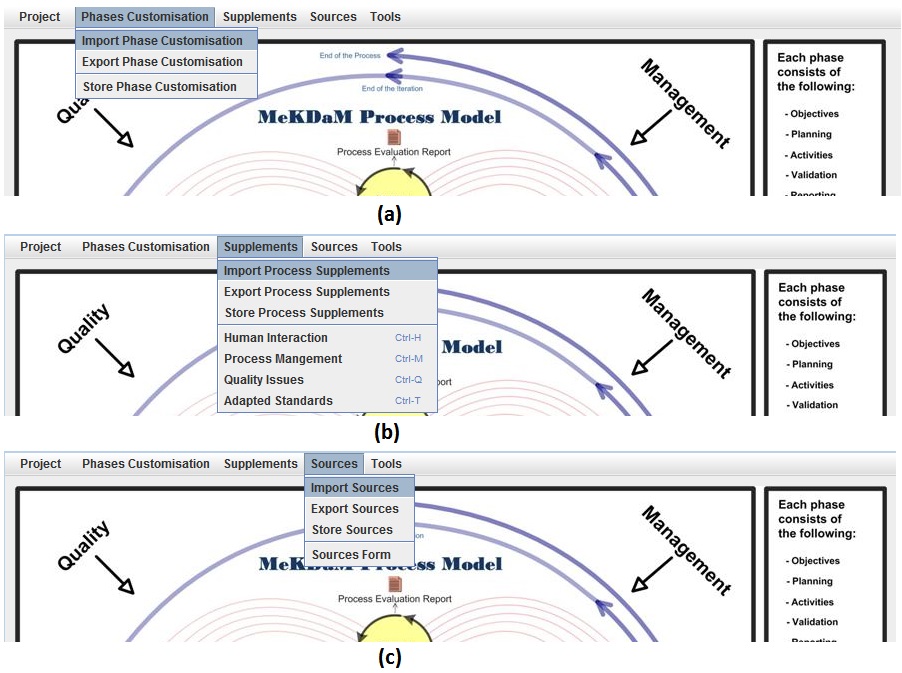}}
\centering
\caption{\textbf{Process Customisation:} A snapshot of the process customisation facilities (a) Phase customisation (b) Customisation of the process practical supplements (c) Traceability customisation}
\label{fig:Process_Customisation}
\end{figure*}

\end{landscape}

\subsection{Appendix: Persistence of the Process Data} \label{Persistence}
XML is a format that is both human and machine readable. Therefore, it was chosen as a mechanism for persisting the process data and reporting its generated deliveries. The following XML files have been generated by the process during the demonstration of the process execution as discussed earlier. The structure of these files reflects the process hierarchic. The (+) indicates a folded information, which might be expanded later as different parts of the same XML file might be illustrated in a different context.

\subsubsection*{Process Hierarchy}\label{Persistence_Process_Hierarchy}
The process is realised as a structured project which holds the process iterations and its performed phases. The XML file illustrates the process structure and provides an example for the process execution. The first XML files provide an example for the process instantiation as a project, while the second file provides an example for the process execution, flow and iteration. The third file provides an example for the process supplements including management constraints and resources, human interaction, standards and quality assurance policies.
\includepdf{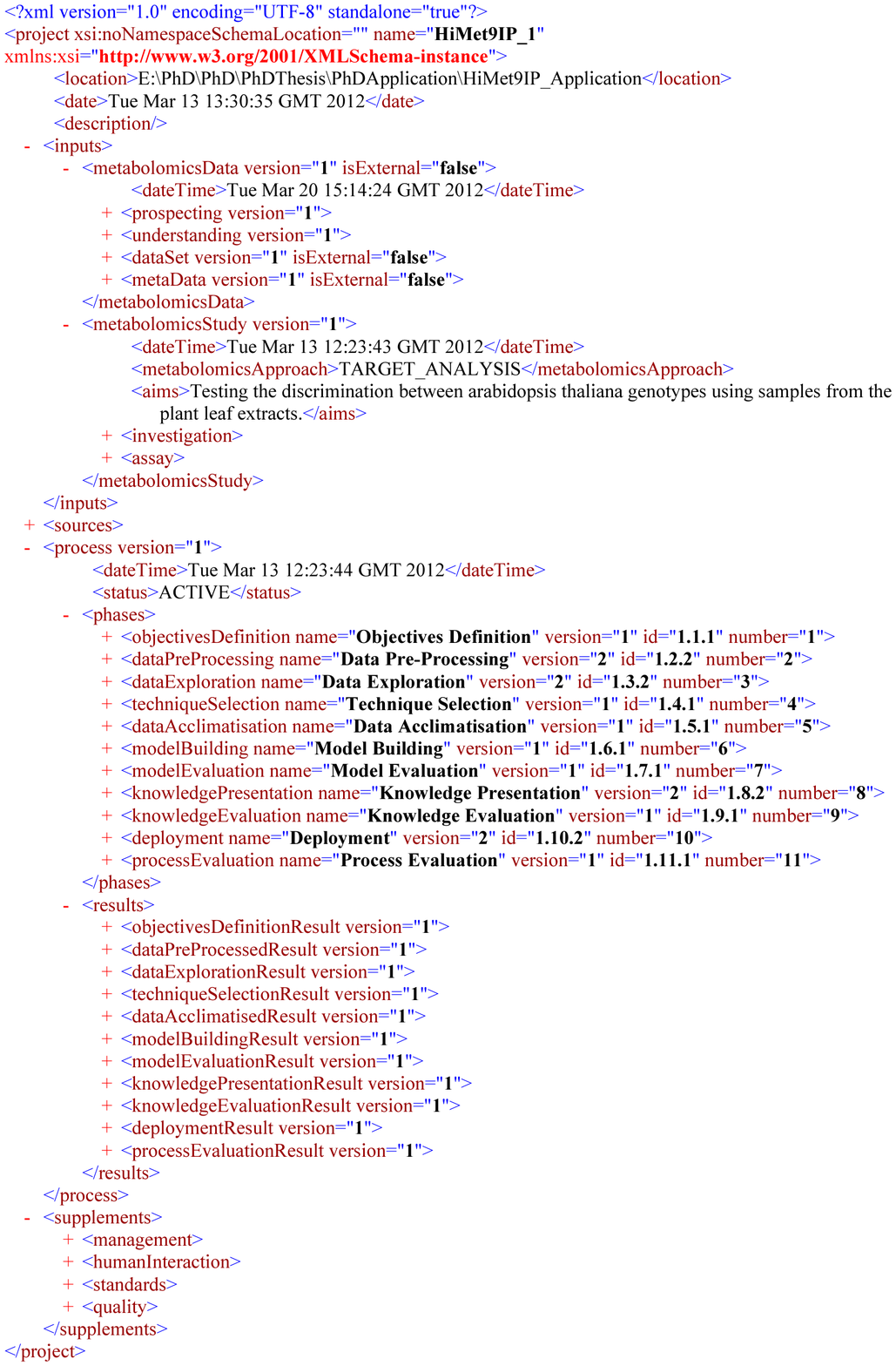}
\includepdf{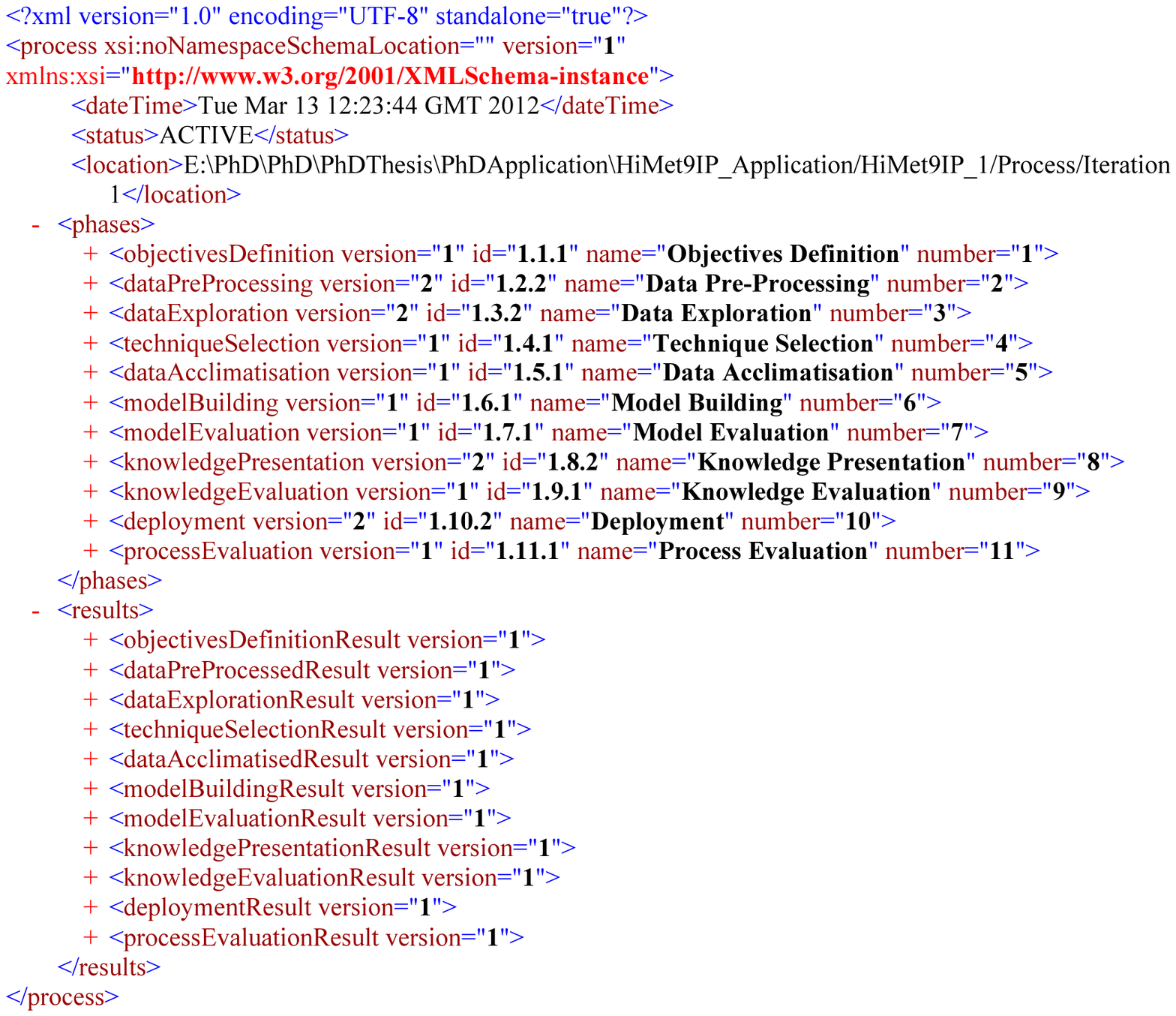}
\includepdf{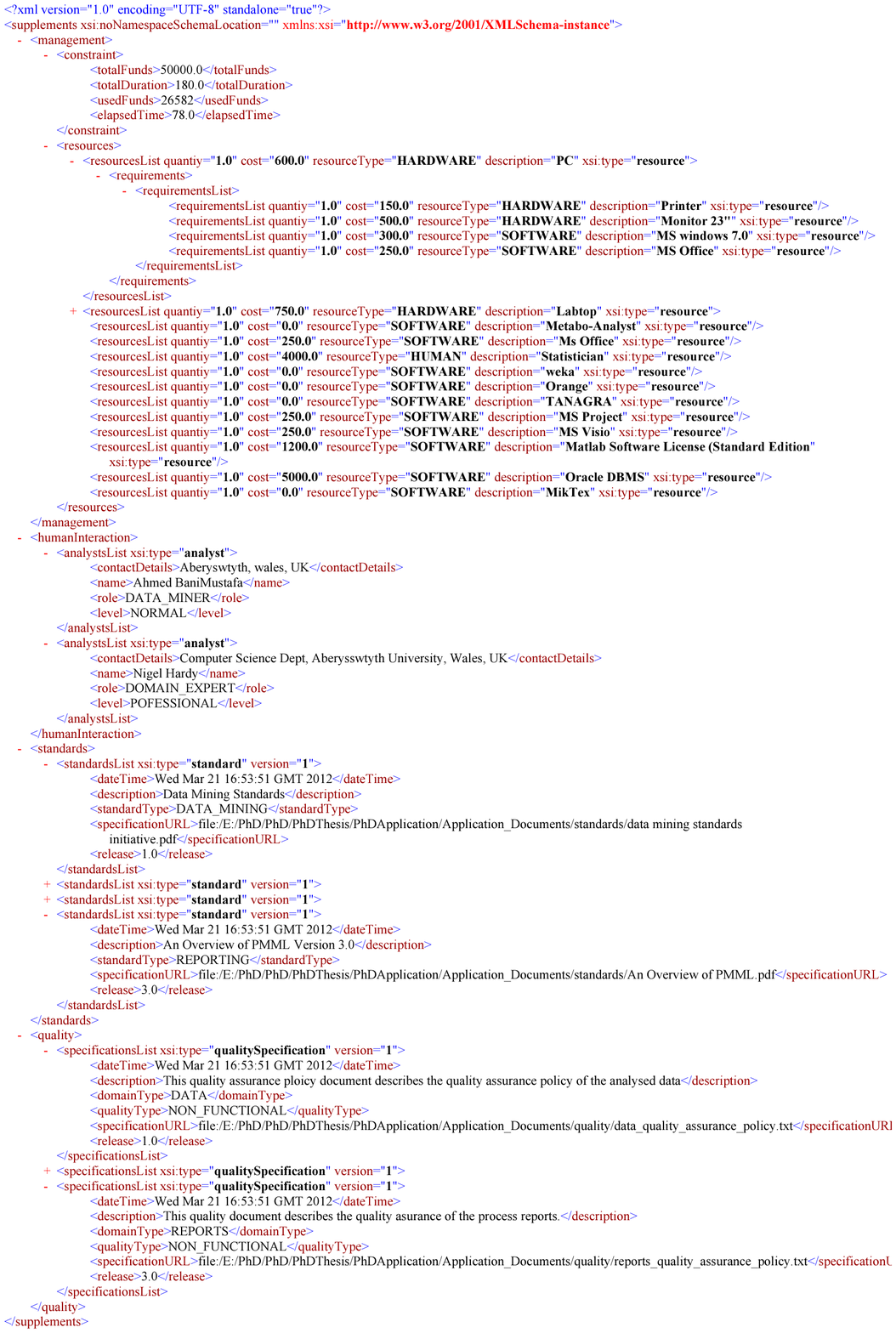}
\subsubsection*{Process Inputs}\label{Persistence_Process_Inputs}
The following three XML files have been generated by MeKDDaM process model in order to persist process inputs including the targeted data and the aims of the study.
\includepdf{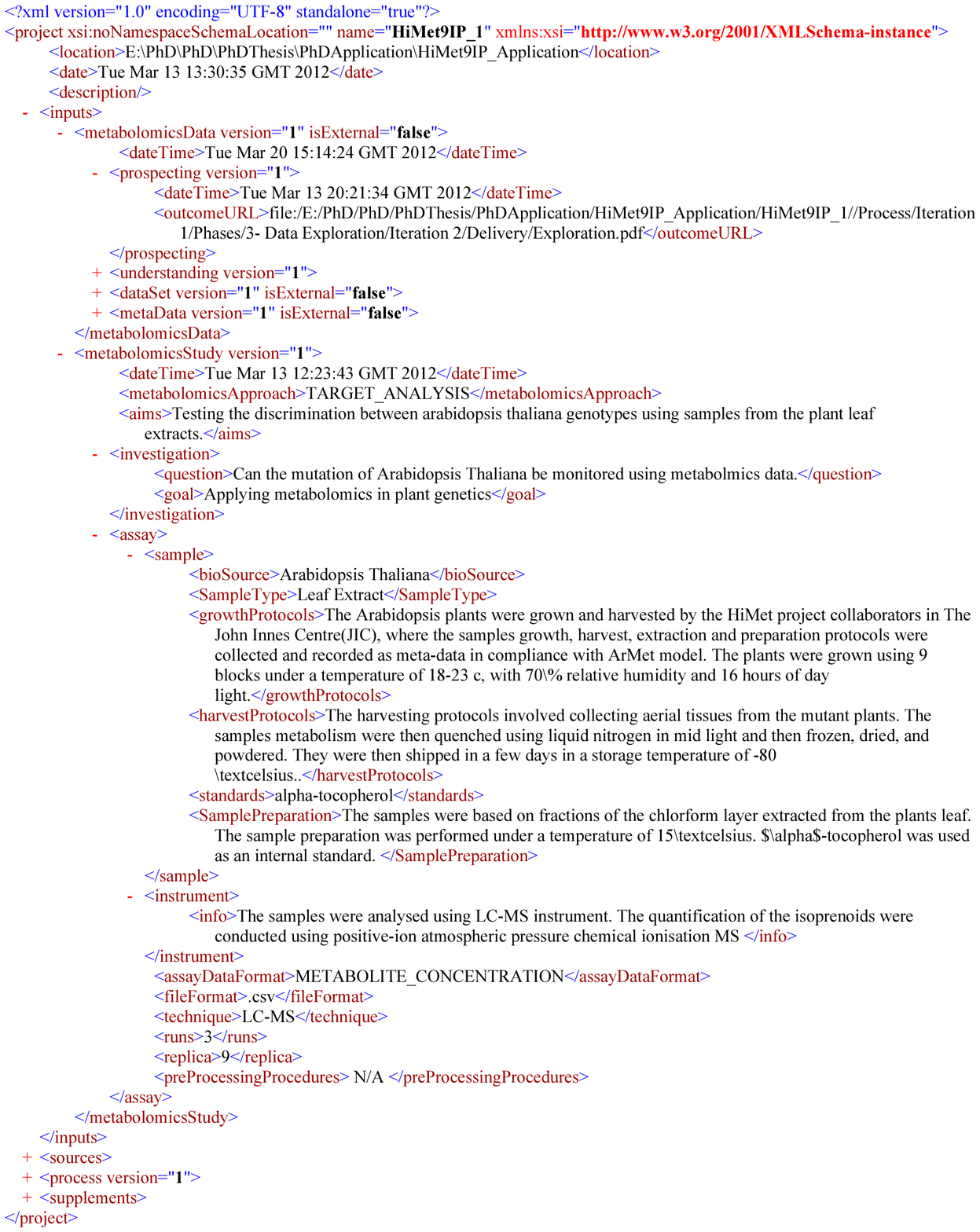}
\includepdf{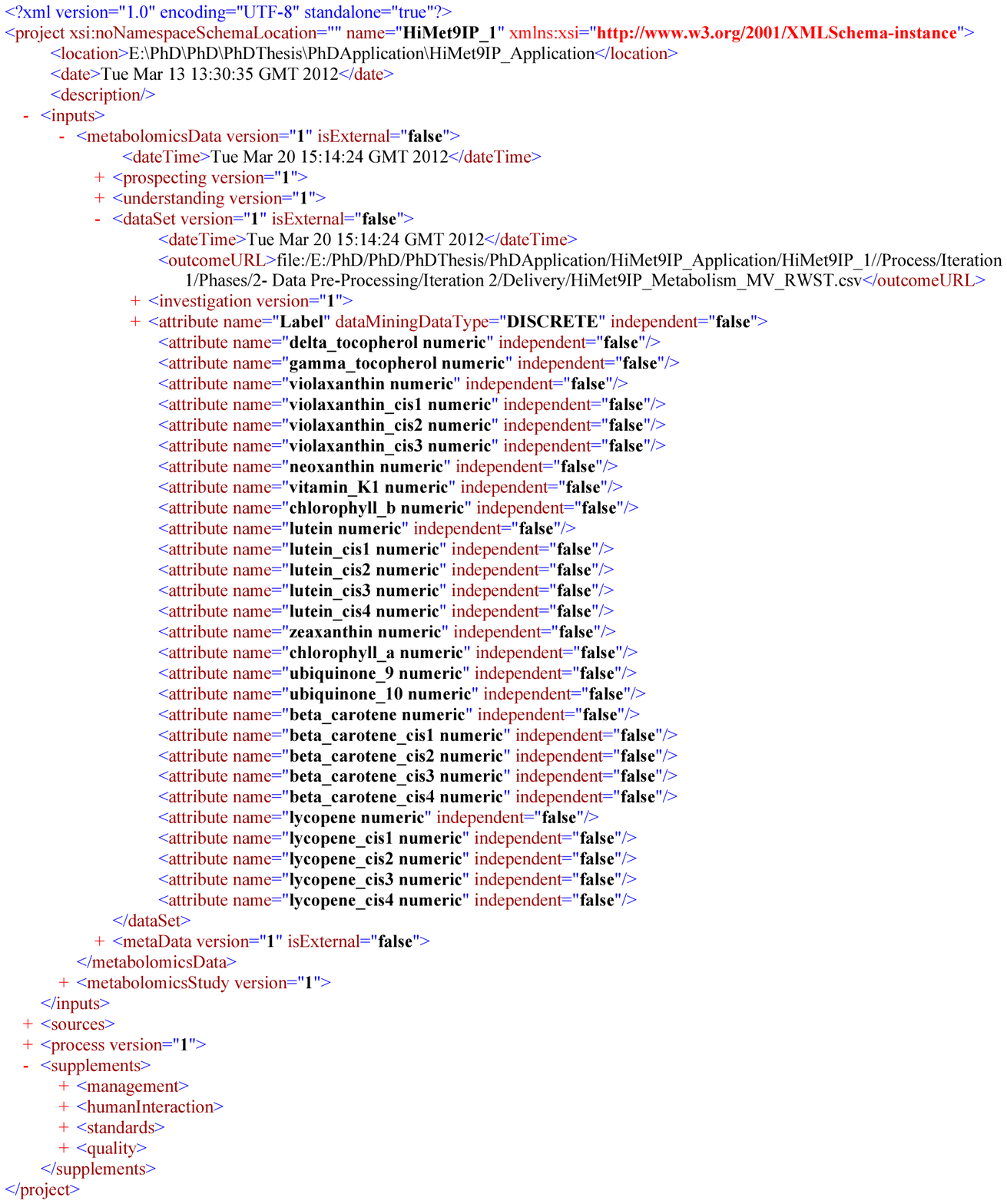}
\includepdf{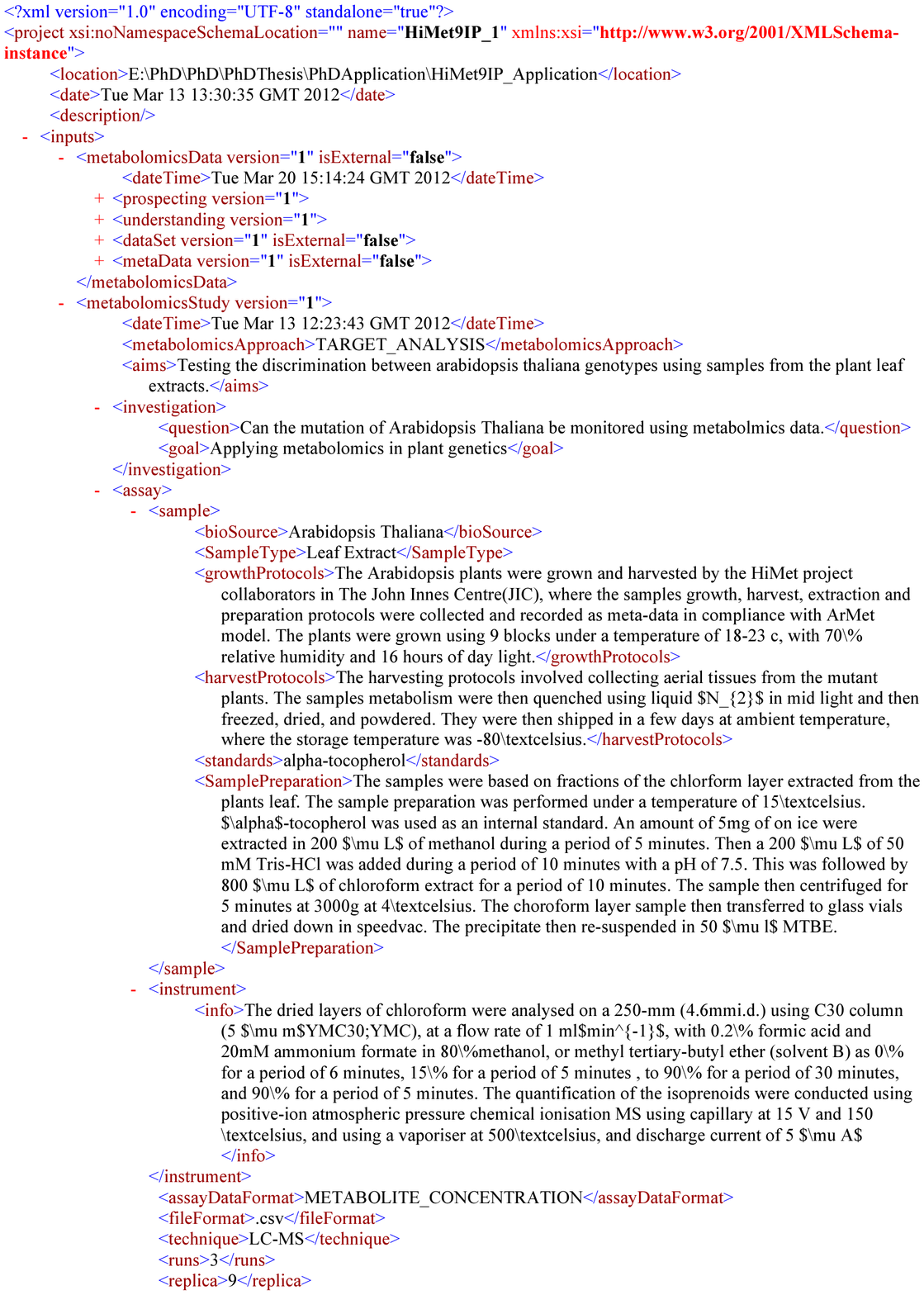}
\subsubsection*{Tasks within the Process Phases}\label{Persistence_Process_Phases}
The following five XML files demonstrate the persistence of MeKDDaM process model phases. The structure of these example XML files reflect the hierarchy of the process model phases and holds information regarding their iteration, internal tasks,
and deliveries. The first XML file illustrates an example of phase prerequisites, while the second provides examples of the objectives. The third file shows an example of planning of a phase, while the fourth shows an example of its performing showing examples of an activity justification and traceability. The fifth XML file shows an example of phase validation.
\includepdf{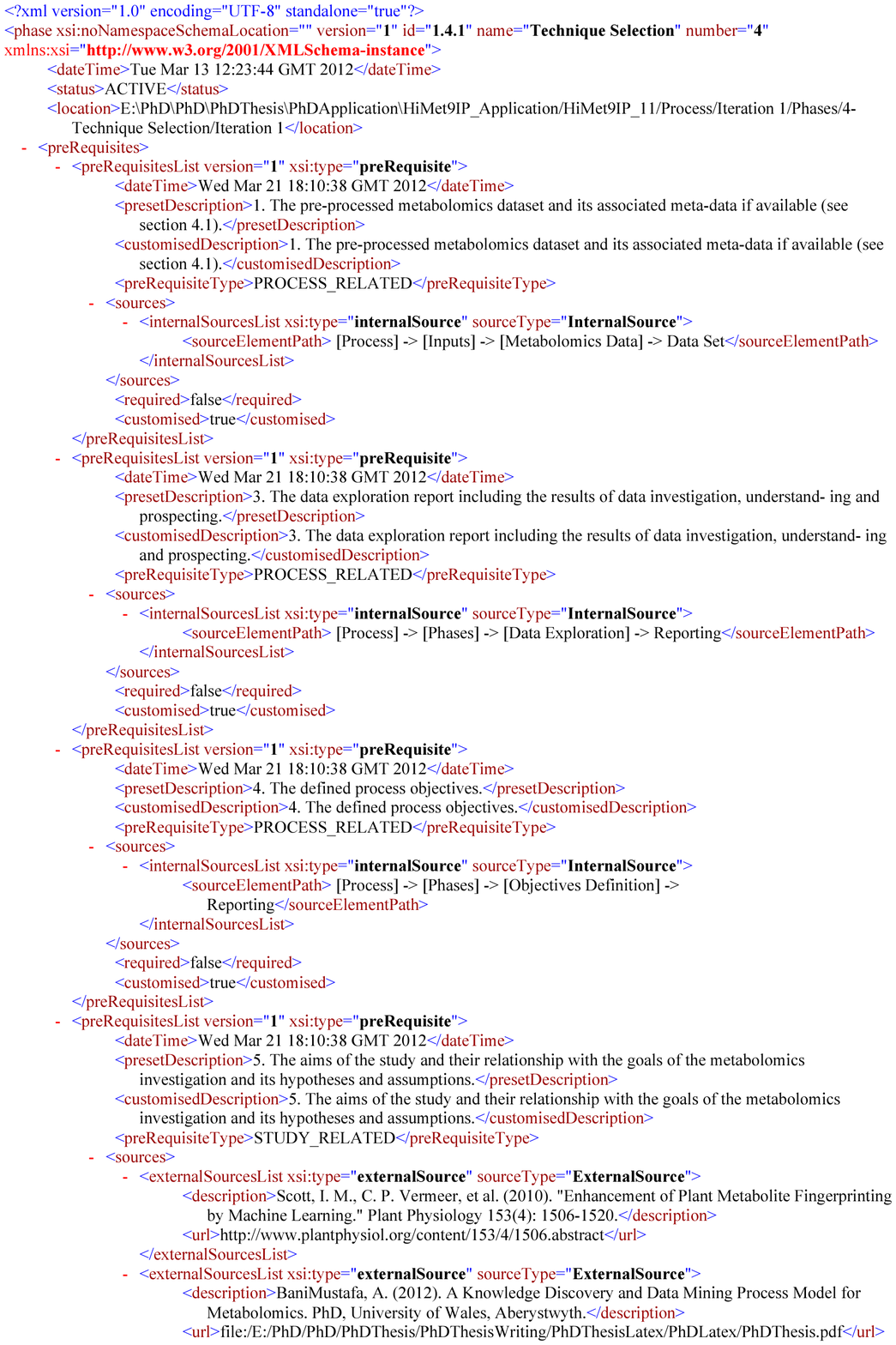}
\includepdf{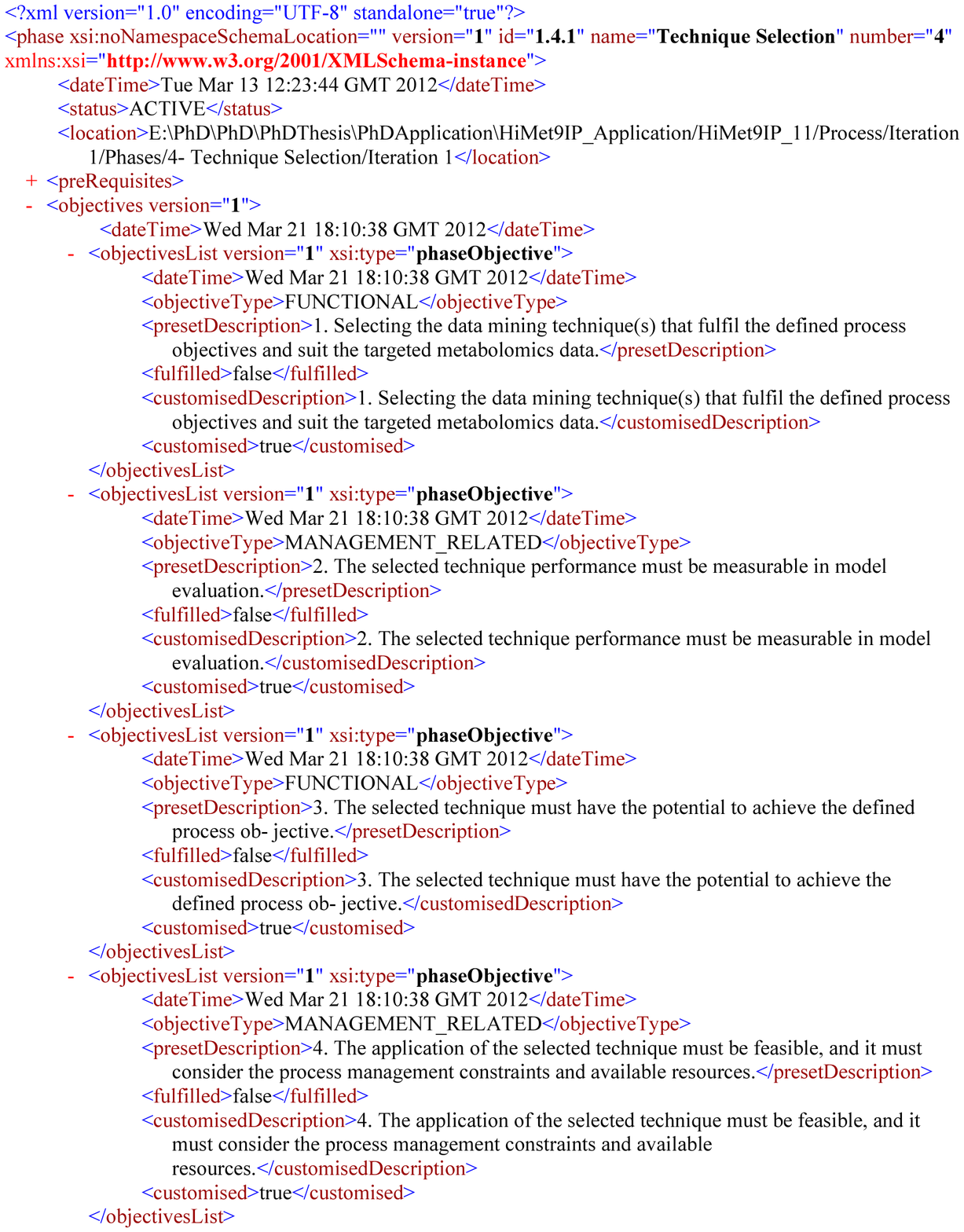}
\includepdf{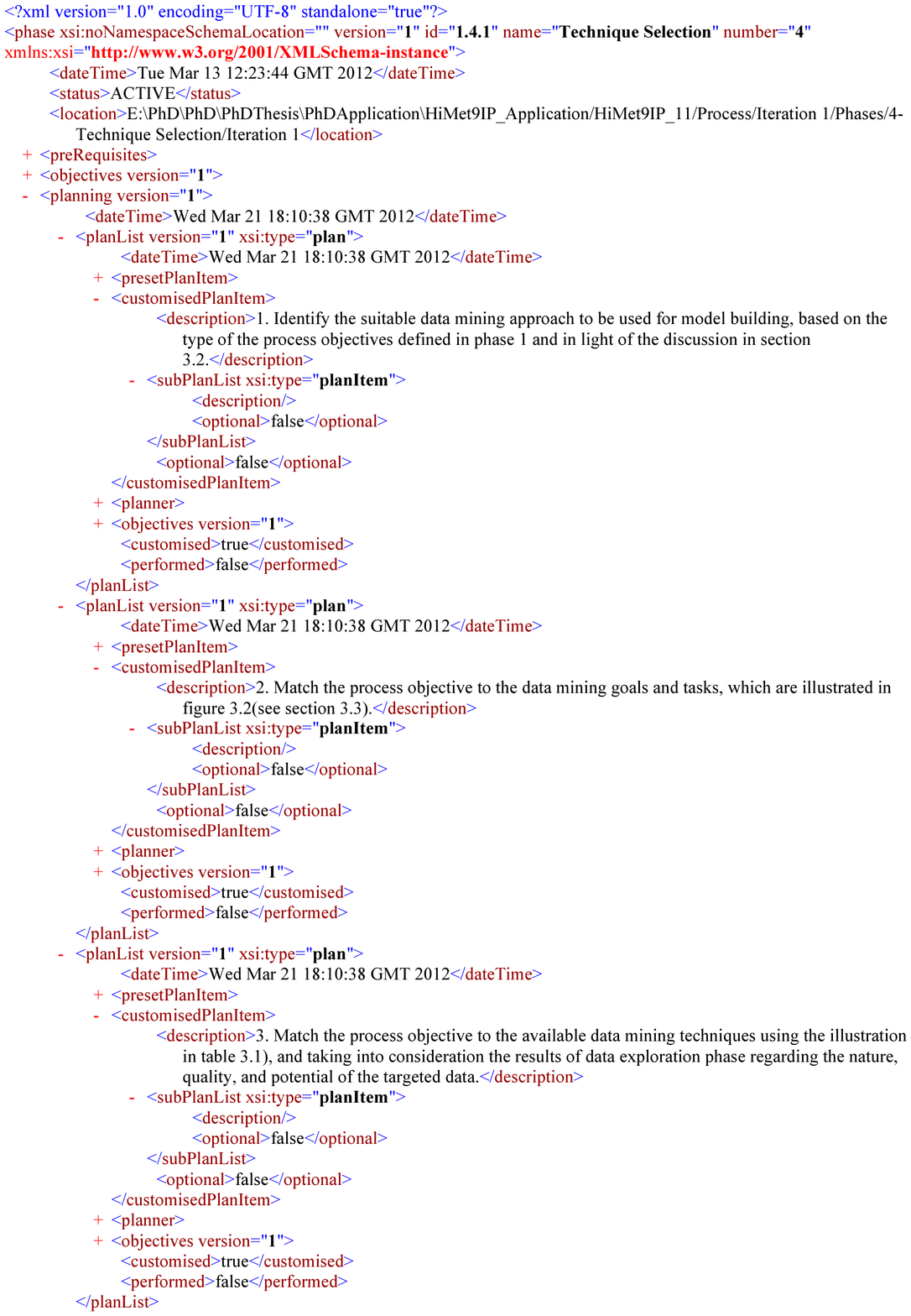}
\includepdf{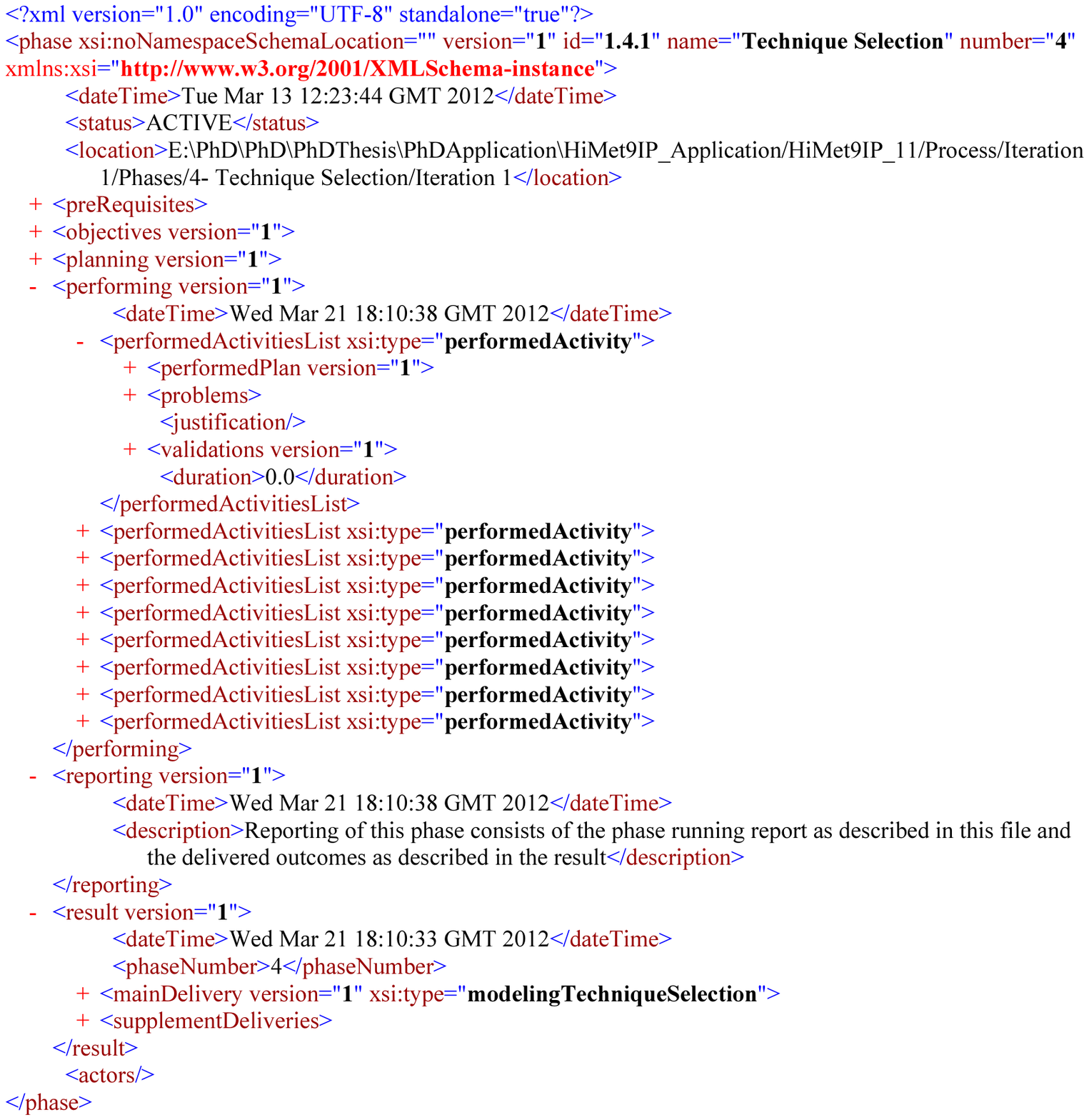}
\includepdf{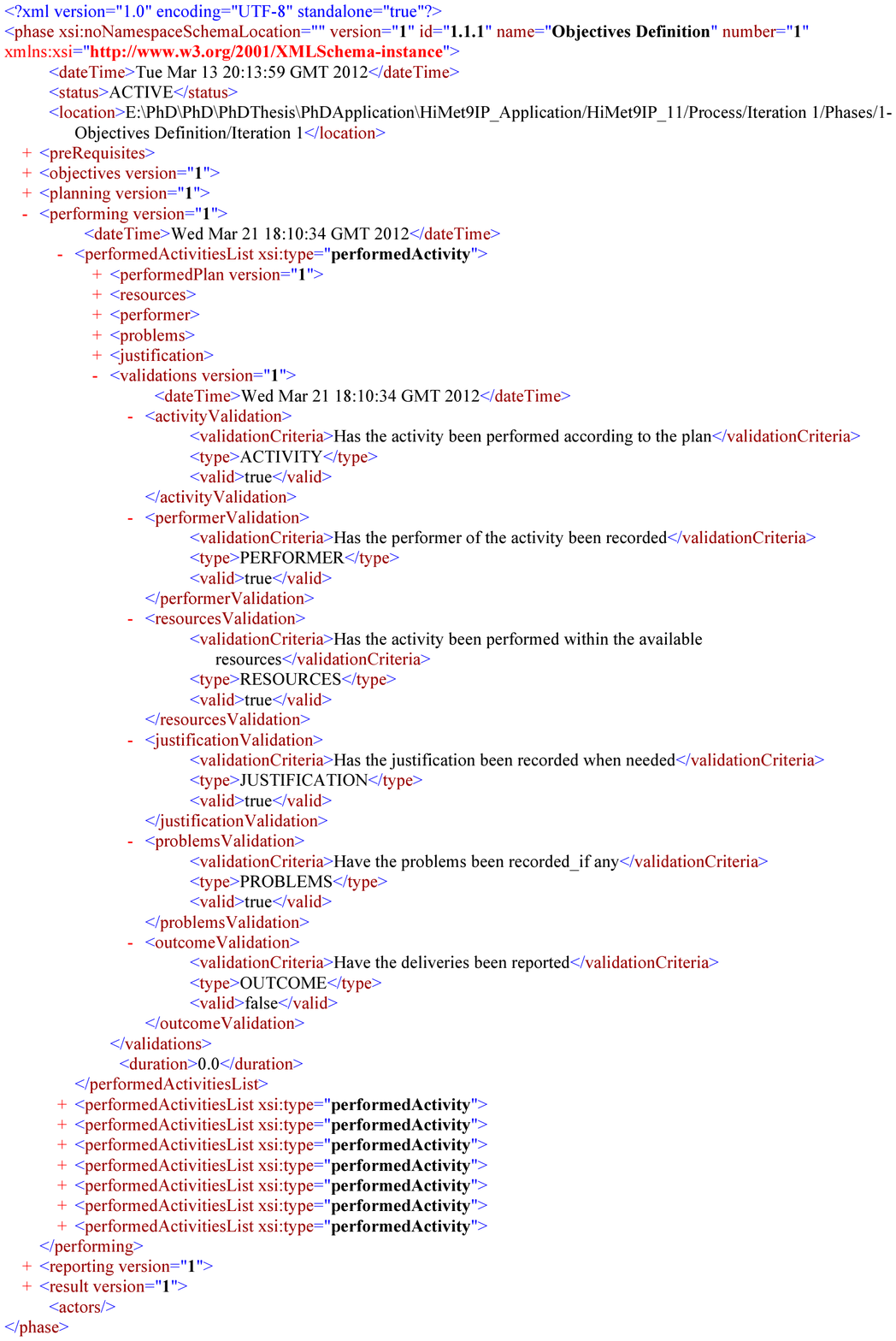}
\subsubsection*{Phase Deliveries}\label{Persistence_Process_Deliveries}
The following five files demonstrate examples of some of the informative deliveries generated by some of the process phases using XML format.
\includepdf{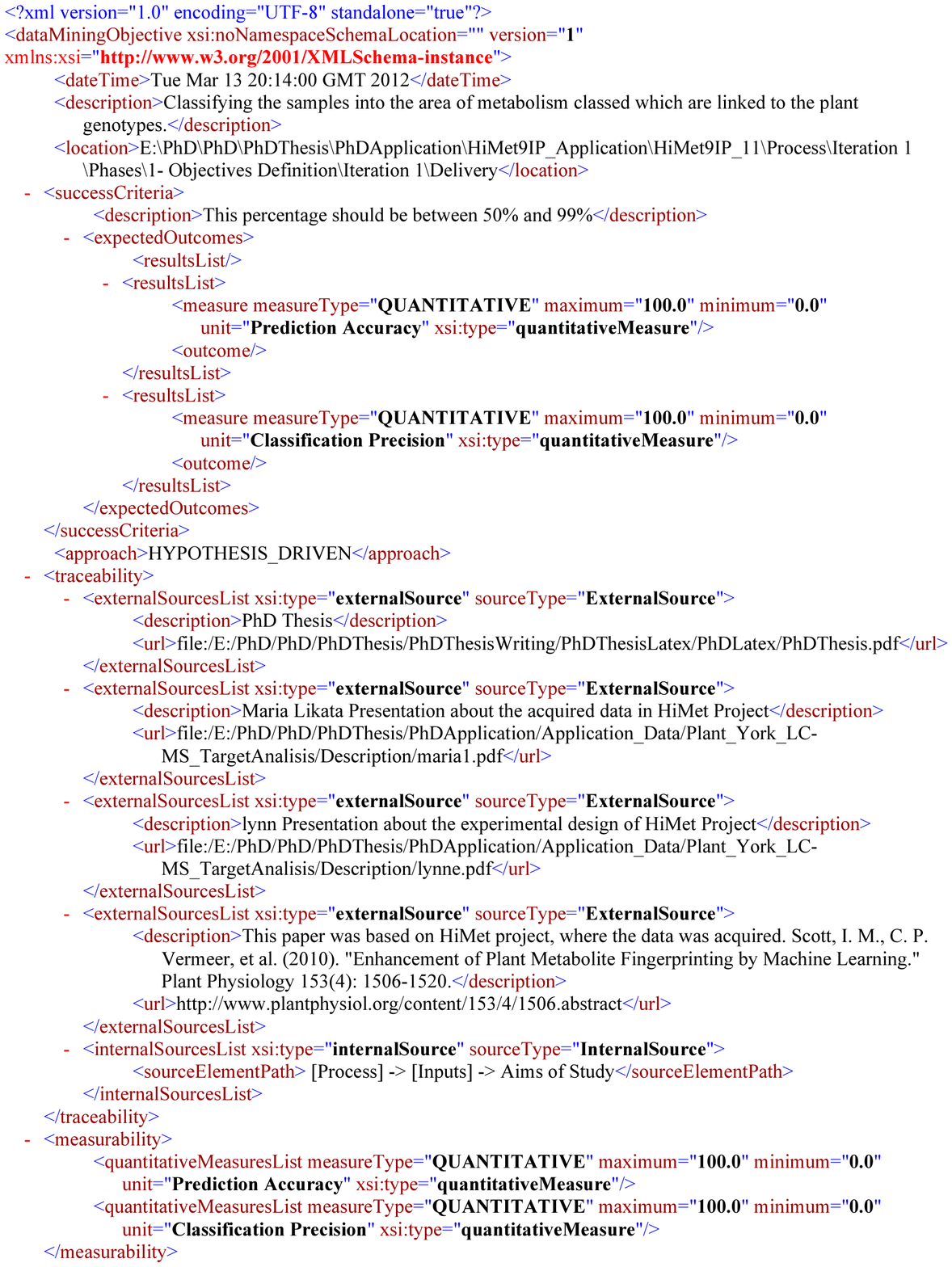}
\includepdf{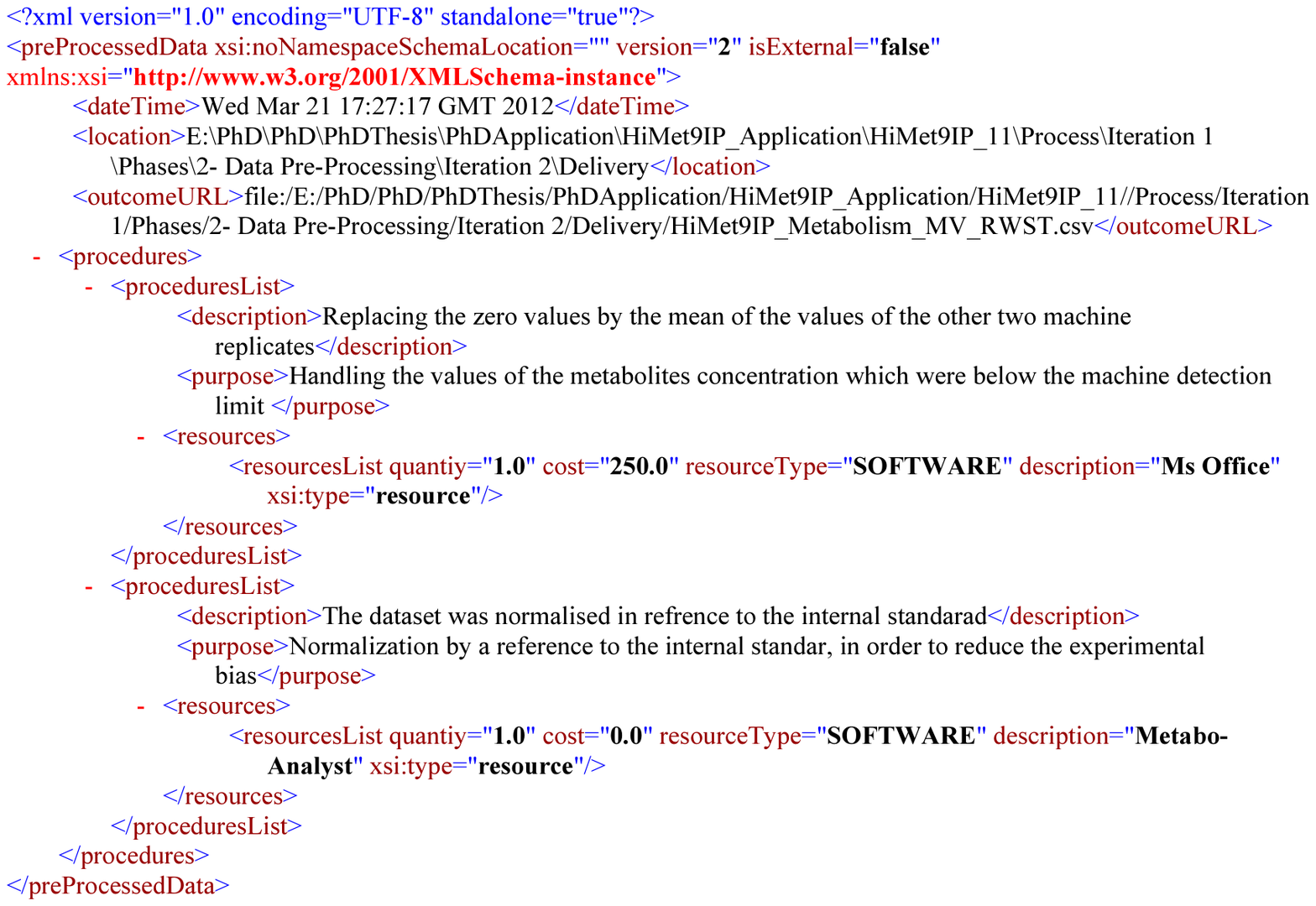}
\includepdf{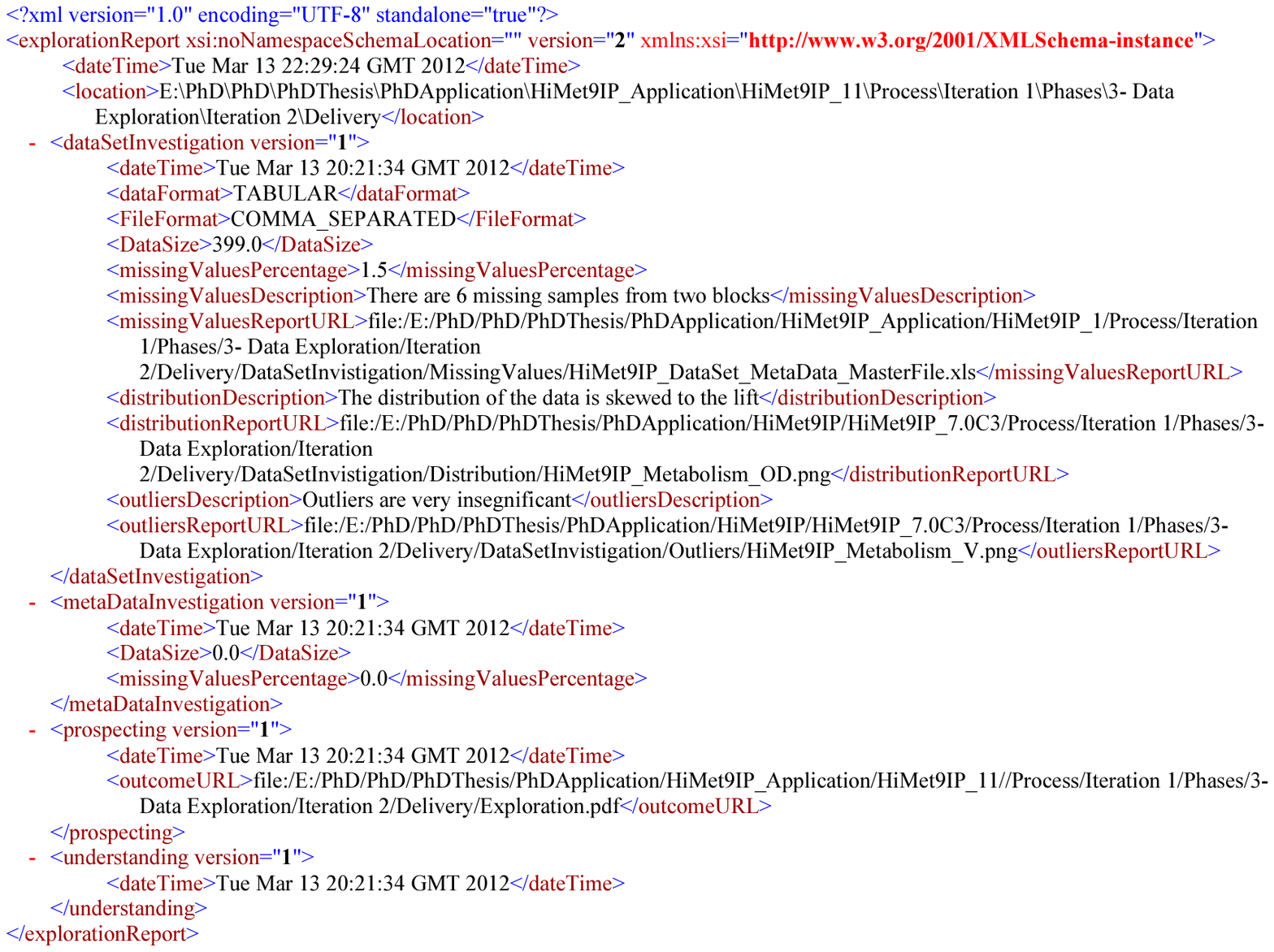}
\includepdf{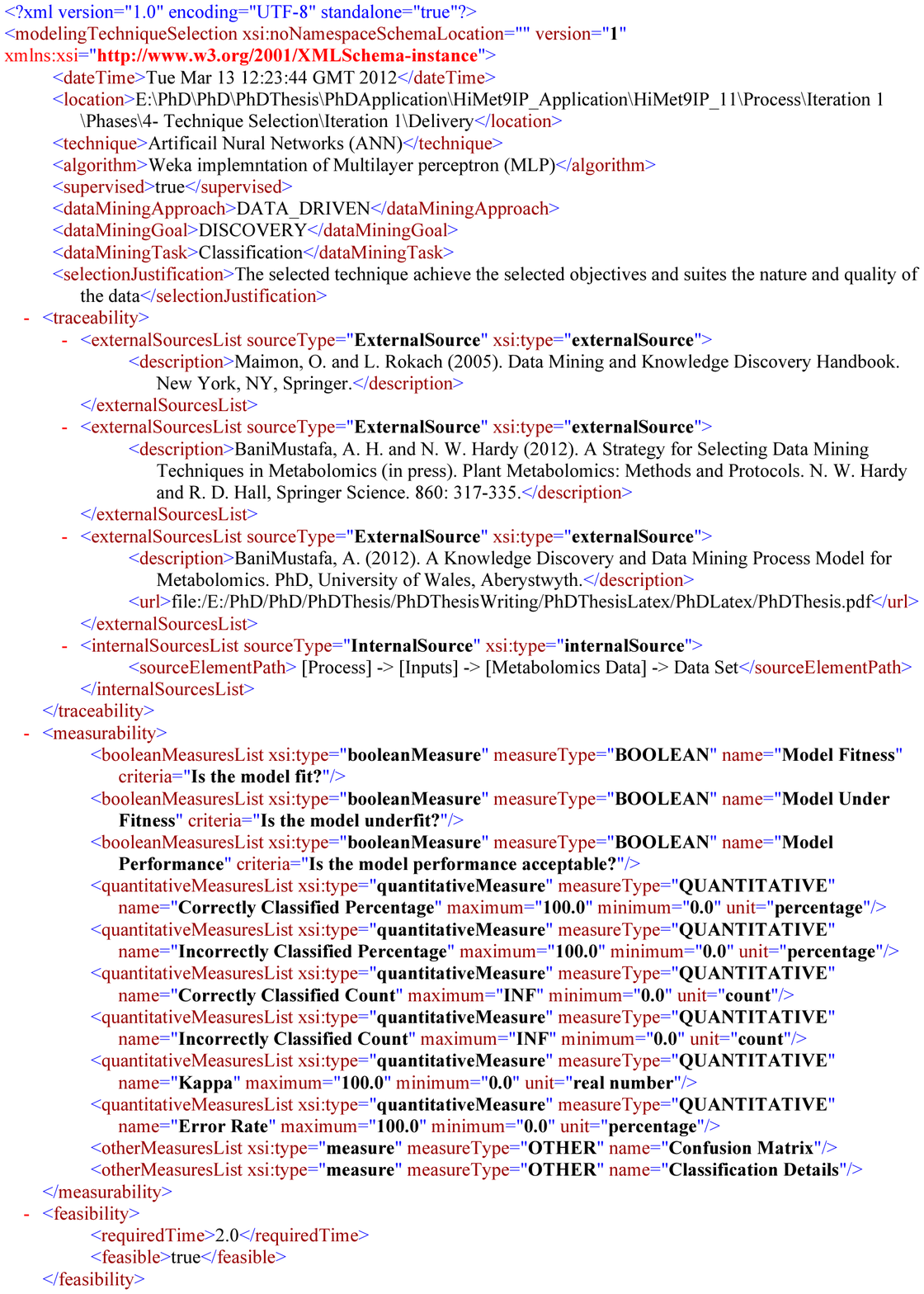}
\includepdf{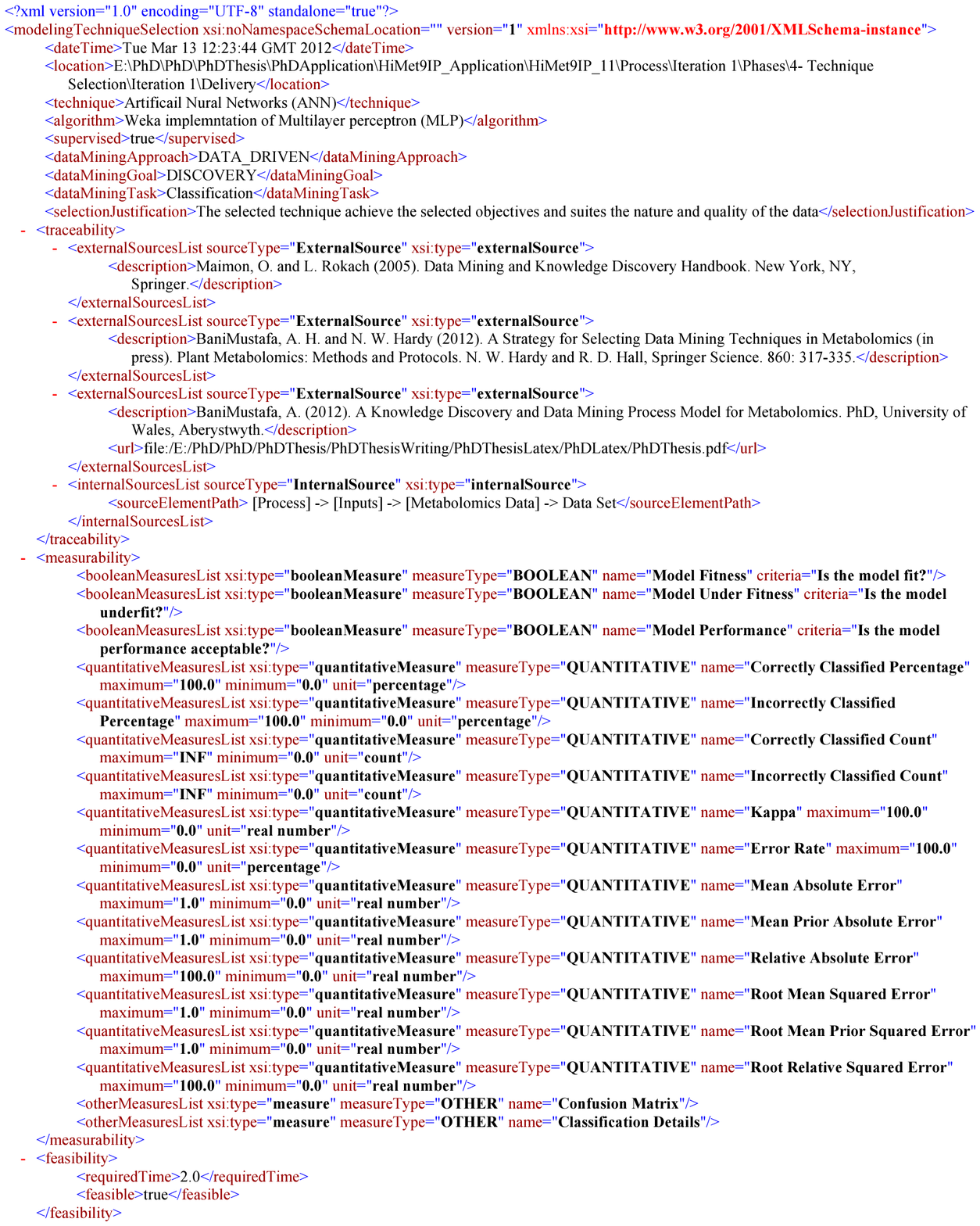}
\includepdf{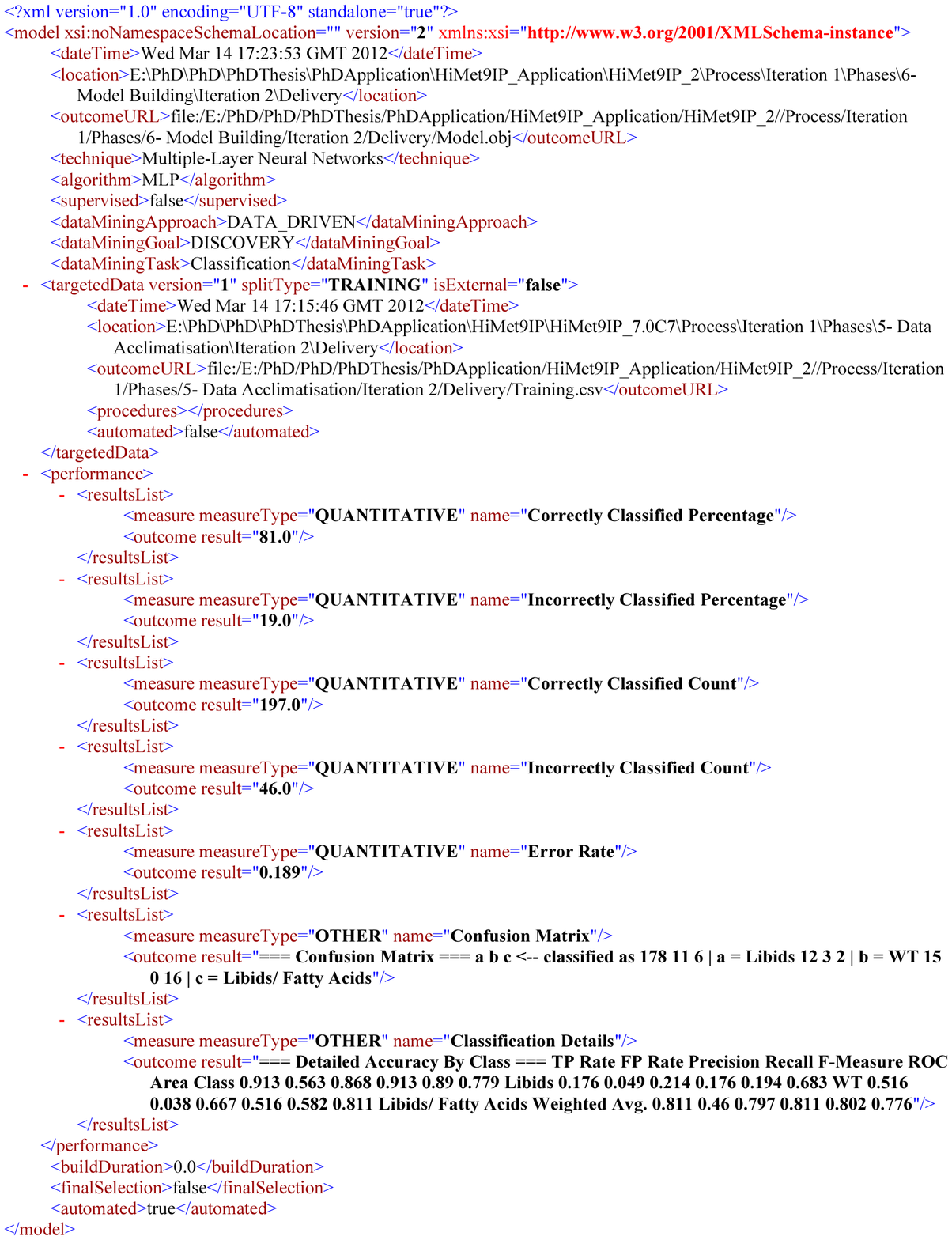}
\includepdf{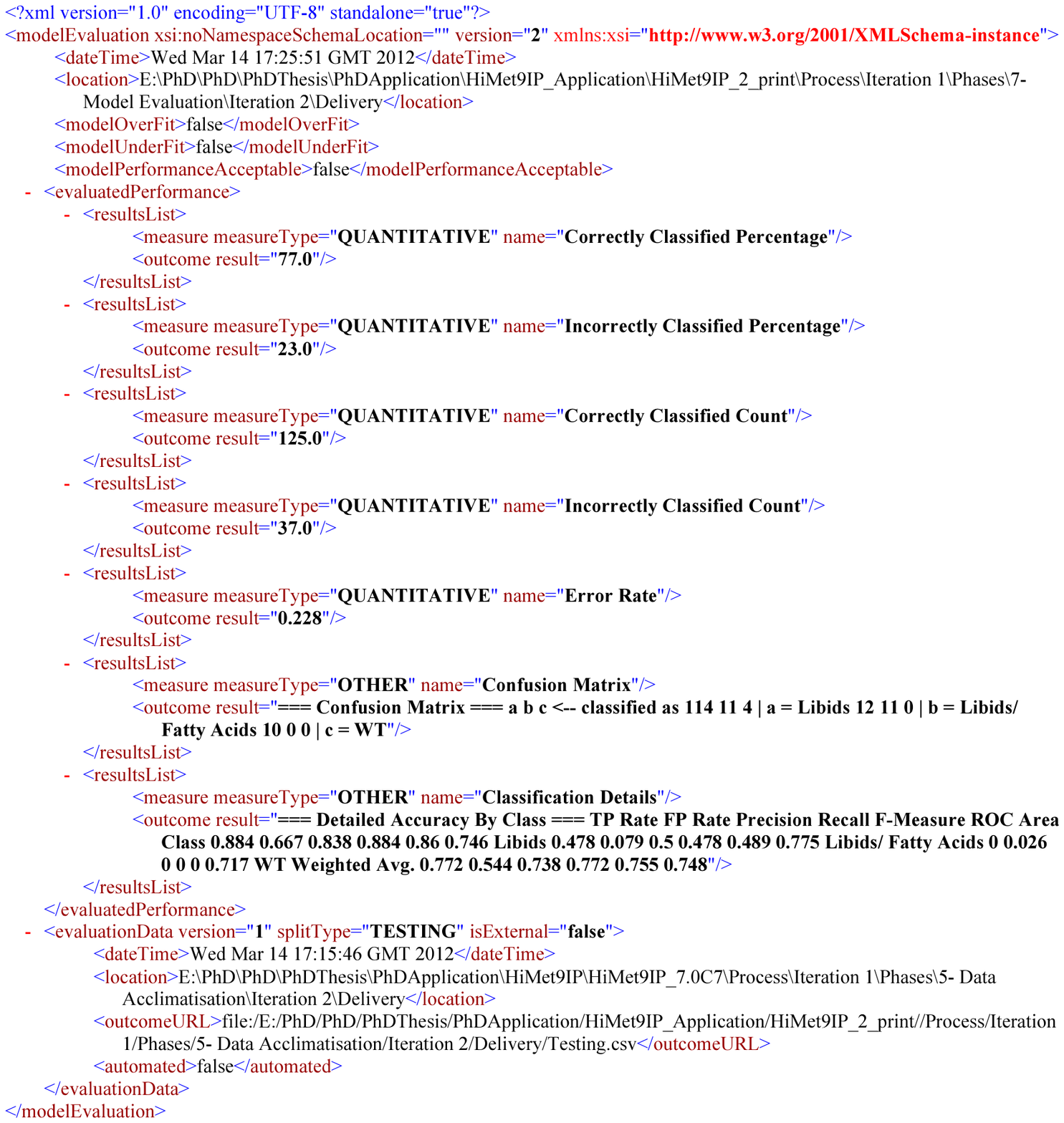}
\includepdf{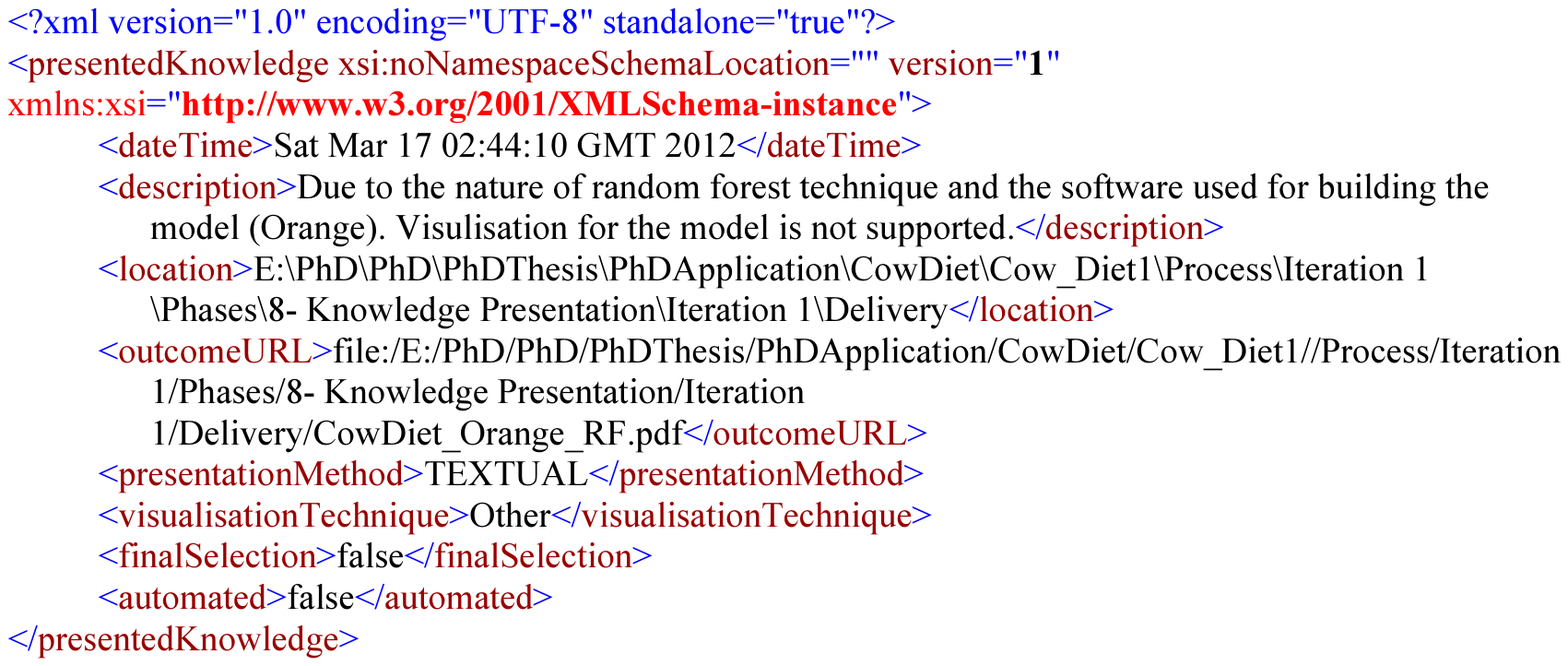}
\includepdf{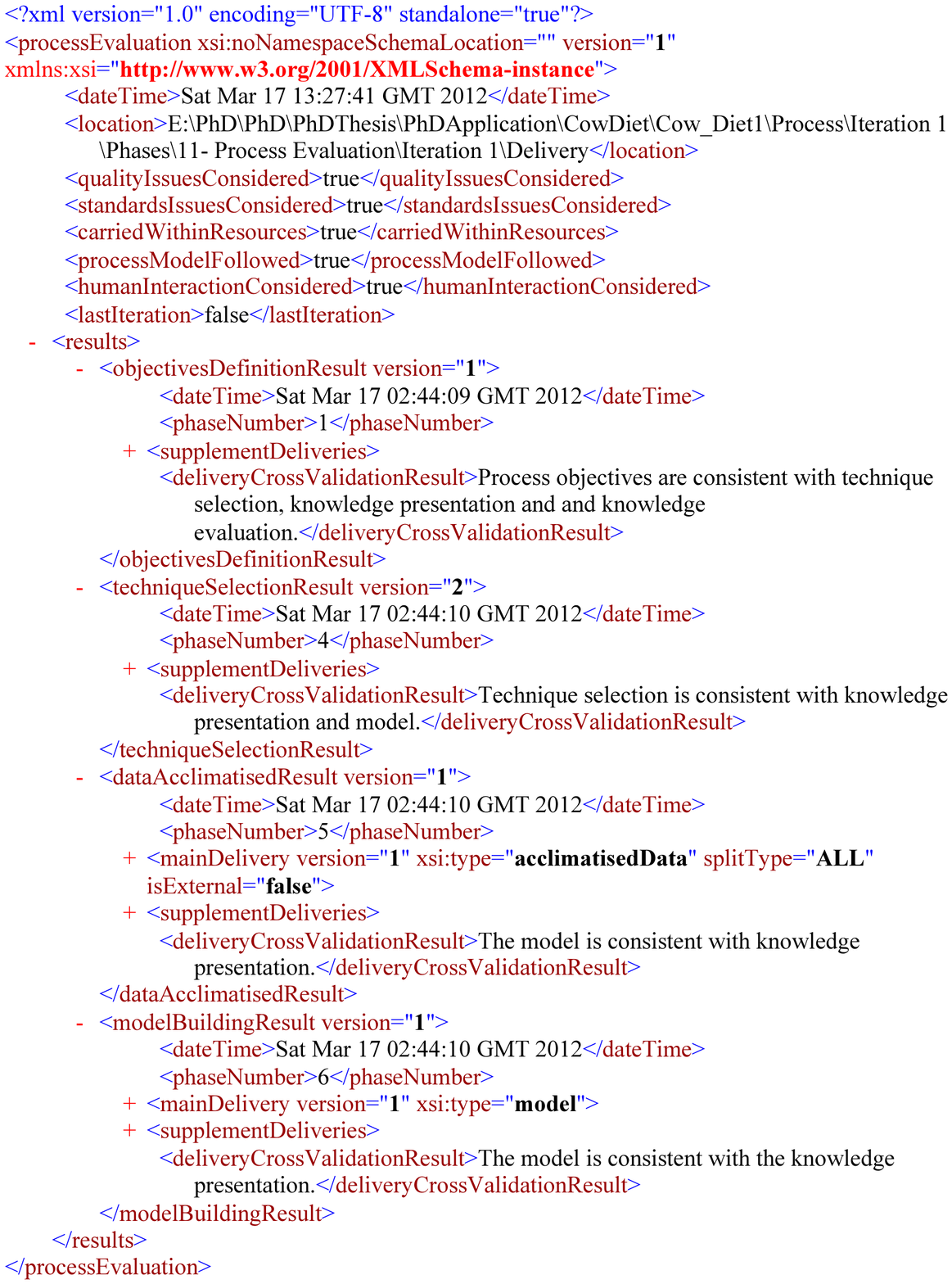}
\includepdf{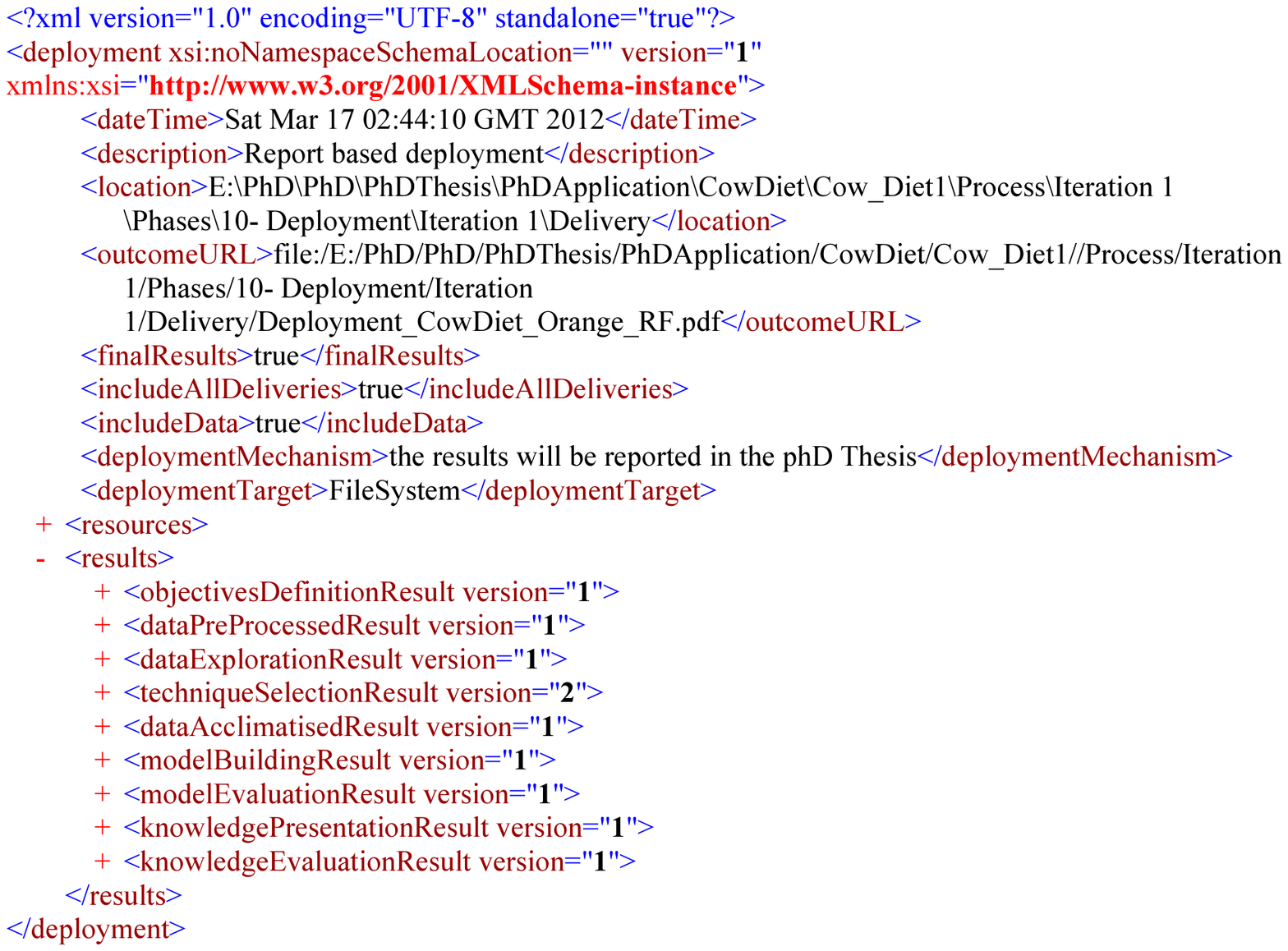}

\end{document}